\documentclass[useAMS,usenatbib]{mn2e}
\usepackage{amssymb,latexsym,graphicx,natbib,times}

\title[NGC~588 with PMAS]{A 2D multiwavelength study of the ionized gas and stellar population in the Giant HII Region NGC~588}

\author[A. Monreal-Ibero et al.]
{A. Monreal-Ibero$^{1,2}$\thanks{E-mail:
ami@iaa.es (AM-I)}\thanks{Based on observations
   collected at the German-Spanish Astronomical Center, Calar Alto,
    jointly operated by the Max-Planck-Institut f\"ur Astronomie
    Heidelberg and the Instituto de Astrof\'{\i}sica de Andaluc\'{\i}a
    (CSIC).},
M. Rela\~no$^{3}$,
C. Kehrig$^{2}$,
E. P\'erez-Montero$^{1}$,
 \newauthor
J. M. V\'{\i}lchez$^{1}$,
A. Kelz$^{2}$,
M. M. Roth$^{2}$,
O. Streicher$^{2}$\\
$^{1}$\textbf{Instituto de Astrof\'{\i}sica de Andaluc\'{\i}a (CSIC), Glorieta de la Astronom\'{\i}a, s/n, 18008 Granada, Spain}\\
$^{2}$\textbf{Astrophysikalisches Institut Potsdam, innoFSPEC Potsdam, An der Sternwarte 16, D-14482, Potsdam, Germany}\\
$^{3}$\textbf{Dpto. de F\'{\i}sica Te\'orica y del Cosmos, Universidad de Granada, Campus Fuentenueva, Granada, Spain}\\
}

\newcommand{\gsim}{\hbox{\rlap{\lower.55ex\hbox{$\sim$}} \kern-.3em
\raise.4ex \hbox{$>$}}}
\newcommand{\lsim}{\hbox{\rlap{\lower.55ex\hbox{$\sim$}} \kern-.3em
\raise.4ex \hbox{$<$}}}

\newcommand{\nha}{\textsc{[N\,ii]}$\lambda$6584/H$\alpha$}
\newcommand{\sha}{\textsc{[S\,ii]}$\lambda\lambda$6717,6731/H$\alpha$}

\newcommand{\ohb}{\textsc{[O\,iii]}$\lambda$5007/H$\beta$}
\newcommand{\oiii}{\textsc{[O\,iii]}$\lambda$5007}

\newcommand{\hb}{H$\beta$}
\newcommand{\ha}{H$\alpha$}
\newcommand{\nii}{\textsc{[N\,ii]}$\lambda$6584}

\newcommand{\sii}{\textsc{[S\,ii]}$\lambda\lambda$6717,6731}
\newcommand{\oii}{\textsc{[O\,ii]}$\lambda\lambda$3726,3729}
\newcommand{\oiiioii}{\textsc{[O\,iii]}$\lambda\lambda$4959,5007/\textsc{[O\,ii]}$\lambda\lambda$3726,3729}
\newcommand{\niisii}{\textsc{[N\,ii]}$\lambda$6584/\textsc{[S\,ii]}$\lambda\lambda$6717,6731}
\newcommand{\niioii}{\textsc{[N\,ii]}$\lambda$6584/\textsc{[O\,ii]}$\lambda\lambda$3726,3729}

\begin{document}

\date{revised version}

\pagerange{\pageref{firstpage}--\pageref{lastpage}} \pubyear{2010}

\maketitle

\label{firstpage}

\begin{abstract}
Giant \textsc{H\,ii} regions (GHIIRs) in nearby galaxies  are a local sample in which we can
study in detail processes in the interaction of gas, dust, and newly formed stars
which are analagous to those which occurred in episodes of higher intensity in
which much of the current stellar population was born.
%
Here, we present an analysis of NGC~588, a GHIIR in M33,  based on optical Integral Field Spectroscopy (IFS) data obtained with the PMAS instrument at the 3.5~m telescope of Calar Alto Observatory, CAHA, together with \emph{Spitzer} infrared images at 8~$\mu$m and 24~$\mu$m.
The extinction distribution measured in the optical shows
complex structure, with three maxima which correlate in position with those of the emission at 24~$\mu$m and 8~$\mu$m. Furthermore, the \ha\ luminosity absorbed by the dust within the \textsc{H\,ii} region reproduces the structure observed in the 24~$\mu$m image, supporting the use of the 24~$\mu$m band as a valid tracer of recent star formation.
A velocity difference of $\sim$50~km~s$^{-1}$  was measured between the areas of high and low surface brightness, which would be expected if NGC~588 were an evolved GHIIR.
%
We have carefully identified the areas which contribute most to the line ratios measured in the integrated spectrum.
Those line ratios which are used in diagnostic diagrams proposed by \citet{bal81} show a
larger range of variation in the low surface brightness areas.
The ranges are  $\sim$0.5 to 1.2~dex for \nha, 0.7 to 1.7~dex for \sha, and 0.3 to 0.5~dex for \ohb, with higher values of \nha\ and \sha, and lower values of \ohb\ in the areas of lower surface brightness.
Ratios corresponding to large ionization parameter ($U$) are found between the peak of the emission in \hb\ and the main ionizing source decreasing radially outwards within the region. Differences between the integrated and local values of the $U$ tracers can be as high as $\sim$0.8~dex, notably when using \oiiioii\ and in the high surface brightness spaxels. \textsc{[O\,ii]}$\lambda\lambda$3726,3729/\hb\ and \oiiioii\ yield similar local values for the ionization parameter, which are consistent with those expected from the integrated spectrum of an \textsc{H\,ii} region ionized by a single star.
The ratio \sha\ departs significantly from the range predicted
by this scenario, indicating the complex ionization structure in GHIIRs.
There is a significant scatter in derivations  of the metallicity using strong line tracers as a function of position, caused by variations in the degree of ionization. The scatter is smaller for $N2O3$ which points to this tracer as a better metallicity tracer than $N2$.
One interesting result emerges from our comparison between integrated and local line ratio values:
measurements of the line ratios of GHIIR in galaxies at distances $\gsim$25~Mpc may be dominated by the ionization conditions in their low surface brightness  areas.
\end{abstract}


\section{Introduction}

Large areas of ionized gas known as Giant \textsc{H\,ii} regions (GHIIRs)
constitute the most conspicuous places of star formation in normal
galaxies \citep[see][for a review]{shi90}.
Their diameters typically range between $\sim$100~pc to $\sim$800~pc
\cite[e.g.][]{ken84,alo02} while their \ha\ luminosity range expands up
to 3 orders of magnitudes \citep[$\sim10^{38} - 10^{40}$~erg~s$^{-1}$,
  e.g.][]{ken84,roz96,fir05,mon07,rel09}.
Regarding their morphologies, some of them present a compact distribution with high
surface brightness while 
others have a more diffuse emission. Also, they can present 
multiple cores and/or shells or ring-like features.
In the same manner, they present
very different content of gas, varying between $\sim10^3$~M$_\odot$ to
almost $10^7$~M$_\odot$ while they usually have
$\sim10^2-10^5$~M$_\odot$ in stars \citep[e.g.][]{cas02}.
Finally, GHIIRs are small-scale examples of the extreme events of star
  formation occurring in starburst galaxies  \citep[e.g.][]{alo09,gar09}. Thus, a good
  knowledge of these objects is highly valuable 
for a better understanding of these more violent phenomena.

The variety of properties in GHIIRs implies a complex structure, far from the textbook-like 
Str\"omgren sphere. Moreover, modelling of star-forming regions showed
that in case of multiple ionizing sources, geometrical effects affect the physical
properties (i.e electron temperature and ionization structure) of
these regions \citep{erc07,jam08}.
The inhomogeneities of the Interstellar Medium (ISM) have been taken into account within the \textsc{H\,ii} regions using the filling factor. This describes the fraction of the total volume of the \textsc{H\,ii}  region with high dense gas while the remaining volume is considered to be of negligible density \citep{ost59}. Recently, detailed models show that the density variations assuming optically thick high density gas clumps give rise to inhomogeneities in the temperature and ionization parameter \citep{gia04,gia05}.
Thus, a single value per physical
magnitude, usually extracted from a specific area of the region, is
not necessarily representative of its physical conditions.

It is in very nearby  GHIIRs (i.e. at $D\lsim$6~Mpc), where ground based
optical telescopes can achieve high linear 
spatial resolution (i.e. $\lsim$40~pc arcsec$^{-1}$), enough to resolve the different elements
(i.e. star clusters, ionized gas, dust, etc.), playing a role in the interaction 
between the massive stars and its surrounding
environment. Also, it is there where the variations of the physical and chemical properties of the ionized gas can be properly sampled.

From the observational point of view, the study of these regions would
benefit from high quality spectroscopic data that map in an un-biased
way the surface of the GHIIR. 
Nowadays, the technique of integral field
spectroscopy (IFS), able to record simultaneously the spectra of an
extended continuous field, offers the possibility of performing such a mapping.
At present,
studies of GHIIRs based on IFS are still scarce.
An example is provided by \citet{gar10}, who 
analyzed one of the brightest GHIIRs in NGC~6946.
%
Also, \citet{lop10} presented a detailed study of a star-forming region at the lower limit of GHIIRs in terms of \ha\ luminosity and size in IC~10, our closest starburst. 
Recently, we presented a 2D spectroscopic analysis of the second
brightest \textsc{H\,ii} region in M~33, the \emph{Triangulum Galaxy}: NGC~595 \citep{rel10}. There,
we showed how the optical extinction map and
the absorbed \ha\ luminosity  are spatially correlated with the
24$\mu$m emission from \emph{Spitzer}, and how the ionization structure
of the region nicely follows the \ha\ shell morphology.  Moreover, we
evaluated the reliability of different line ratios as metallicity ($Z$) tracers.
In a companion paper, we presented a novel approach to model these complex structures. There, we reproduced our observations by jointly fitting the radial profiles of different optical (i.e. \ha, \oii, \oiii) and infrared (i.e. 8~$\mu$m, 24~$\mu$m) magnitudes to a set of CLOUDY-based photoionization models \citep{per10}.

Our experience with NGC~595 shows the importance of carrying out an
analysis using the combined information of optical and infrared data
together with modelling. However, given their diversity, the sample of
GHIIRs studied by means of this methodology cannot be reduced to only 
one example.
Instead, studies of other regions sampling a different range in the parameter space would be desirable. We present here, the analysis of a region with a different morphology, and relatively lower metallicity and high \ha\ luminosity: NGC~588.
This region is located in the
outskirts of M~33, at the end of a spiral 
arm, at a radius of $\sim$14$^{\prime}$ \citep[i.e. $\sim3.42$~kpc
for a distance to M33 of 840 kpc;][]{fre91}.
This area presents \textsc{H\,i} emission \citep{ver10,gra10}. However, no local H$_2$ (or CO) emission has been detected towards NGC~588 \citep{ver10,isr90}.
With a size of
$\sim30^{\prime\prime}\times50^{\prime\prime}$
(i.e.$\sim$120~pc$\times$200~pc at our assumed distance), NGC~588 has  
been classified within the ring-like class \citep{sab80}. Its stellar
content has been thoroughly studied \citep{jam04,pel06,ube09}. The
different estimates for its
total stellar mass range between $\sim1.3\times10^3$ and
$\sim5.6\times10^3$~M$_\odot$ with an age for the burst of
$\sim3.5-4.2$~Myr and low metallicity (i.e. $Z\sim0.4Z_\odot$),
consistent with direct measurements using 
long-slit \citep[12+log(O/H) = 8.30,][]{vil88}. NGC~588 contains two
Wolf-Rayet (WR) stars. One of them was classified as WNL 
while the other as Ofpe/WN9, an intermediate object between Of and WN
stars \citep{mas96}. 
From the kinematic point of view, TAURUS-2 Fabry-P\`erot data show how
NGC~588 seems a relatively evolved system, with its ionized gas
kinematics dominated by a 
collection of large stellar wind shells \citep{mun96}.
Finally, a relatively faint and point-like X-ray emitting source associated with this region has been detected \citep{plu08}.


In this work, we combine 2D optical spectroscopic
observations with the \emph{Potsdam Multi-Aperture Spectrophotometer}
\citep[PMAS,][]{rot05} and infrared imaging with \emph{Spitzer}. 
These will constitute the first published results from data obtained with
the new PMAS's CCD. A modelling of NGC~588 intending to reproduce the observed magnitudes will be presented in a companion paper (P\'erez-Montero et al. in preparation).
We describe the observations and data processing to create the maps of the relevant magnitudes in Section \ref{secobserva}. The main observational results are presented in Section \ref{secresults}. Finally, we summarize our main conclusions in Section \ref{secondly}.

\begin{figure}
\includegraphics[width=0.45\textwidth, clip=,bbllx=100, bblly=245,
  bburx=480, bbury=535]{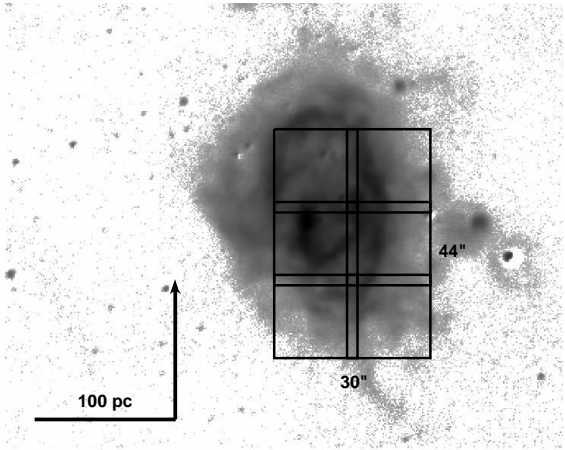}
 \caption{Mosaic to map NGC~588 overplotted on a continuum-subtracted
H$\alpha$ direct image from NOAO Science Archive \citep{mas07}. The
orientation is north up and east to the left. The H$\alpha$ image is shown in
logarithmic stretch to better enhance all the morphological features of the
\textsc{H\,ii} region and covers a range of 3.4~dex. \label{apuntado}}
\end{figure}

\section{Observations and Data Reduction}
\label{secobserva}
 
\subsection{Observations}

The IFS data of NGC~588 were obtained on October 9-10, 2009 during the
commissioning run of the new \emph{Potsdam Multi-Aperture Spectrophotometer}
\citep[PMAS,][]{rot05} CCD at the 3.5 m telescope at the Calar Alto Observatory
(Spain). The new PMAS 4k x 4k CCD is read out in four quadrants which have slightly different gains \citep{rot10}.
Data were taken using the Lens Array Mode (LARR) configuration which
is made out of a $16\times16$ array of microlenses coupled with fibres (hereafter
\emph{spaxels}). We used the 1$^{\prime\prime}$ magnification which provides a
field of view (FoV) of $16^{\prime\prime}\times
16^{\prime\prime}$.  We used the V600 grating and the 2$\times$2
binning mode achieving an effective dispersion of 1.59~\AA~pix$^{-1}$, and a
$\sim$3.4~\AA\ full width half maximum spectral resolution. With the new
PMAS's CCD, the covered spectral range was from 3\,620 to 6\,800~\AA\ for most
of the spaxels. This permits us to observe the main emission lines in the optical
from \oii\ to \sii. However, for $\sim$40 spaxels, always at the edge of the
LARR, this spectral range was slightly reduced due to the vignetting associated
with the 3.5~m telescope. This prevents us from obtaining information for the
\oii, \sii, and \ha\ emission lines in some specific areas (see also section
\ref{reduccion}). 

We made a mosaic of 6 tiles to map most of the surface of NGC~588. The
distribution of the different tiles is shown in Fig. \ref{apuntado}
overplotted on the \ha\ emission-line image from National Optical Astronomy
Observatory (NOAO) Science Archive \citep{mas07}. Contiguous tiles had a
2\farcs0 overlapping, to make easier a common relative flux calibration of the
data. In total, we covered a field of 30\farcs0$\times$44\farcs0 which at the
distance of NGC~588 corresponds to $\sim$120~pc$\times$180~pc.

We obtained three exposures of 400~s per
tile. Atmospheric conditions during the observations were non-photometric and
typical seeing ranged between 1\farcs2 and 1\farcs6.  The sky
transparency during the observing nights presented variations of $\lsim$15 $\%$.
All the data were taken at airmasses \lsim1.1 in order to prevent strong effects
due to differential atmospheric refraction.

In addition to the science frames, continuum and HgNe arc lamp
exposures in order to minimize the effects due to instrument flexures.
Also, a nearby sky background frame was obtained during the second night by moving the IFU off-target.

Finally, exposures of the spectrophotometric standard star
BD+28D4211 were obtained in order to correct for the instrument response and
perform a relative flux calibration.

\subsection{Data reduction and map creation \label{reduccion}}

The first steps of the data reduction were done through the P3d tool that is
designed to be used with fibre-fed integral-field spectrographs \citep{san10}.
After trimmimg, combining the four quadrants and subtracting the bias level, the expected locations of the spectra were
traced on a continuum-lamp exposure obtained before each target exposure.  We
extracted the target spectra by adding the signal from the central traced pixels ant its four neighbours. The spectra were wavelength calibrated using the exposures
of HgNe arc lamps obtained immediately after the science exposures. We checked the accuracy of the wavelength
calibration using the \textsc{[O\,i]}$\lambda$5577~\AA\ sky line, and found standard
deviations of $<$0.14\AA, which allowed us to determine the centroid of line with an accuracy of $\sim$7~km~s$^{-1}$ at $\sim$5\,000~\AA.

Fibres have different transmissions that may depend on the wavelength. The continuum-lamp exposures were used to determine the differences in the fibre-to-fibre transmission and to obtain a normalized fibre-flat image, including the wavelength dependence. This step was carried out by running the \texttt{fiber\_flat.pl} script from the R3D package \citep{san06}.  In order to homogenize the response of all the fibres, we divided our wavelength calibrated science images by the normalized fibre-flat. 
To estimate the accuracy of the fibre-to-fibre response correction, we fitted a Gaussian to four emission lines distributed along the whole spectral range in an extracted, wavelength calibrated and flat-field corrected arc exposure. We used the ratio between the standard deviation and the mean flux in each line as a proxy for the accuracy of the fibre-to-fibre response correction. For those lines in the central part of our spectral range, this correction was very good, with ratios of $\sim$2\%. In the blue and red edges
these ratios reached values of $\sim$12\%, due to the contribution of fibres affected by vignetting.

In the next step, the three exposures taken for the same pointing were combined in order to remove cosmic rays, using the \texttt{imcombine} routine
in IRAF.\footnote{IRAF is distributed by the National Optical Astronomical
Observatories, which are operated by the Association of Universities for
Research in Astronomy, Inc., under cooperative agreement with the National
Science Foundation.} Flux calibration was performed using the IRAF tasks
\texttt{standard}, \texttt{sensfunc} and \texttt{calibrate}. We co-added the
spectra of the central fibres of the standard star exposure to create 
a one dimensional spectrum that was used to obtain the sensitivity function.
For wavelengths larger than 4\,000~\AA, the uncertainty in the flux calibration is $\sim$
1\%, while across the bluer spectral range (i.e. $<$4\,000 \AA), the
associated error can reach $\sim$ 5\%.

Given the large size of the new PMAS's CCD, the four corners of each exposure suffer from telescope vignetting \citep{rot10}. Due to the way the fibres of the LARR are arranged at the entrance of the spectrograph, a maximum of two columns of fibres at the east and west sides of the LARR were affected by this. The wavelength range affected was larger for those fibres located more towards the edge of the spectrograph and could be up to $\sim$3\,898~ \AA\ in the blue end and from $\sim$6\,498~\AA\ in the red end. These parts of the spectra were masked  and then we used the offsets commanded to the telescope and the PMAS acquisition images to construct a mosaic datacube.

After creating this datacube, maps for the different observables were derived following the methodology presented in \citet{rel10}. Basically, we performed a Gaussian fit to the emission lines using the IDL-based routine \texttt{mpfitexpr} \citep{mar09} and derived the quantities of interest for each individual spaxel. Then, we used these together with the position of the spaxels within the datacube to create an image suitable to be manipulated with standard astronomical software. Hereafter, we will refer to this with both terms: \emph{map} and \emph{image}. 

Even after having allowed for 2$^{\prime\prime}$ of overlap between tiles, vignetting prevents us from deriving information for some lines in the central columns of our mosaic. In those maps where (at least) one line affected by vignetting was involved, those spaxels were masked and interpolated using \texttt{fixpix} within IRAF, for presentation purposes.

\section[]{Results }
\label{secresults}

\subsection{Integrated properties \label{sec_integrado}}

\begin{figure*}
\includegraphics[width=0.96\textwidth]{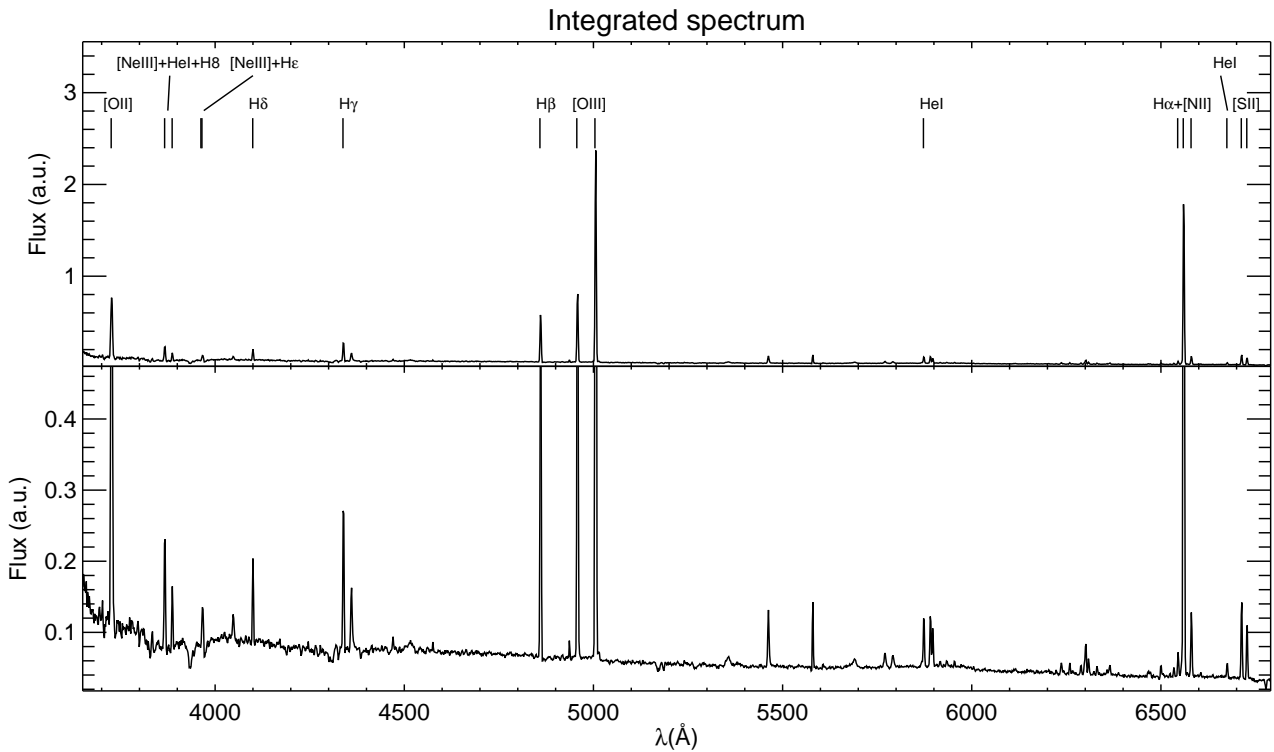}
 \caption{Integrated spectrum of NGC~588 obtained by co-adding the
   signal of all the spaxels in the field of view with two different
   normalizations to better visualize the different observed emission
   lines. Fluxes are in arbitrary units.  The sky spectrum was created by combining the spectra
corresponding to the vigneting-free fibres of the background frame. \label{espectro_total} } 
\end{figure*}

\begin{table}
\begin{center}
\caption{Observed and extinction-corrected emission line fluxes, normalized to I(\hb)=1
  from the integrated spectrum of NGC~588.
f($\lambda$) is the reddening function normalised to \hb\ from \citet{flu94}. } 
\scriptsize
\begin{tabular}{cccccccccc}
\hline\noalign{\smallskip}
Emission Line &  $\lambda_{\mathrm{obs}}$ & f($\lambda$ ) &  I($\lambda$)/I(\hb) &  I($\lambda$)/I(\hb)$_{\mathrm{corr}}$ \\
\hline\noalign{\smallskip}
3727 \textsc{[O\,ii]}      &   3725.52 & 0.255         & 1.922$\pm$0.048 & 2.317 $\pm$ 0.091       \\
3868 [Ne\,\textsc{iii]}    &   3866.01 & 0.227         & 0.389$\pm$0.013 & 0.461 $\pm$ 0.020       \\
3889 H8+He\textsc{i}       &   3886.21 & 0.223         & 0.423$\pm$0.014 & 0.501 $\pm$ 0.021    \\
3970 H$\epsilon$+[Ne\,\textsc{iii]}  &3966.03  &0.208  & 0.152$\pm$0.011 & 0.179 $\pm$ 0.013   \\
4101 H$\delta$    &   4099.02 & 0.180                  & 0.199$\pm$0.013 & 0.274 $\pm$ 0.016  \\
4340 H$\gamma$    &   4338.19  & 0.133                 & 0.397$\pm$0.010 & 0.451 $\pm$ 0.020    \\
4861 H$\beta$     &  4859.46   & 0.000                 & 1.000$\pm$0.016 & 1.000 $\pm$ 0.021    \\
4959 \textsc{[O\,iii]}  &  4957.02   &  -0.027         & 1.500$\pm$0.021 & 1.524 $\pm$ 0.046     \\
5007 \textsc{[O\,iii]}  &  5004.90  &  -0.041          & 4.400$\pm$0.049 & 4.428 $\pm$ 0.131 \\
5876 He\,\textsc{i}   & 5872.40  &  -0.217             & 0.154$\pm$0.004 & 0.140 $\pm$ 0.007 \\
6563 \ha   &  6559.34  &     -0.314                    & 3.508$\pm$0.034 & 2.938 $\pm$ 0.114  \\
6584 \textsc{[N\,ii]}  &  6579.40  & -0.316            & 0.246$\pm$0.004 & 0.210 $\pm$ 0.008  \\
6678 He\,\textsc{i}    &  6674.29  & -0.329            & 0.036$\pm$0.002 & 0.031 $\pm$ 0.002  \\
6717 \textsc{[S\,ii]}  &  6712.40  & -0.332            & 0.210$\pm$0.004 & 0.178 $\pm$ 0.007 \\
6731 \textsc{[S\,ii]}  &  6726.76  &  -0.335           & 0.147$\pm$0.003 & 0.124 $\pm$ 0.005  \\
\hline\noalign{\smallskip}
\end{tabular}
\label{lambdaint}
\end{center}
\end{table}

\begin{table}
\begin{center}
\caption{Main diagnostic emission line ratios and integrated
  physical properties of NGC~588. All the physical
  properties were estimated  using our extinction corrected integrated fluxes.
\label{tablapropsinte}
} 
\begin{tabular}{ccc}
\hline\noalign{\smallskip}
Parameter &  Value$_{\mathrm{corr}}$\\ \hline\noalign{\smallskip}
log (\nha)   & -1.146$\pm$0.077\\
log (\sha)   & -0.988$\pm$0.078\\
log (\ohb)   & 0.646$\pm$0.050\\
log (\textsc{[O\,iii]}$\lambda\lambda$4959,5007/\textsc{[O\,ii]}$\lambda$3727) & 0.406$\pm$0.069\\
log $R_{\rm 23}$         & 0.915$\pm$0.053\\
N2O3        &  -1.792$\pm$0.127\\
\textsc{[S\,ii]}$\lambda$6717/\textsc{[S\,ii]}$\lambda$6731       &1.435$\pm$0.143 \\
log(\niioii)   &  -1.043$\pm$0.077\\
log(\niisii)   &  -0.271$\pm$0.078\\

\hline
 c(\hb)                       &  0.26$\pm$0.03 \\
 n$_{\rm e}$ (cm$^{-3}$)      &  $<100$ \\
 q$_{\rm eff}$ (cm~s$^{-1}$)  &  $3.9\times10^7$ \\
 T$_{\rm e}$ (K)              & 11\,140$\pm$180$^{(\ast)}$\\
 12+log(O/H)                  & 8.16$\pm$0.02 \\
 \hline\noalign{\smallskip}
\end{tabular}
\end{center}
$^{(\ast)}$ Value derived from the \textsc{[O\,iii]} $\lambda $4363/$\lambda $5007 line ratio by \citet{jam05}.
\end{table}

Here we analyze the integrated spectrum for NGC~588, obtained after co-adding the signal of $\sim$1\,000~spaxels. The
sky substracted
spectrum is shown in Fig. \ref{espectro_total}, where we used two different normalization factors in order to better display all the observed emission lines. The positions of the detected nebular emission lines were marked with labels. In addition, several sky line residuals are clearly visible. Flux for the main emission lines was measured using \texttt{splot} within IRAF, which integrates the line intensity over a locally fitted continuum. As was shown in \citet{rel10}, both \texttt{splot} and \texttt{mpfitexpr} give similar results for the high signal-to-noise integrated spectrum.
We derived the reddening coefficient, c(H$\beta$), from the F(H$\delta$)/F(H$\beta$), F(H$\gamma$)/F(H$\beta$) and F(H$\alpha$)/F(H$\beta$) line ratios. We performed a linear fit (by minimizing the chi-square error statistic) to the difference between the theoretical and observed Balmer
decrements vs. the reddening law \citep{flu94}, while simultaneously
solving for the effects of underlying Balmer absorption with
equivalent width, EW$_{abs}$.  We assumed that the EW$_{abs}$
is the same for all Balmer lines \citep[e.g.][]{kob99}. The theoretical Balmer line intensities were obtained from
\citet{sto95} assuming Case B, T$_{\rm e}$ = 10$^{4}$ K,
$n_{\rm e}$=100 cm$^{-3}$ \citep[typical values found in \textsc{H\,ii} regions,][]{ost06}.


%
We derived a mean extinction for the region of $A_V = 0.548$. Given that the Galactic extinction in the line of sight to NGC~588 is $A_V^{Gal} = 0.146$ \citep{sch98} most of it is intrinsic to NGC~588 and agrees within the uncertainties with previous measurement on its brightest parts \citep[$A_V=0.49-0.81$,][]{mel79,via86,mel87}.

%
Table \ref{lambdaint} presents both the measured and extinction-corrected fluxes. Errors were estimated using the formula presented in \citet{cas02b}.

\begin{equation}
\sigma_{\mathrm{line}} = \sigma_{\mathrm{cont}} \times N^{1/2}  (1 +  \frac{EW}{N\Delta\lambda})^{1/2}
\end{equation}
where $\sigma_{\mathrm{cont}}$ is the standard deviation in a continuum close to the line of interest, $N$ is the number of pixels sampling the line, $EW$ is its equivalent width and $\Delta\lambda$ is the dispersion in \AA~pix$^{-1}$.
The quoted uncertainties of the reddening corrected line fluxes take into account the measurement and reddening errors.

\begin{figure*}
\includegraphics[width=0.48\textwidth, clip=,bbllx=20, bblly=45,
  bburx=595, bbury=750]{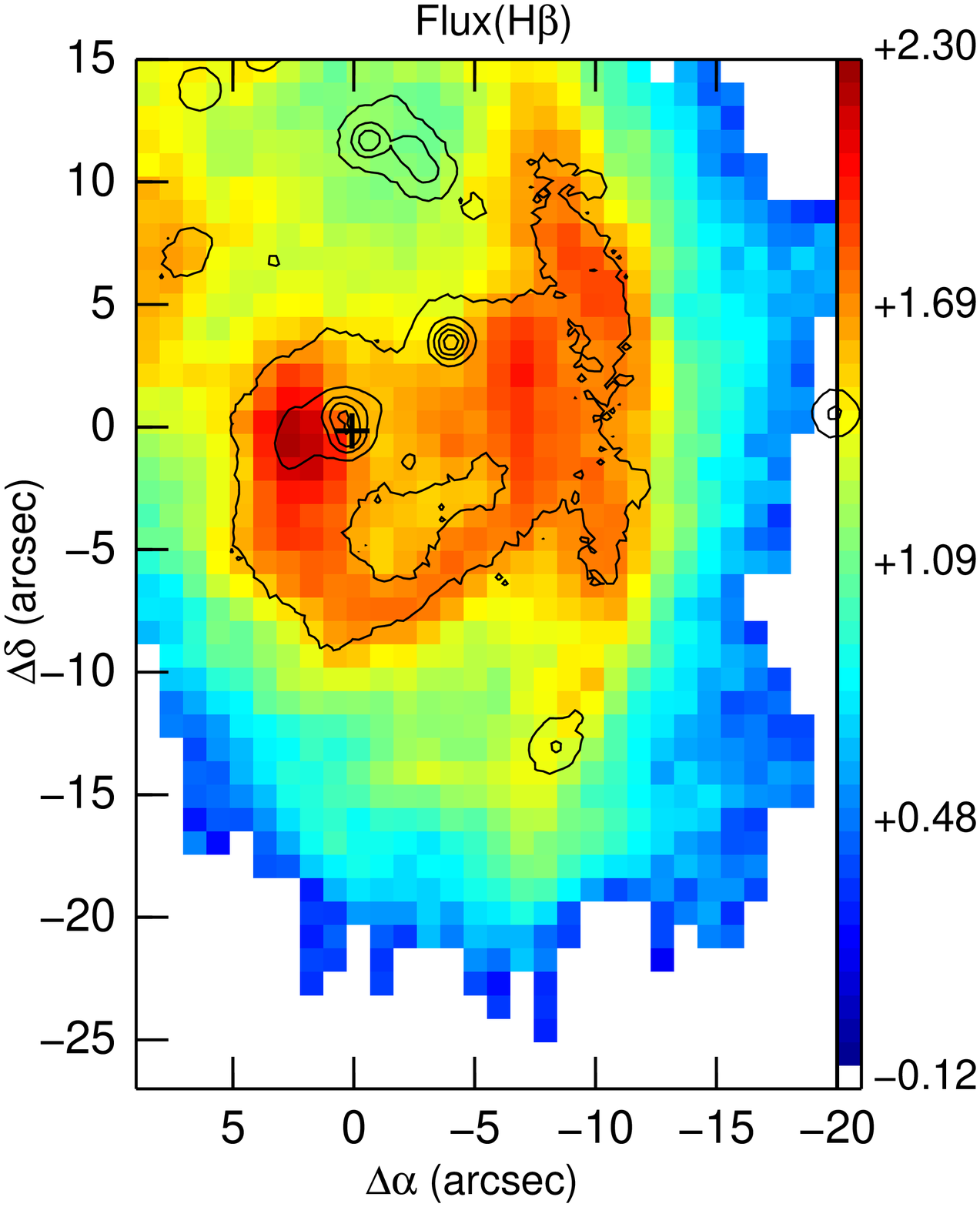}
\includegraphics[width=0.48\textwidth, clip=,bbllx=20, bblly=45,
  bburx=595, bbury=750]{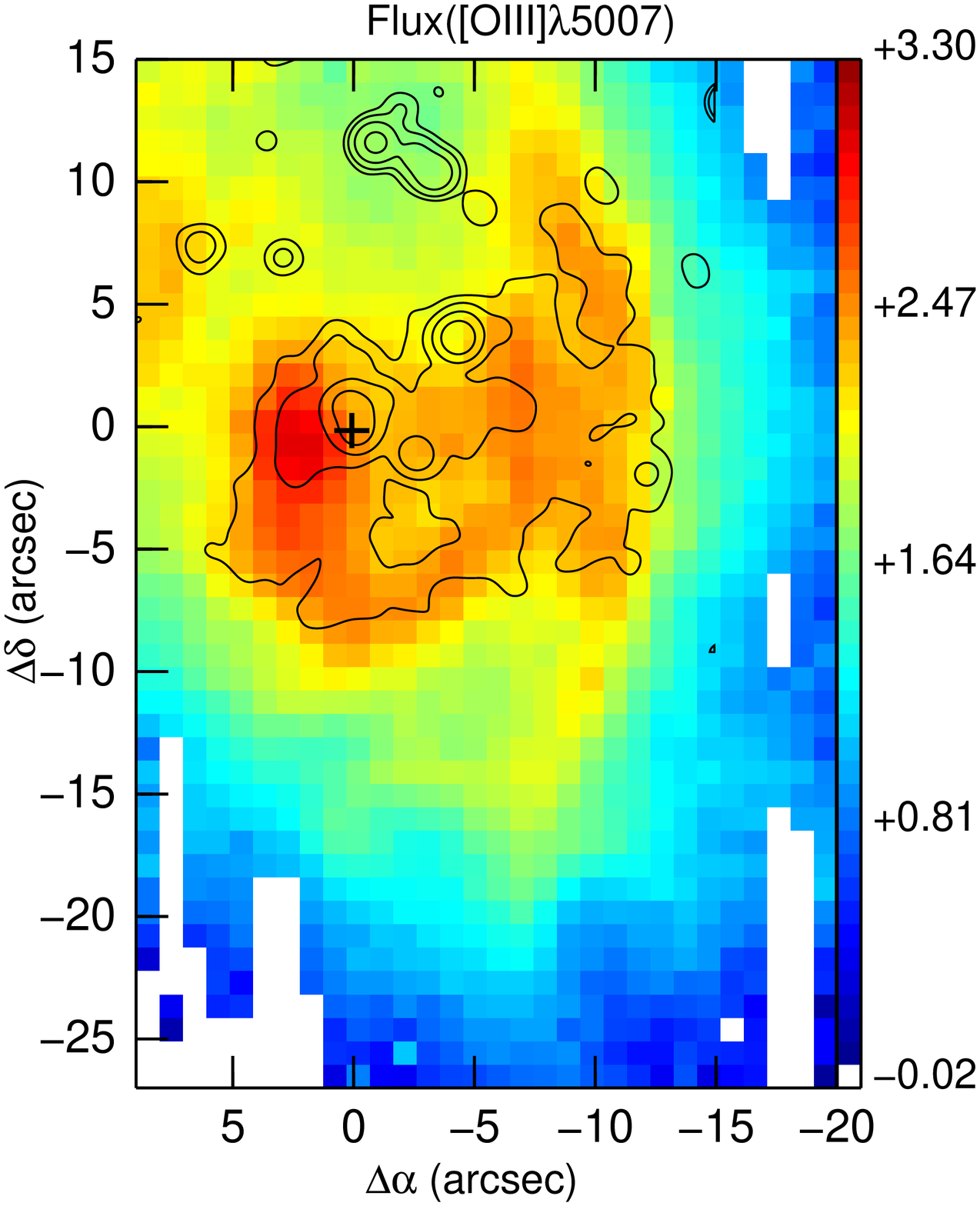}
 \caption{\emph{Left:} Map of the observed \hb\ flux derived from our PMAS
   data.
Each spaxel has a  $1^{\prime\prime}\times1^{\prime\prime}$ size. A logarithmic stretch
   covering 2.4~dex was used to better enhance all the
   morphological features. Units are
   arbitrary. Contours correspond to the $R$ band direct image from
   NOAO Science Archive \citep{mas07} and show the location of the ionizing stars within the region. The orientation is north up and 
   east to the left.
   The peak in this continuum image, at coordinates RA(J2000): 1h32m45.7s,
   Dec.(J2000): +30d38m55.1s, marks the origin of our coordinate system
   and will appear as a cross in the following figures for
   reference. \emph{Right:} Similar map for the observed \oiii\ flux. The logarithmic stretch covers 3.3~dex and contours represent the HST-WFPC2 image with the $F336W$ filter (program 5384; P.I. W. William), convolved with a 0\farcs5 Gaussian filter.
    \label{fluhb} } 
\end{figure*}

\begin{figure*}
\begin{tabular}{cc}
\includegraphics[width=0.48\textwidth, clip=,bbllx=50, bblly=85,
  bburx=435, bbury=590]{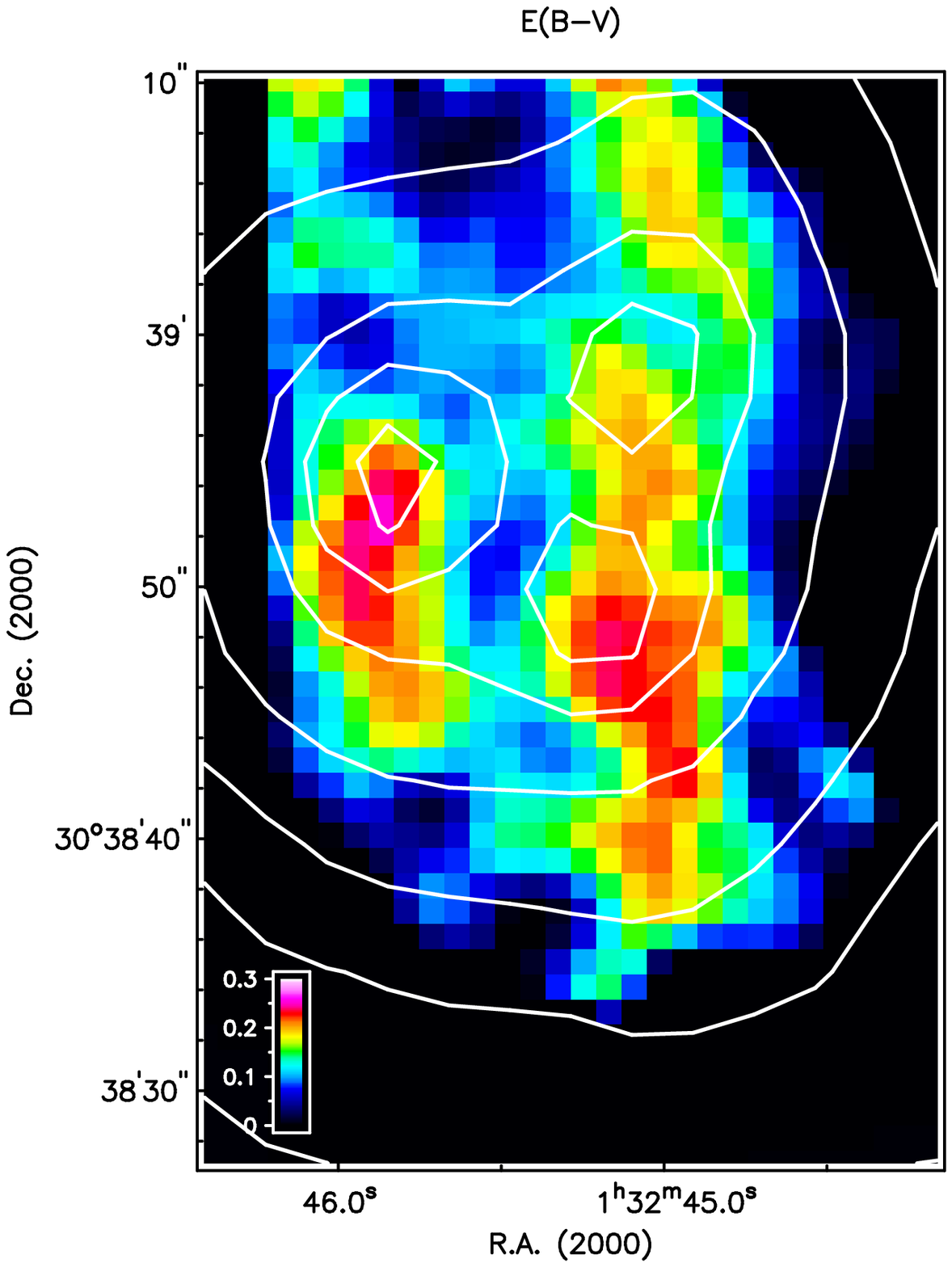} &
\includegraphics[width=0.48\textwidth, clip=,bbllx=50, bblly=85,
  bburx=435, bbury=590]{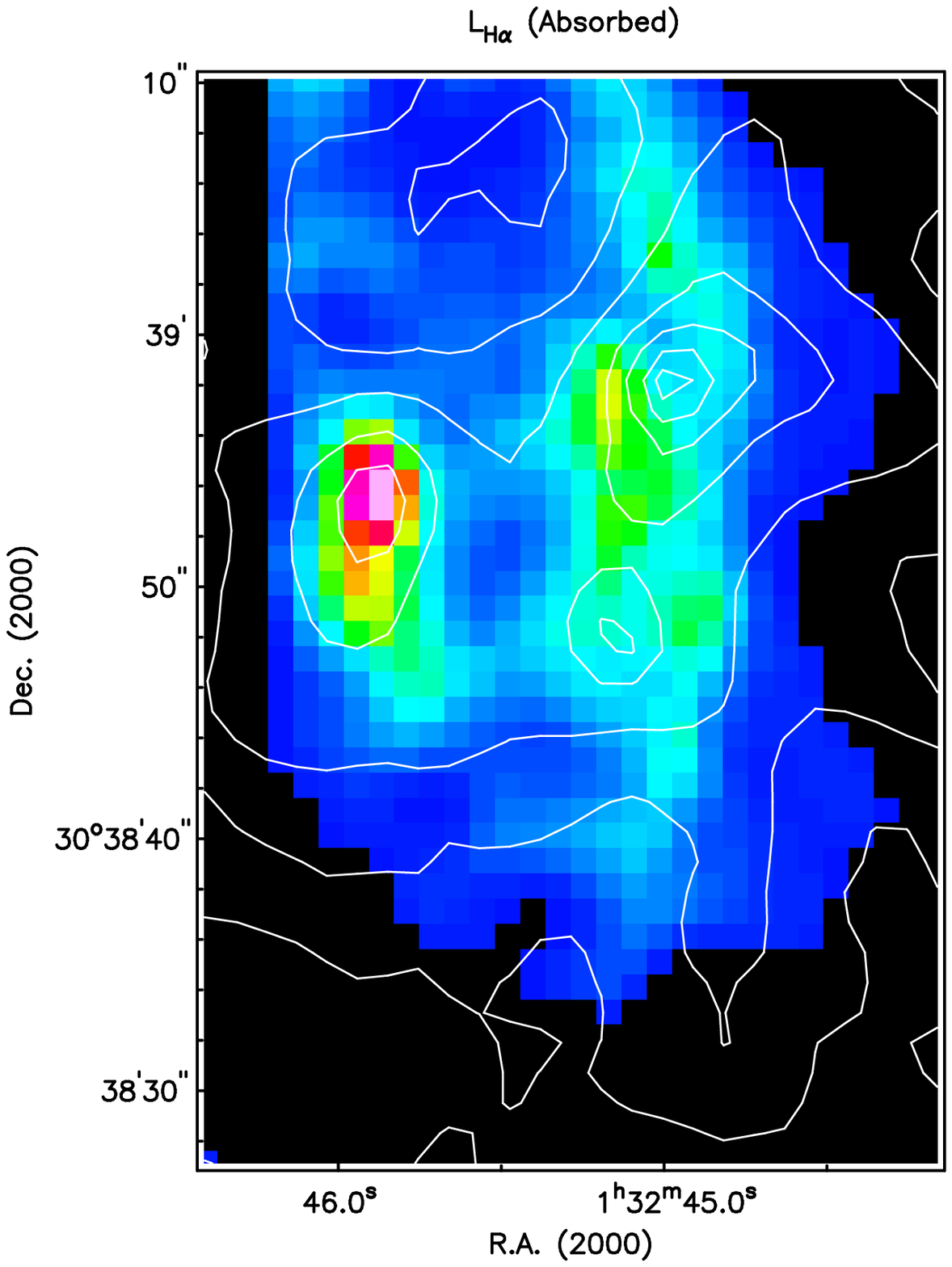} \\
\end{tabular} 

\caption{\emph{Left:} Reddening map for NGC~588 obtained assuming an intrinsic \ha/\hb\ = 2.86 and a correction of 1~\AA\ in absorption for \hb. The extinction law of \citet{flu94} and $E(B-V)=A_V/3.1$
\citep{rie85} were utilized.  The map has been convolved with a Gaussian
filter with $\sigma$=1$^{\prime\prime}$ to better trace the extinction structure. The 24~$\mu$m emission was overplotted with contours.  
The intensity contours are at 2, 5, 10, 20, 40, 60, 80, 95 per cent of the maximum intensity within the region. A 1\% contour level corresponds to a 3$\sigma$ value.
\emph{Right:} {\it Absorbed} \ha\ luminosity of NGC~588 with 8~$\mu$m emission contours overplotted. The contour levels correspond to the same percentage as for the 24~$\mu$m case. Here, a 1\% contour level corresponds to a 1$\sigma$ value.}
\label{ext+ir}
\end{figure*}

To investigate if the nebular properties derived from long-slit are representative of the whole \textsc{H\,ii} region, we compared our flux measurements from the integrated spectrum of NGC~588 to the values presented by \citet{jam05} and \citet{vil88} who used long slit at position angle (P.A.) $\sim-45^\circ$ including the main ionizing cluster in NGC~588 \citep[e.g. see Fig. 1 in ][]{jam05}.
Most of our measurements present a difference with respect to \hb\ when comparing with previously reported measurements that can range between $\sim$8\% (e.g. \oiii\ and He\,\textsc{i}$\lambda$6678) up to $\sim$64\% in the case of H8+He\,\textsc{i}$\lambda$3889.
This indicates that 
it is not trivial how the line ratios measured at the brightest knots trace those for the \textsc{H\,ii}  region as a whole.
This issue will be explored in more detail in Sec. \ref{secgas}.

Table \ref{tablapropsinte} contains the principal diagnostic emission-line ratios measured from the integrated spectrum as well as the physical parameters derived from them.
%
Given the blue-shift for M~33 and the presence in the sky substracted
spectrum of strong residuals for the Hg\textsc{i}$\lambda$4358 sky-line, it was not possible to measure the \textsc{[O\,iii]}$\lambda$4363 nebular line and thus determine the electron temperature ($T_e$) by means of the \textsc{[O\,iii]} $\lambda$4363/$\lambda$5007 line ratio. Another possibility would have been using \textsc{[S\,ii]}$\lambda$4067 and/or \textsc{[N\,ii]}$\lambda$5755 together with \sii\ and/or \nii. To search for these lines, in addition to the total spectrum, we created a spectrum by co-adding  the spectra in an aperture of 5$\times$9 spaxels centred at the peak of emission for the ionized gas. In this way, we increased the signal-to-noise ratio -- and thus, our detection limit -- since only the brightest spectra were included. However, these features were detected neither in the total spectrum nor in the one involving the brightest spaxels. Thus, we assumed a $T_e$ of 11\,140~K, as derived by \citet{jam05} from the  \textsc{[O\,iii]} $\lambda$4363/$\lambda$5007 line ratio  to estimate the electron density ($n_e$) and used the task \texttt{temden}, based on the \texttt{fivel} program \citep{sha95}
included in the IRAF package \texttt{nebular}. The derived $n_e$ was consistent   with being below the low density limit.

Another quantity quoted in Table \ref{tablapropsinte} is the ionization parameter, defined as:

\begin{equation}
q_{\mathrm{eff}} = \frac{Q(H_o)}{4\pi R_s^2 n_e}
\end{equation}
where $Q(H_o)$ is the number of ionizing photons per second emitted by the stars, $R_s$ is  the Str\"omgren  radius of the \textsc{H\,ii} region and $n_e$ is the electron density. We estimated the number of ionizing photons using the expression provided by \citet{ken98}:

\begin{equation}
Q(H_o) (\mathrm{s^{-1}}) = 7.31\times10^{11} L(H\alpha)     (\mathrm{erg~ s^{-1}})
\end{equation}
and the extinction corrected \ha\ luminosity reported by \citet{rel09}. Since the sulfur line ratio is consistent with being below the low density limit regime, we assumed a  face value of $n_e=20$~cm$^{-3}$. Also, we utilized a Str\"omgren radius  of 80~pc which is an approximated value inferred from the \ha\ image.

Table \ref{tablapropsinte} also includes several metallicity sensitive line ratios. The most widely used is probably the $R23$=(\oii + \textsc{[O\,iii]$\lambda\lambda$4959,5007})/\hb\ index \citep{pag79}.
However, the $Z-R23$ relation is two valued and thus, independent metallicity tracers are needed to determine which of the two branches is appropiate for NGC~588. 
One possibility would be the $N2=\log$(\nii/\ha) line ratio. According to the empirical parametrization proposed by \citet{per09}, we derived a metallicity of $12+\log(O/H) = 8.16$. This result does not point in a conclusive manner to either the upper or the lower branch. Instead, the metallicity of NGC~588 falls in the \emph{knee} region of the $Z-R23$ relation, where uncertainties can be as high as 0.7~dex. Alternatively, one can use the N2O3=$\log$((\nha)/(\ohb)).
We derived a metallicity of $12+\log(O/H) = 8.15$ and $8.18$ using the parametrization proposed by \citet{pet04} and \citet{per09}, respectively.
%
Thus for the purpose of this section, we will consider as the characteristic metallicity of the region, the mean of those derived from the N2 and N2O3 parameters: 8.16$\pm$0.02.
This value agree within the uncertainties with the $12+\log(O/H) = 8.17$ metallicity reported by \citet{jam05}, is slightly lower than the expected value assuming that this region follows the metallicity grandient for M~33 \citep[8.28$\pm$0.08,][]{ros08} and $\sim$0.15~dex lower than the value reported by \citet{vil88}. 

\subsection{Structure of the ionized gas and the stellar component}

In Figure \ref{fluhb}, we present the \hb\  and \oiii\ flux maps for NGC~588
derived from our PMAS data. Contours reproducing archive continuum images in the red and blue spectral bands have been overplotted for reference (see caption of Fig. \ref{fluhb} for details). 
%
Our PMAS data cover the whole southern part of the region plus most of
the northern one.
This GHIIR is dominated by emission from a broken elongated ring-like structure with major and minor axes of $\sim40^{\prime\prime}$ and $\sim25^{\prime\prime}$ (i.e. $\sim$160~pc and $\sim$100~pc), respectively and at P.A.$\sim$10$^\circ$ and $\sim$100$^\circ$.
In addition, there is a bridge of ionized gas emission joining the two halves of the ring-like structure from $\sim$[1\farcs0,$-$7\farcs0] to $\sim$[$-$7\farcs0,$-$3\farcs0]\footnote{Hereafter, the reported positions will refer the relative coordinates to the main ionizing cluster.} at P.A.$\sim-70^\circ$.
%
%
The morphologies of the \hb\ and the \oiii\ maps are very similar but show some rather subtle differences which indicate the complex ionization structure, which will be explored in more detail in section  \ref{secgas}.
The main ionizing cluster, as depicted by the archive continuum images, is not at the centre of the ring like structure, but at $\sim$2\farcs0 from the peak of emission in \hb.
Finally, there are also several secondary peaks of emission, most of them in the northern half of the region, which are associated with very massive (i.e. $30-45$~M$_\odot$) individual ionizing stars \citep{jam04}.

\subsection{Extinction distribution and dust}

The distribution of the extinction was derived by means of the \ha\ and \hb\ emission line maps. We assumed an intrinsic Balmer emission line ratio of \ha/\hb = 2.86 \citep{ost06} for a case B aproximation and $T_e =$ 11\,150~K  and used the extinction curve of \citet{flu94}. We included a 1~\AA\ correction to take into account the \hb\ absorption line due the underlying stellar population.
This absorption feature was clearly visible in a co-added spectrum extracted in an rectangular area with low surface brightness in the emission lines located at $\sim[-12\farcs0,-17\farcs0]$ of about $10^{\prime\prime}\times8^{\prime\prime}$.
However, it was not detected, and thus impossible to be fitted for individual spaxels.

The reddening map was created assuming $E(B-V) = A_V/3.1$ \citep{rie85} and is displayed in the left panel of Fig. \ref{ext+ir} with the 24~$\mu$m image from \emph{Spitzer} overplotted with contours. This map shows how irregular the extinction distribution is with low values of reddening ($\sim0.00-0.25$). 
This strongly contrasts with the findings of \citet{jam04} who, using long-slit, reported an almost constant extinction of $E(B-V) = 0.11\pm0.02$ and reinforces the need of 2D unbiased spectral mapping to characterize the physical properties of GHIIRs.
%
The optical reddening map presents three maxima that spatially correlate very well with the maxima of dust emission in the \emph{Spitzer} 24~$\mu$m and 8~$\mu$m bands (see Fig. \ref{ext+ir}).
%
This is consistent with the idea of extinction caused by absorption of dust associated with the GHIIR.
Other dust-gas configurations would have caused a different set of maps. For example, if dust were behind the region, there would not have been a counterpart in the $E(B-V)$ map to the peaks in the  map at 24~$\mu$m.
In the right panel of Fig. \ref{ext+ir} we present the absorbed \ha\ luminosity
- defined as the difference between the total extinction-corrected \ha\ luminosity and the \ha\
luminosity corrected for the foreground Galactic extinction, $E(B-V)$=0.044, \citep{sch98} - obtained using our derived reddening map with the 8~$\mu$m contours overlaid. A comparison of these two panels shows how the 24~$\mu$m emission presents a more compact distribution towards the centre of the 
region and correlates better with the absorbed \ha\ luminosity map than the 8~$\mu$m emission. This was seen in other \textsc{H\,ii} regions like NGC~604 \citep{rel09} and NGC~595 \citep{rel10}.
The 24~$\mu$m emission has proved to be a good tracer of the recent star formation for a large range of \ha\ luminosities, ranging from moderate \textsc{H\,ii} regions to dusty powerful starbursts \citep[e.g.][]{rie09,cal10}. 
The spatial correlation between the \ha\ luminosity and 24~$\mu$m maps indicates the infrared band as a star-formation tracer also in low dust environments, such as NGC~588. 
%
Due to the low spatial resolution of the 70~$\mu$m and 160~$\mu$m \emph{Spitzer} bands we were not able to compare the emission at these wavelengths with our reddening map. \emph{Herschel} data, which will be available soon to the scientific community, will permit a proper comparison to be made  in the future. 

\begin{figure}
\includegraphics[width=0.48\textwidth, clip=,bbllx=-20, bblly=5,
  bburx=595, bbury=750]{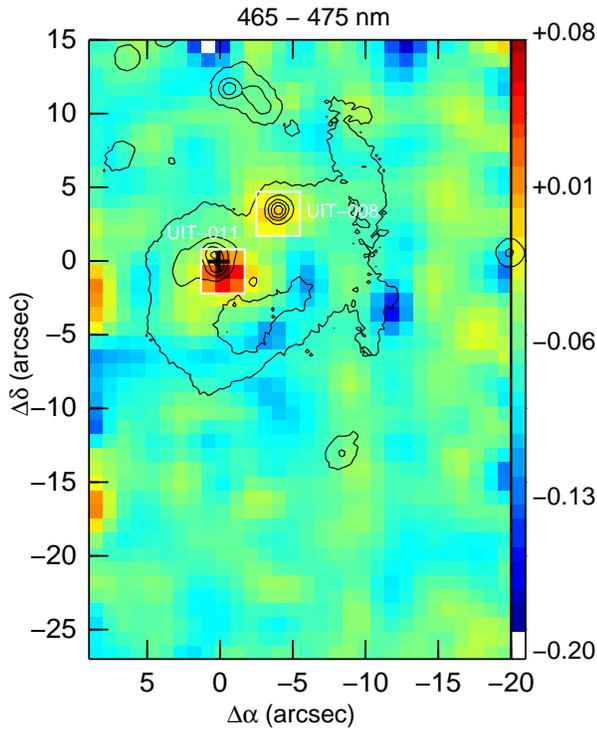}
\caption{4650$-$4750~\AA\ map derived from the PMAS data after
  convolving with a Gaussian of $\sigma$=1\farcs0. Continuum was
  subtracted by averaging the spectral ranges of 4490$-$4540~\AA\ and
  4755$-$4805~\AA. A logarithmic stretch covering $\sim$0.3~dex was used to better enhance the
   peaks of emission. 
Labels with the detected WR are included and white contours delineate
the spaxels utilized to create the extracted spectra. Contours
  correspond to the continuum subtracted 
  $R$ broad-band  direct image from 
   NOAO Science Archive \citep{mas07}.
The orientation is north up and east to the left.
The main ionizing cluster, at coordinates RA(J2000): 1h32m45.7s, 
   Dec.(J2000): +30d38m55.1s, marks the origin of our coordinates
   system.   \label{mapwr} }
  \end{figure}  

\subsection{WR stellar population}
\label{secstar}

\citet{rel10} presented a novel and simple technique to detect WR stars in a swift way and compared its results to classical, more time-consuming techniques. Using the same set of observations (i.e. the datacube), one can identify the WR candidates by simulating the action of narrow filters and creating continuum subtracted maps at the emission bump at 4\,700~\AA\ (the \emph{blue bump}) and at 5\,700~\AA\  (the \emph{red bump}), characteristic of WR stellar emission, 
and localizing the peaks of emission, afterwards. Then, the candidates can be confirmed by extracting the spectra of the associated spaxels. Following this methodology, we confirmed previous cataloged WR stars in NGC~595 and discovered a new one further away from the main ionizing clusters. Here, we apply the same methodology to NGC~588.


There are two known WR stars in NGC~588 that have been widely studied in the past.
The first one, named UIT-011 by \citet{mas96} - MC~3 by \citet{dri08} - was detected for the first time using narrow band imaging by \citet{con81} and was spectroscopically confirmed by \citet{mas83} later on. It was classified as WNL with $M_V=-7.9$~mag.
The second one, named UIT-008 is a transition Of/WN9 star.
%
Both stars were modelled using multi-band photometry with the HST by \citet{ube09}. They derived effective temperatures of 57\,000~K and 32\,000~K  and bolometric luminosities of $\log (L/L_\odot) = 6.48$ and 5.97, for UIT-011 and UIT-008 respectively.

\begin{figure}
\includegraphics[width=0.44\textwidth, clip=,bbllx=40, bblly=65,
  bburx=760, bbury=315]{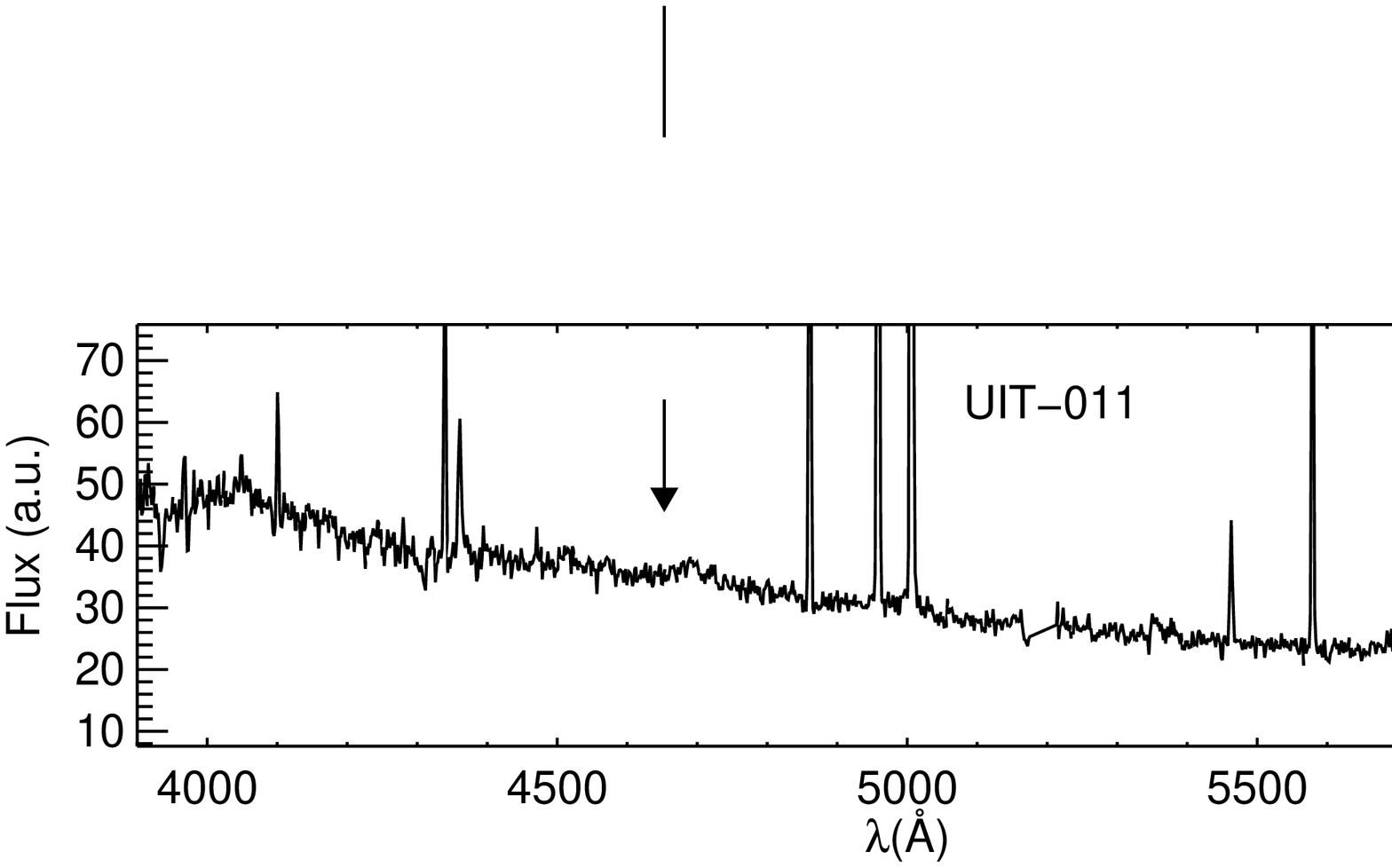}
\includegraphics[width=0.44\textwidth, clip=,bbllx=40, bblly=65,
  bburx=760, bbury=315]{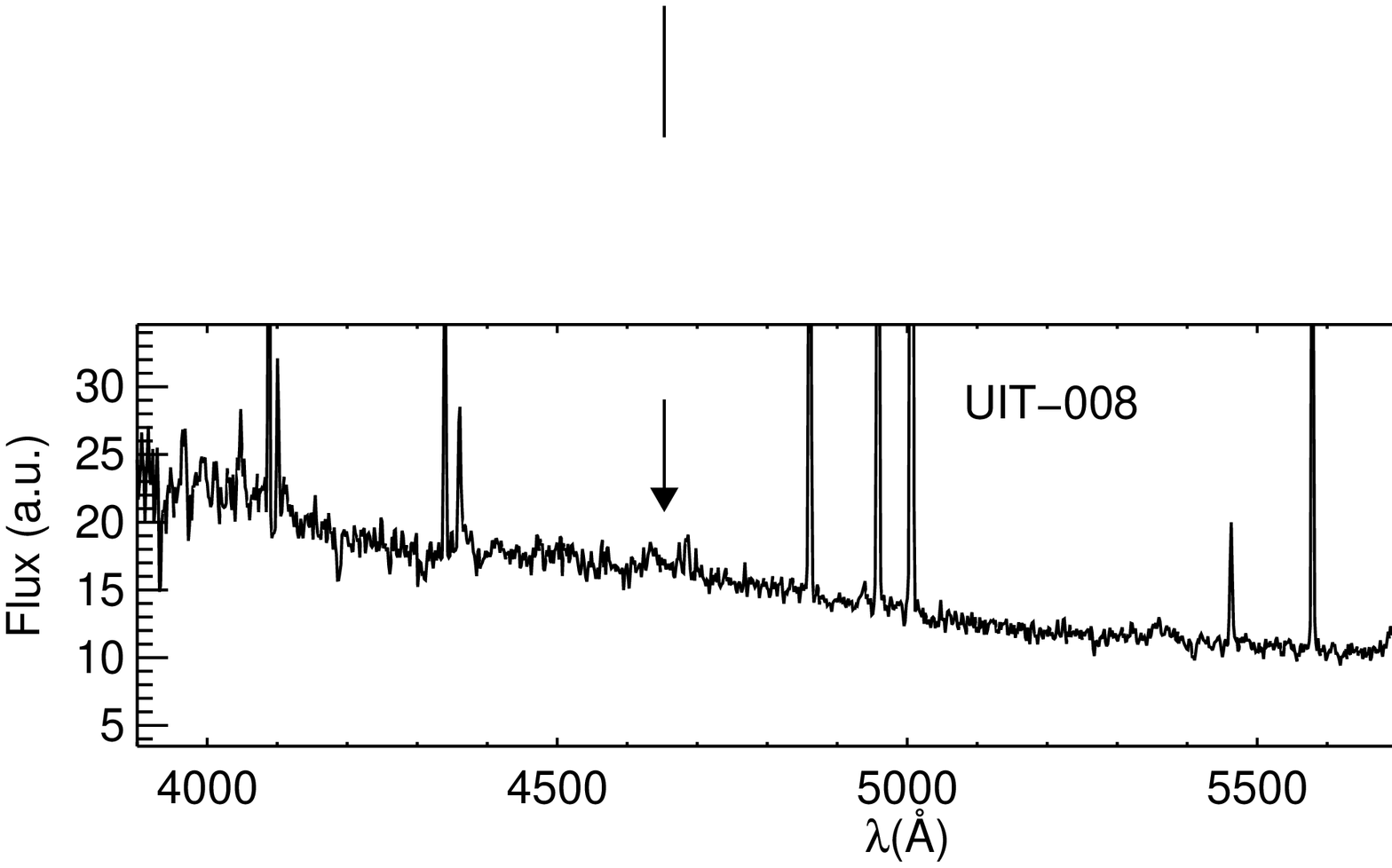}
\caption{Extracted spectra for two detected peak of emission in the PMAS continuum subtracted 4650$-$4750~\AA. Vertical arrows mark the position of the expected
  \emph{blue} and \emph{red bumps}. \label{espec_wr}}
\end{figure}

Fig. \ref{mapwr} presents the continuum substracted \emph{blue bump} map after convolving with  a 1$^{\prime\prime}$-Gaussian. We identify  two main peaks of emission whose positions agree well with those previously reported for UIT-011 and UIT-008 \citep{dri08}. No aditional WR in NGC~588 was found. The extracted spectra for the two stars is presented in Fig. \ref{espec_wr} and show clearly both the \emph{blue} and \emph{red} bump.

The cases of NGC~595 and NGC~588, in \textsc{H\,ii} regions, as well as existing ones for starburst galaxies \citep[e.g. the Anntenae, II~Zw~70, Mrk~996, NGC~5253, IC~10,][]{bas06,keh08,jam09,mon10,lop10} illustrate the effectiveness of IFS in the finding and characterization of the WR population.
At this stage, the possibility of using this technique routinely should be taken into account. In particular, it would suit perfectly in the case of WR finding in galaxies at larger distances and, more important, with large gradients in their velocity fields. Here, the traditional technique of search for candidates via imaging first, and spectroscopic confirmation afterwards, might well miss some of the WR populations since the \emph{blue/red bump} might move outside the spectral range of the narrow filter. On the contrary, the methodology presented here can be easily modified and implemented to define what can be called \emph{synthetic tunable} filters that take into account the movements of the galaxy and thus preventing these losses.
An additional advantage of using an Integral Field Unit, especially if it has a large field of view
is the detection of runaway WR stars ejected by the central star cluster
of the region \citep{dra05}. 
In particular, IFS-based instruments with relatively large field of view like PPak/PMAS \citep{kel06}, or MUSE \citep{bac10} are (or will be soon) under operation. They open the possibility of carrying out surveys of large samples of galaxies where this kind of simple techniques could be particularly useful.

\begin{figure}
\includegraphics[width=0.48\textwidth, clip=,bbllx=-20, bblly=5,
  bburx=595, bbury=750]{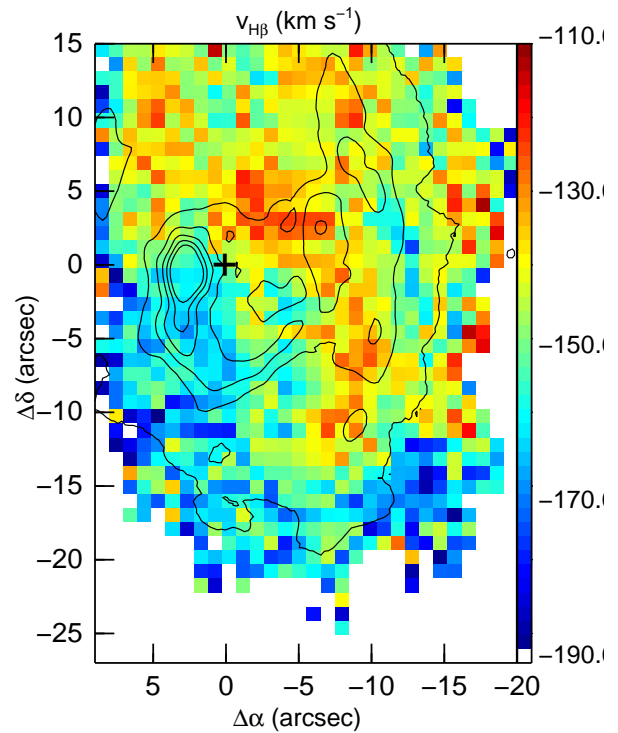}
\caption{Velocity field derived from the \hb\ emission line.
Contours correspond to the continuum subtracted \ha\ direct
  image from  NOAO Science Archive \citep{mas07}.
The orientation is north up and east to the left.
The main ionizing cluster, at coordinates RA(J2000): 1h32m45.7s, 
   Dec.(J2000): +30d38m55.1s, marks the origin of our coordinates
   system.   \label{mapcinema} }
\end{figure}  

\subsection{Kinematics of the ionized gas}
\label{kinematic}

The superior spectral resolution of the present observations, twice that of the observations of  NGC~595, allowed us to derive the velocity field maps from the strongest emission lines. In Fig. \ref{mapcinema} we show the map corresponding to \hb. No relevant differences were found 
from the map derived using \oiii. The velocity field has a complex 
structure with values ranging between $-190$~km~s$^{-1}$ and $-110$~km~s$^{-1}$. The north-west part of the region seems to be more redshifted than the south-east,  which is the region with higher \hb\ surface brightness (see Fig. \ref{fluhb}). In the surroundings of the location of the stellar cluster, marked in 
Fig. \ref{mapcinema}  as a black cross, there is a pronounced velocity gradient: the north-west part 
has velocities of $-120$~km~s$^{-1}$ while the south-east has velocities of $-170$~km~s$^{-1}$. 
The velocity separation of $\sim$25~km~s$^{-1}$ between these 
zones and the location of the stellar cluster, corresponding to a velocity of 
$\sim$30~km~s$^{-1}$ in the galaxy plane \citep[inclination of M~33, $i=56$~deg][]{van00}, and 
the symmetry of the velocity field suggest the existence of a shell expanding 
in the interior of the region.
The shell expansion velocity, $\sim$30~km~s$^{-1}$, is slightly lower than the values observed 
in high luminosity  \textsc{H\,ii} regions of a set of spiral galaxies
\citep[v$_{exp}\sim40-90$~km~s$^{-1}$,][]{rel05} and also in NGC~604 
\citep[v$_{exp}\sim$40~km~s$^{-1}$,][]{yan96}. However, based on a kinematic study of 
NGC~588 and NGC~604, \citet{mun96} suggest that NGC~588 is 
more evolved than NGC~604 and therefore, we would expect lower velocities 
for the shells in the first region than in NGC~604, consistent with the result found here. 

In order to check whether the winds coming from the stellar population 
within the \textsc{H\,ii} region could produce the expansion of the observed shell, we have 
made a crude estimation of the kinetic energy involved in the shell and compared 
with the input kinetic energy from the stars.
Using the \ha\ luminosity of the region from \citet{rel09}, we predict an emission measure (EM) of 4000 (pc~cm$^{-6}$)
for an \textsc{H\,ii}  
region radius of 140~pc, corresponding to the aperture radius used to obtained 
the \ha\ luminosity. The EM is then used to derive a $<n_e>_{rms}$ of 5~cm$^{-3}$ and 
integrating over the \textsc{H\,ii}  region volume we derive a total ionized mass of $\sim$6$\times10^5$~M$_\odot$.
Assuming, as an upper limit that the whole mass has been swept up by the shell 
we obtain a kinetic energy for the shell of $3.8\times10^{51}$~erg. Starburst99 models \citep{lei99}
using ranges of values for the stellar mass of $1-6\times10^3$~M$_\odot$ and age of $3.5-4.2$~Myr, a Salpeter initial mass function, and metallicities of Z=0.004 and 0.008 
give a range of kinetic energy input of $5.9-64.2\times10^{51}$~erg, at least twice as high as
the upper limit 
of the kinetic energy of the shell. This crude calculation shows that the winds 
from the stellar cluster within NGC~588 are able to produce the observed shell 
in the \textsc{H\,ii} region and create the observed \hb\ morphology with holes and 
filaments shown in Fig. \ref{fluhb}.
Moreover, the lower energy associated with the shell, 30~km~s$^{-1}$, compared with values closer to 100~km~s$^{-1}$ for younger, less evolved regions \citep{rel05}, is consistent with
with the findings by \citet{mun96}.

\begin{figure*}
\includegraphics[width=0.48\textwidth, clip=,bbllx=20, bblly=45,
  bburx=595, bbury=750]{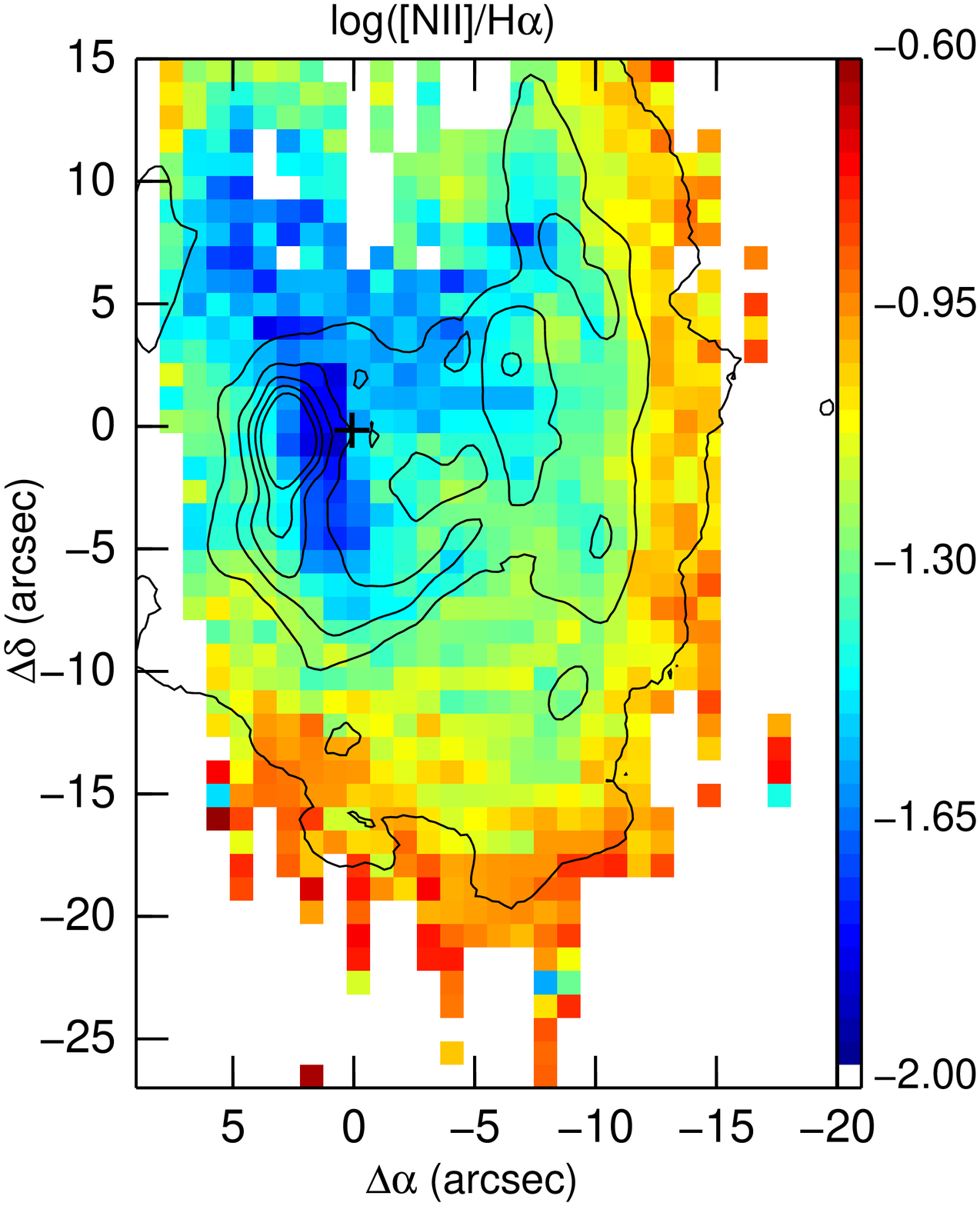}
\includegraphics[width=0.48\textwidth, clip=,bbllx=20, bblly=45,
  bburx=595, bbury=750]{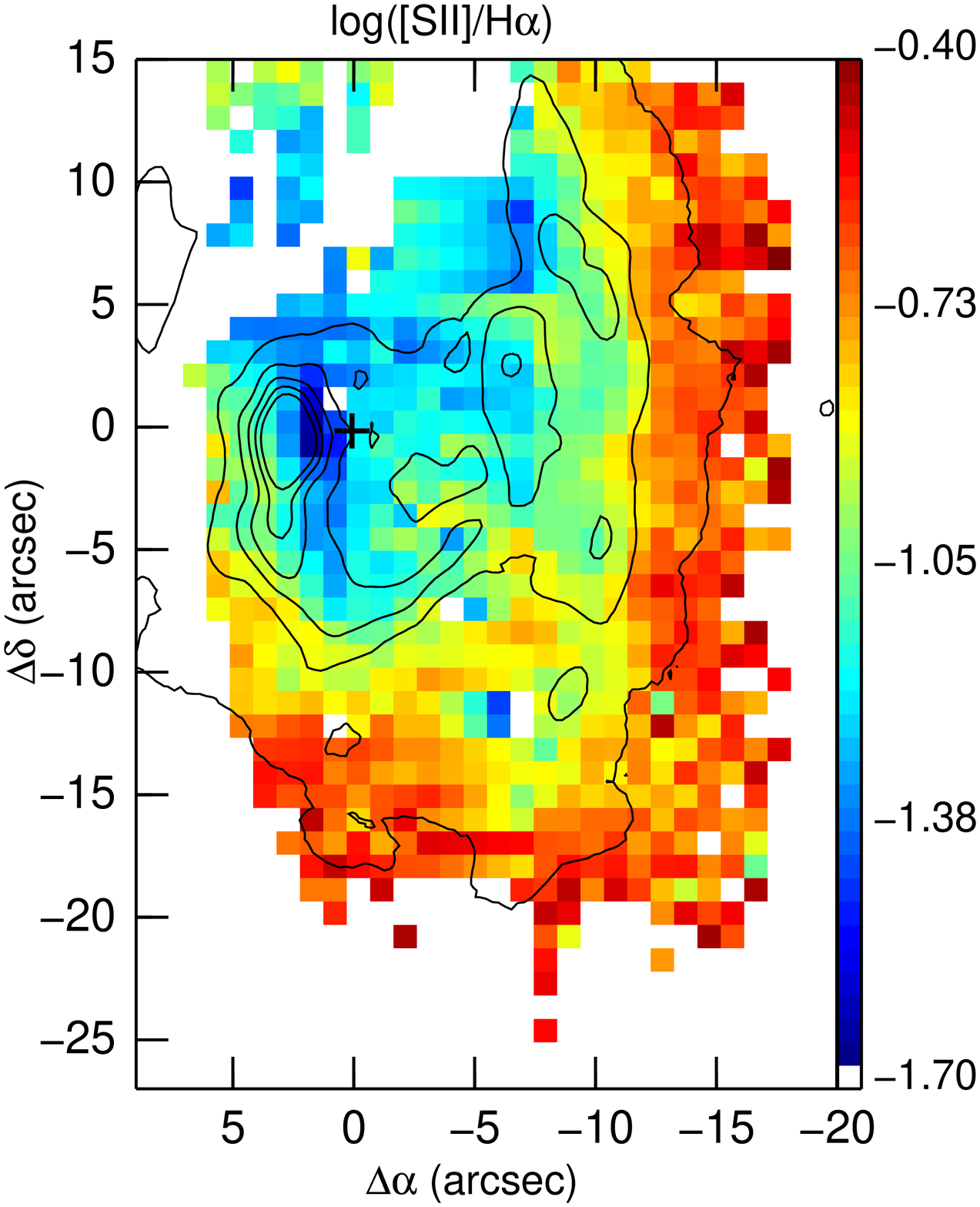}
\includegraphics[width=0.48\textwidth, clip=,bbllx=20, bblly=45,
  bburx=595, bbury=750]{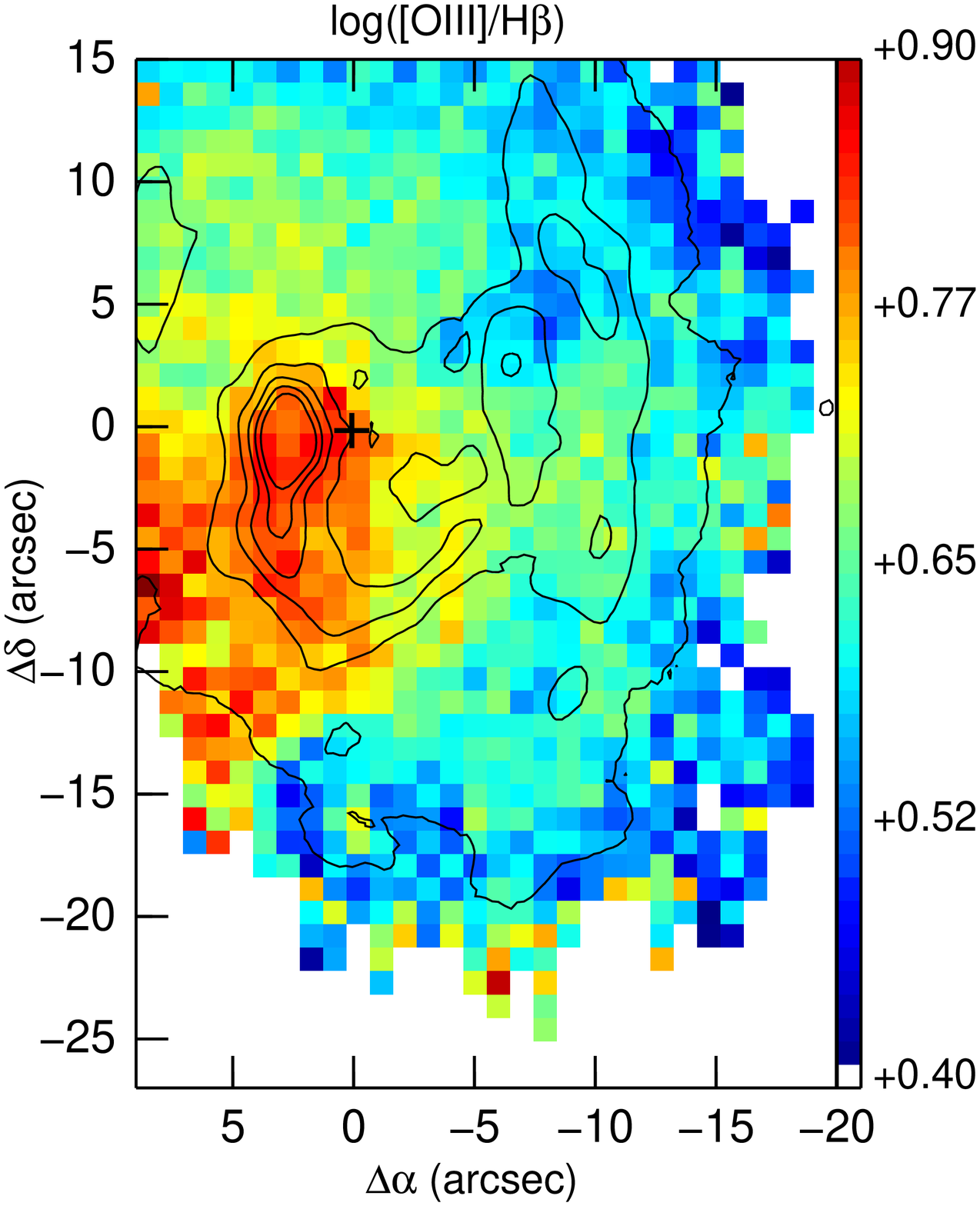} 
\includegraphics[width=0.48\textwidth, clip=,bbllx=20, bblly=45,
  bburx=595, bbury=750]{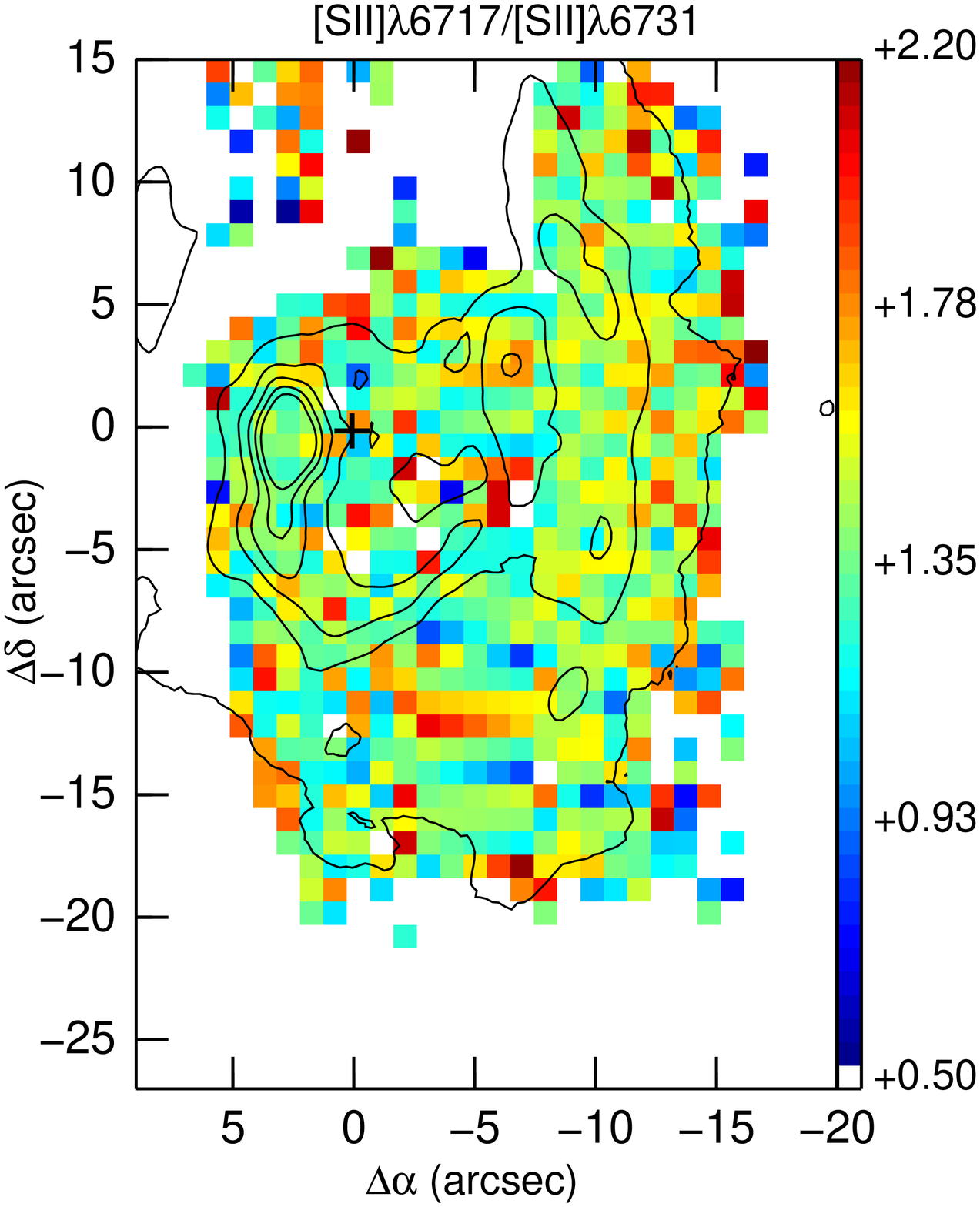}
\caption{Map of the observed \nha\ (upper left), \sha\ (upper right), \ohb\ (bottom
  left) and  \textsc{[S\,ii]}$\lambda$6717/\textsc{[S\,ii]}$\lambda$6731 (bottom
  right) line ratios. Contours correspond to the continuum subtracted \ha\ direct
  image from 
   NOAO Science Archive \citep{mas07}. The orientation is north up and
   east to the left. The main ionizing cluster, at coordinates
   RA(J2000): 1h32m45.7s, 
   Dec.(J2000): +30d38m55.1s, marks the origin of our coordinate system.
  \label{mapasionizacion} } 
\end{figure*}

\subsection{Characterization of the ionized gas}

\label{secgas}

\subsubsection{Density structure}

Electron density can be determined from the ratio between two lines of the same ion emitted by different levels with similar excitation energies.
We used the \textsc{[S\,ii]}$\lambda$6717/\textsc{[S\,ii]}$\lambda$6731 ratio in our analysis.
As is shown in the lower right map of Fig. \ref{mapasionizacion}, no structure for the $n_e$ was found.
We measured a mean($\pm$standard deviation) \textsc{[S\,ii]}$\lambda$6717/\textsc{[S\,ii]}$\lambda$6731 value of 1.20($\pm0.17$), which is consistent with the value derived from the integrated spectrum (see Table \ref{tablapropsinte}).
For the assumed temperature, this implies a $n_e$ of $\sim$250~cm$^{-3}$
and agrees within the uncertainties with the values reported by \citet{vil88}  and \citet{jam05}.
This value for the electron density corresponds to the density of the clumps within the region and 
differs from the r.m.s. electron density derived in the previous section using the \ha\ surface brightness 
of the region. The ratio of both density estimates is a measure of the volume fraction occupied by dense clumps.

\subsubsection{Line ratios in the BPT diagnostic diagrams \label{sec_bpt}}

\begin{figure*}
\includegraphics[width=0.33\textwidth]{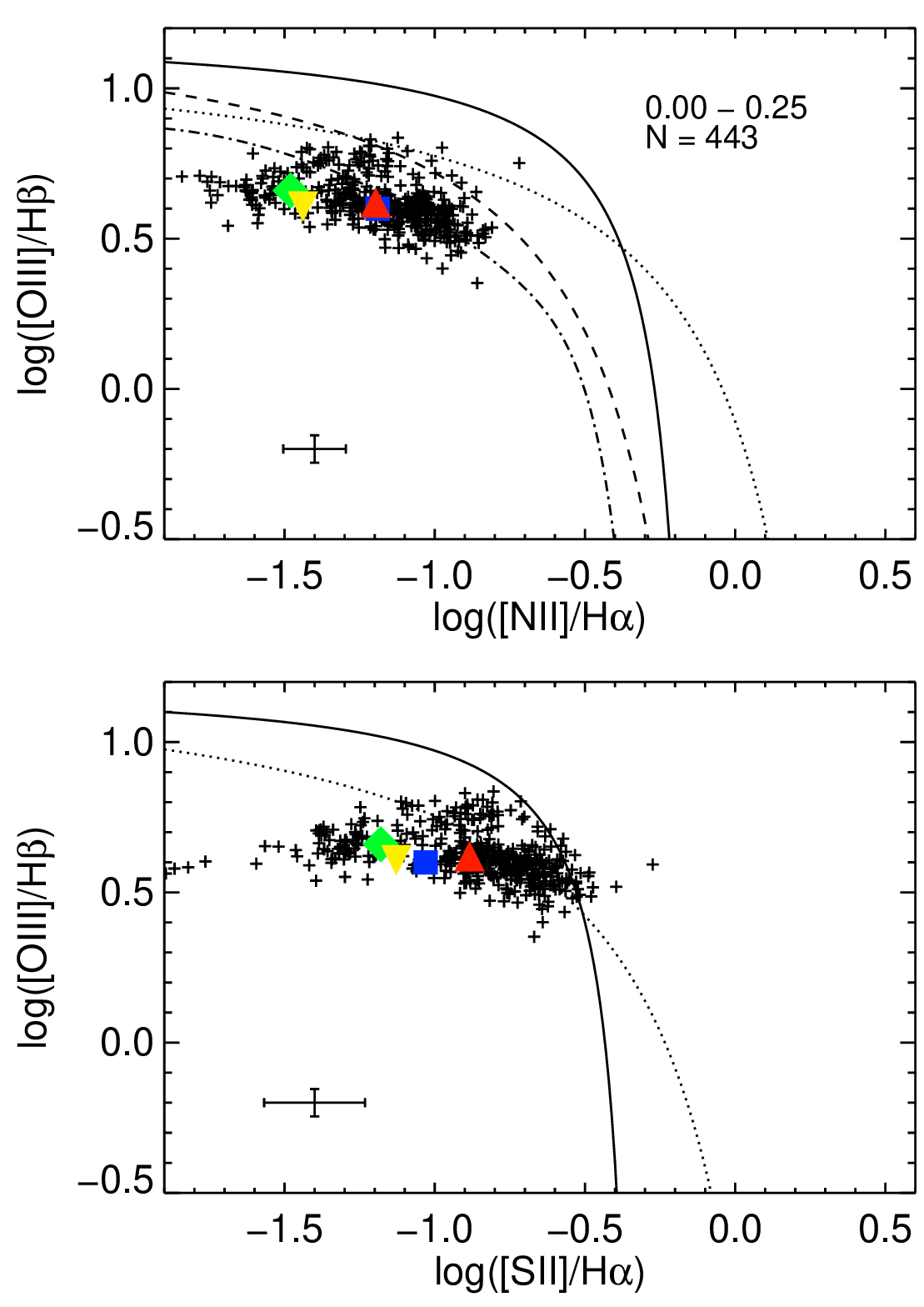}
\includegraphics[width=0.33\textwidth]{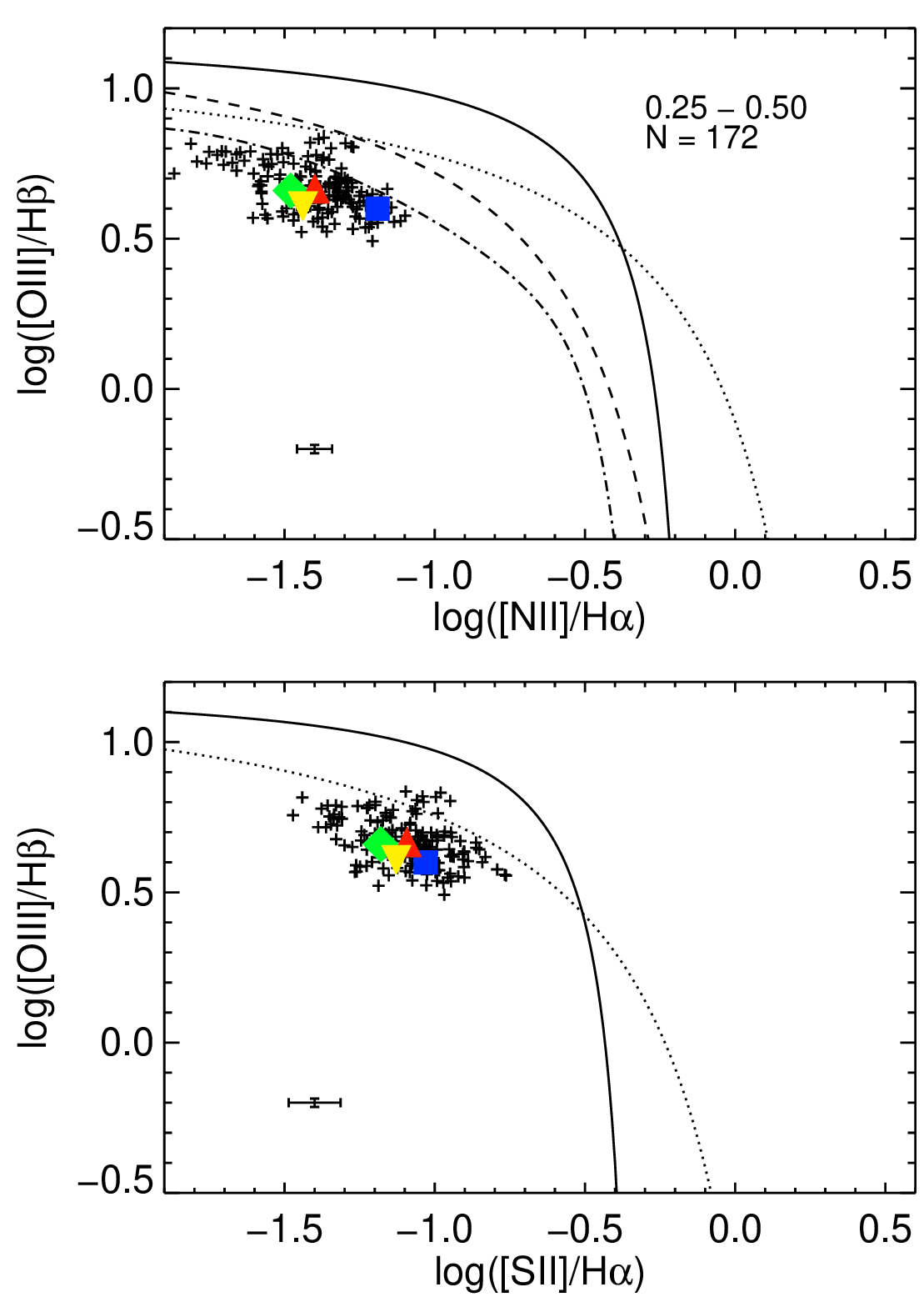}
\includegraphics[width=0.33\textwidth]{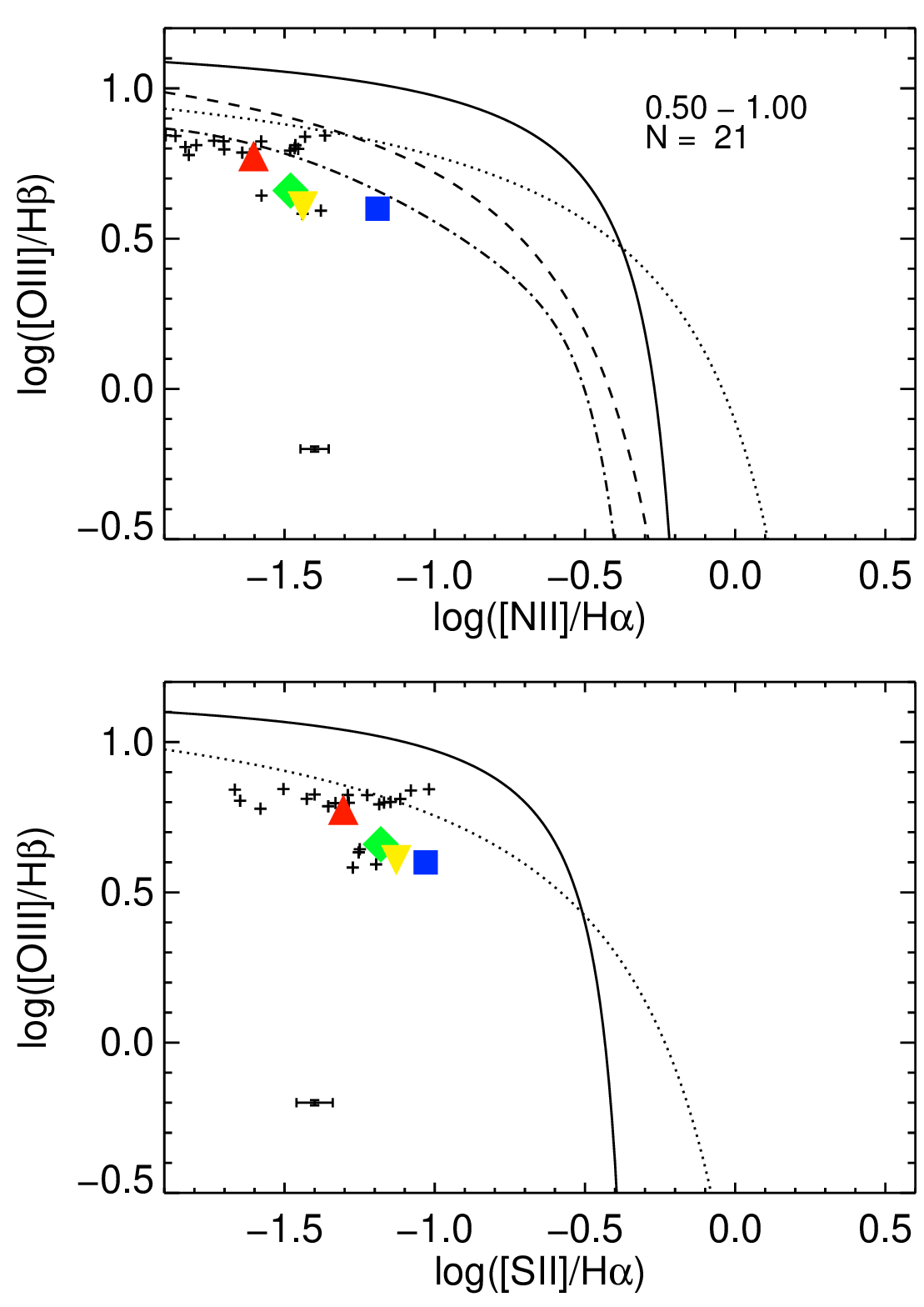}
\caption{Position of the individual spaxels in NGC~588 in the BPT
     diagnostic diagrams separated in three L(\hb)/L(\hb)$_{max}$ bins (\emph{Left column:} 0.00-0.25; \emph{Middle column:} 0.25-0.50; \emph{Right column:} 0.50-1.00).
Solid curves show
     the empirical borders found by \citet{vei87} between ionization
     caused by different mechanisms, while dotted lines show the
     theoretical borders proposed by \citet{kew01a} to delimit the area
     where the line ratios can be explained by star formation. Black
     dashed and dot-dashed lines show the revised borders by 
     \citet{kau03} and \citet{sta06}, respectively.
These were empirically determined using SLOAN data.
Green diamonds mark the values reported by \citet{vil88} while yellow inverted triangles are used for those of \citet{jam05}. Red triangles and blue squares indicate the mean values for a given bin and the values derived from the integrated spectrum as reported in Table \ref{tablapropsinte}, respectively. 
The number of considered data points as well as the  L(\hb)/L(\hb)$_{max}$ range are indicated in the right upper corner of the diagrams involving \nha. Typical errors are shown in the left lower corner of each diagram. \label{diagdiag}}
\end{figure*}

\begin{figure*}
\includegraphics[width=0.48\textwidth, clip=,bbllx=20, bblly=45,
  bburx=595, bbury=750]{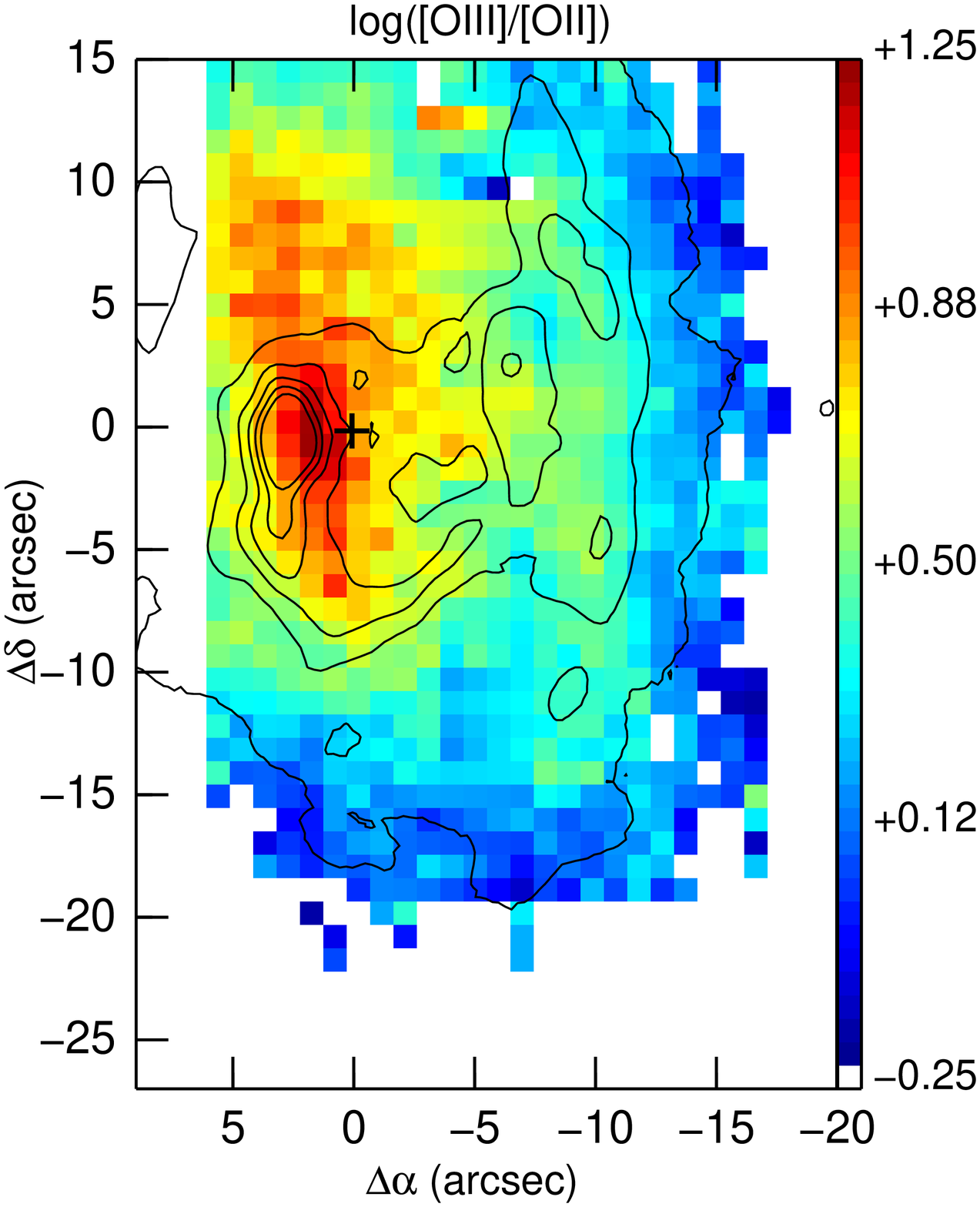}
\includegraphics[width=0.48\textwidth, clip=,bbllx=20, bblly=45,
  bburx=595, bbury=750]{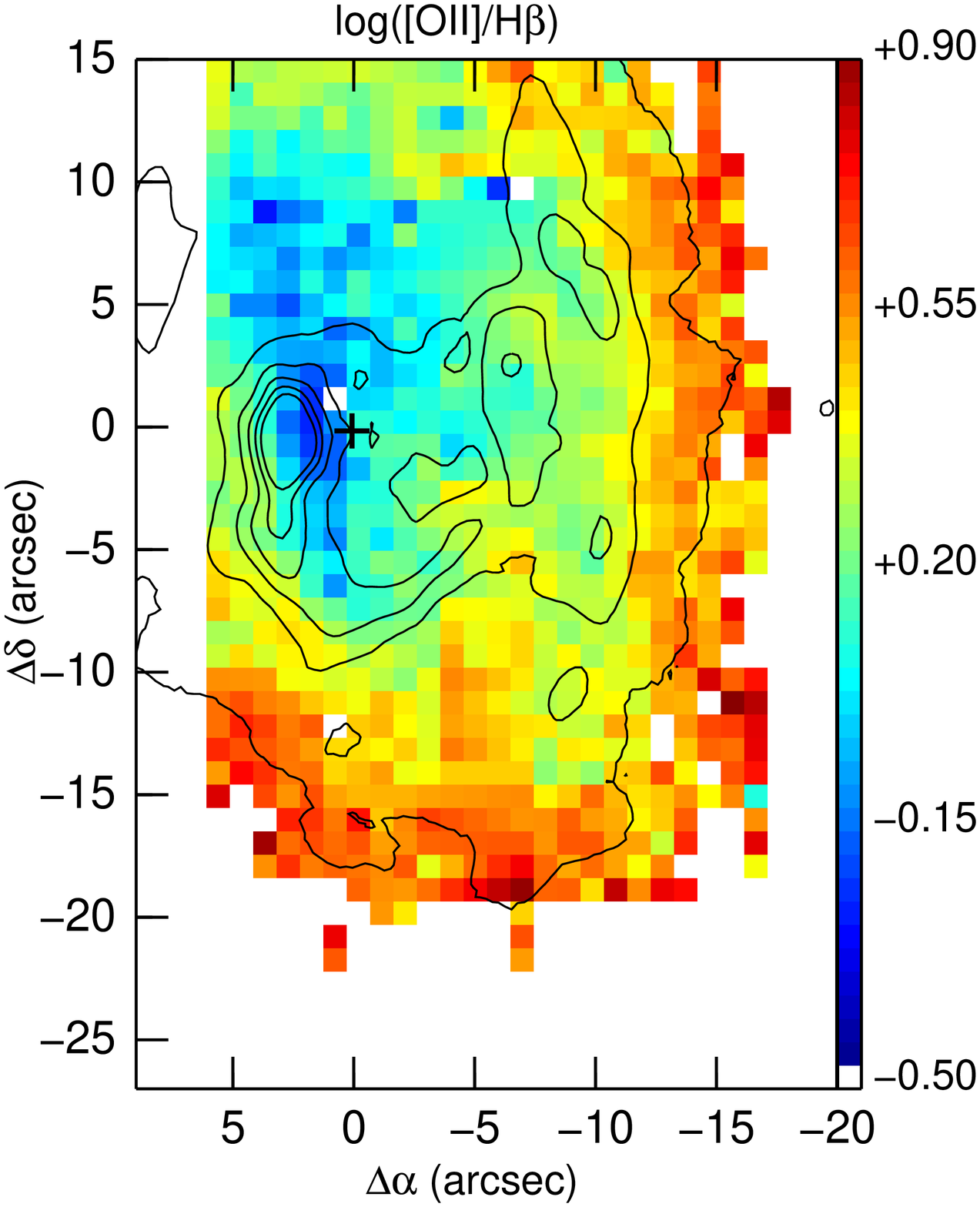}
\caption{Maps of the line ratios sensitive to the ionization parameter: \oiiioii\ (\emph{left}) and \oii/\hb\ (\emph{right}).  
Contours correspond to the continuum subtracted  \ha\ direct image from 
   NOAO Science Archive \citep{mas07}. The orientation is north up and
   east to the left. The main ionizing cluster, at coordinates
   RA(J2000): 1h32m45.7s, 
   Dec.(J2000): +30d38m55.1s, marks the origin of our coordinates
   system. \label{umap}}
\end{figure*}

\begin{figure*}
\includegraphics[width=0.33\textwidth]{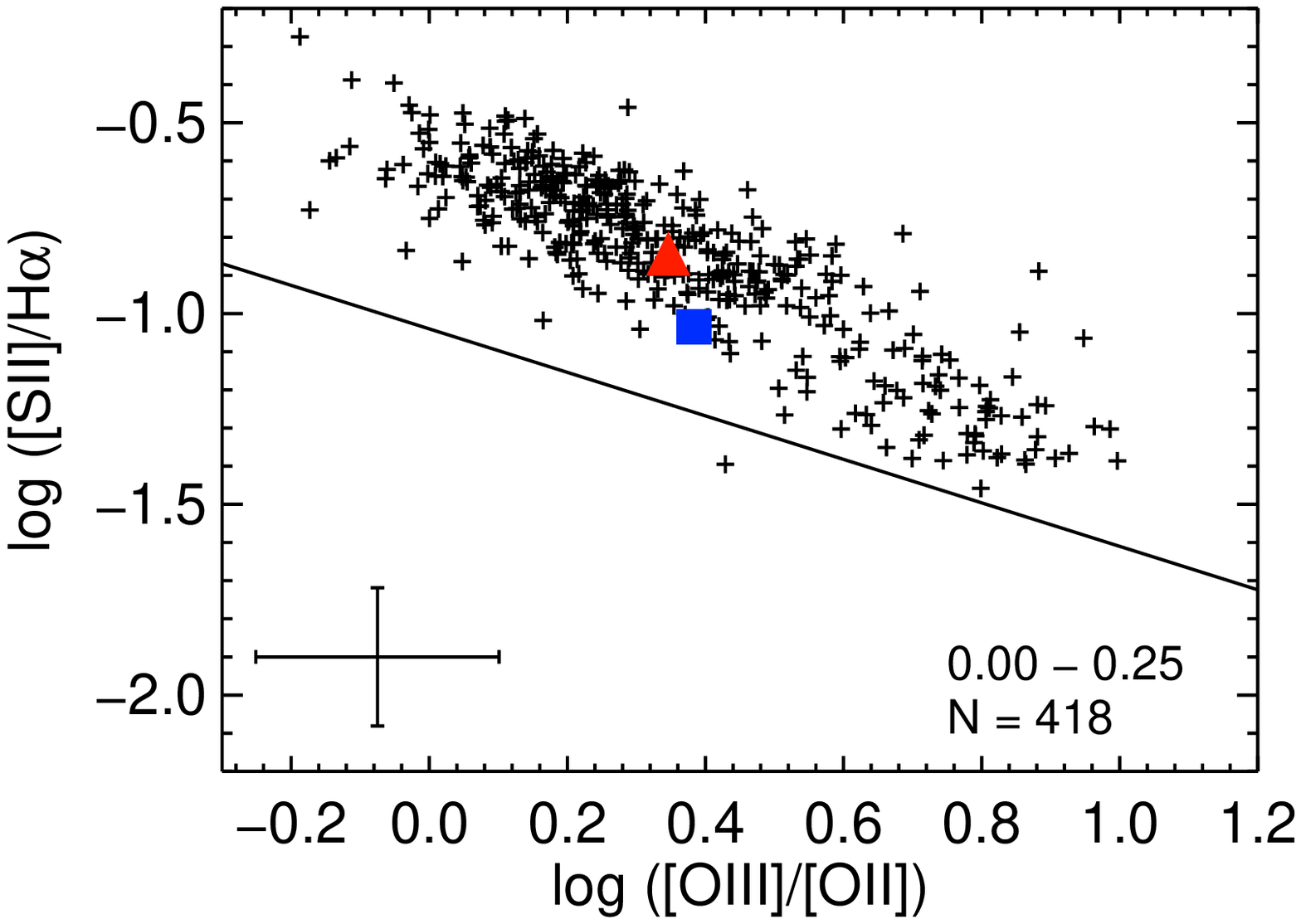}
\includegraphics[width=0.33\textwidth]{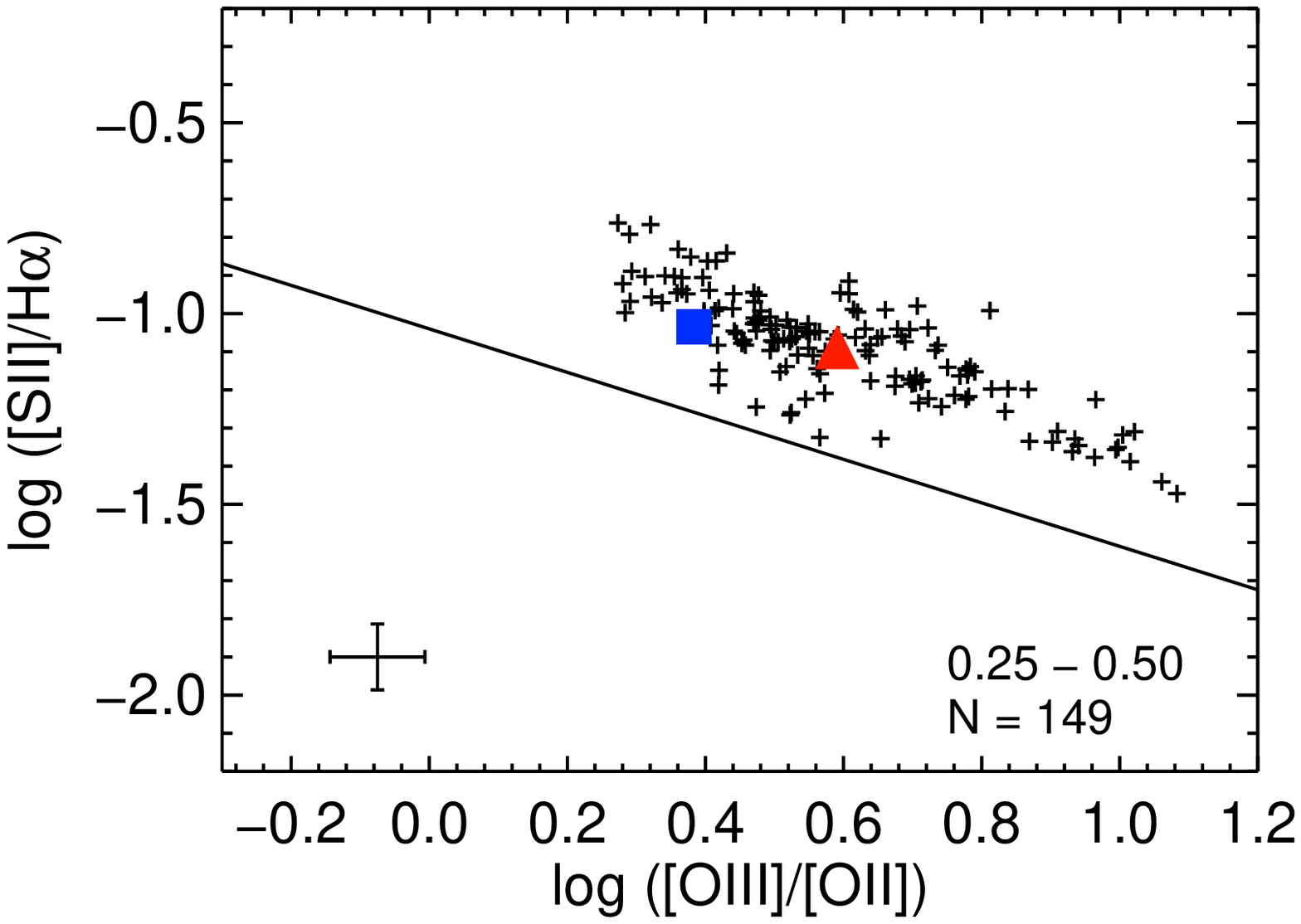}
\includegraphics[width=0.33\textwidth]{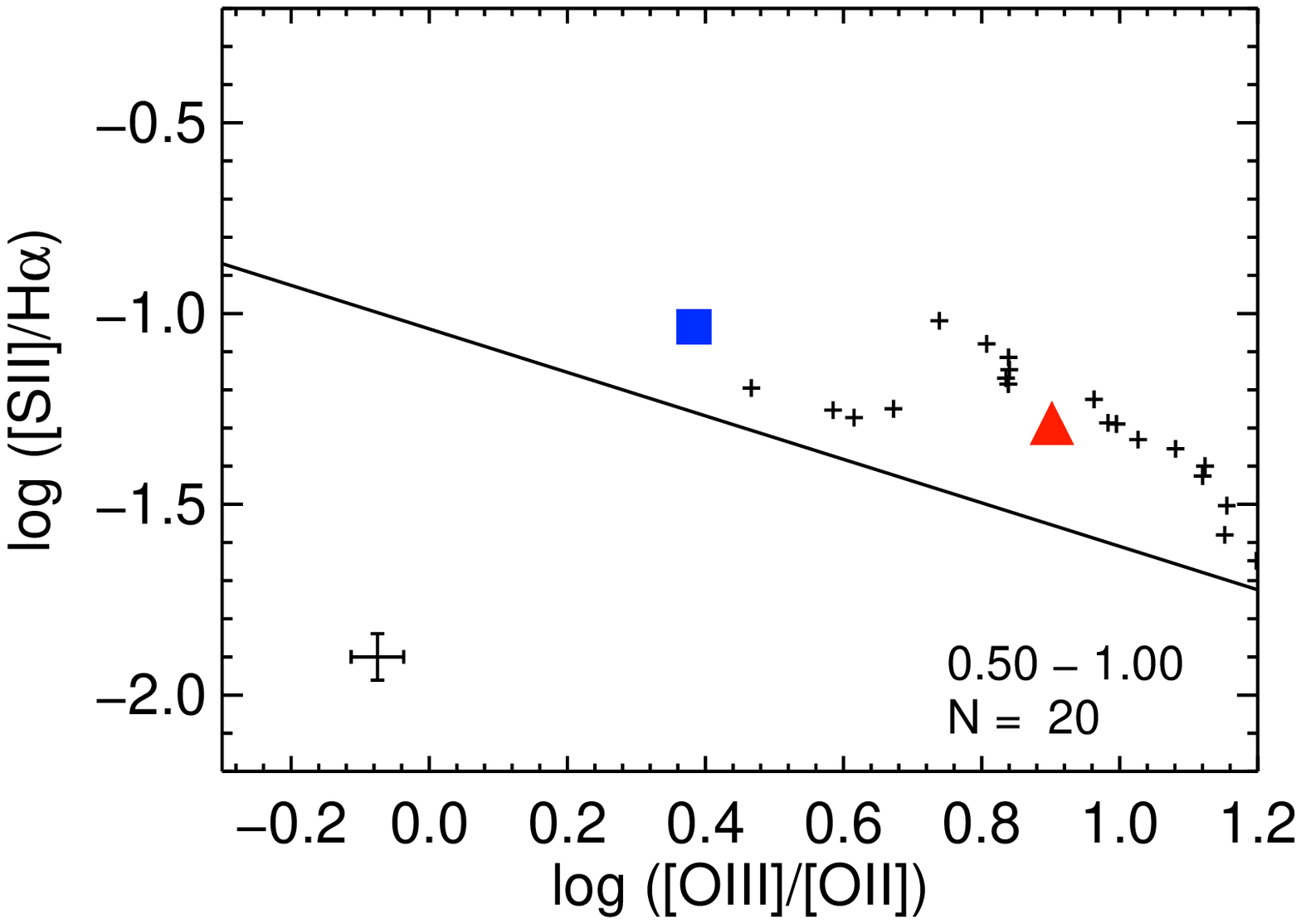}

\includegraphics[width=0.33\textwidth]{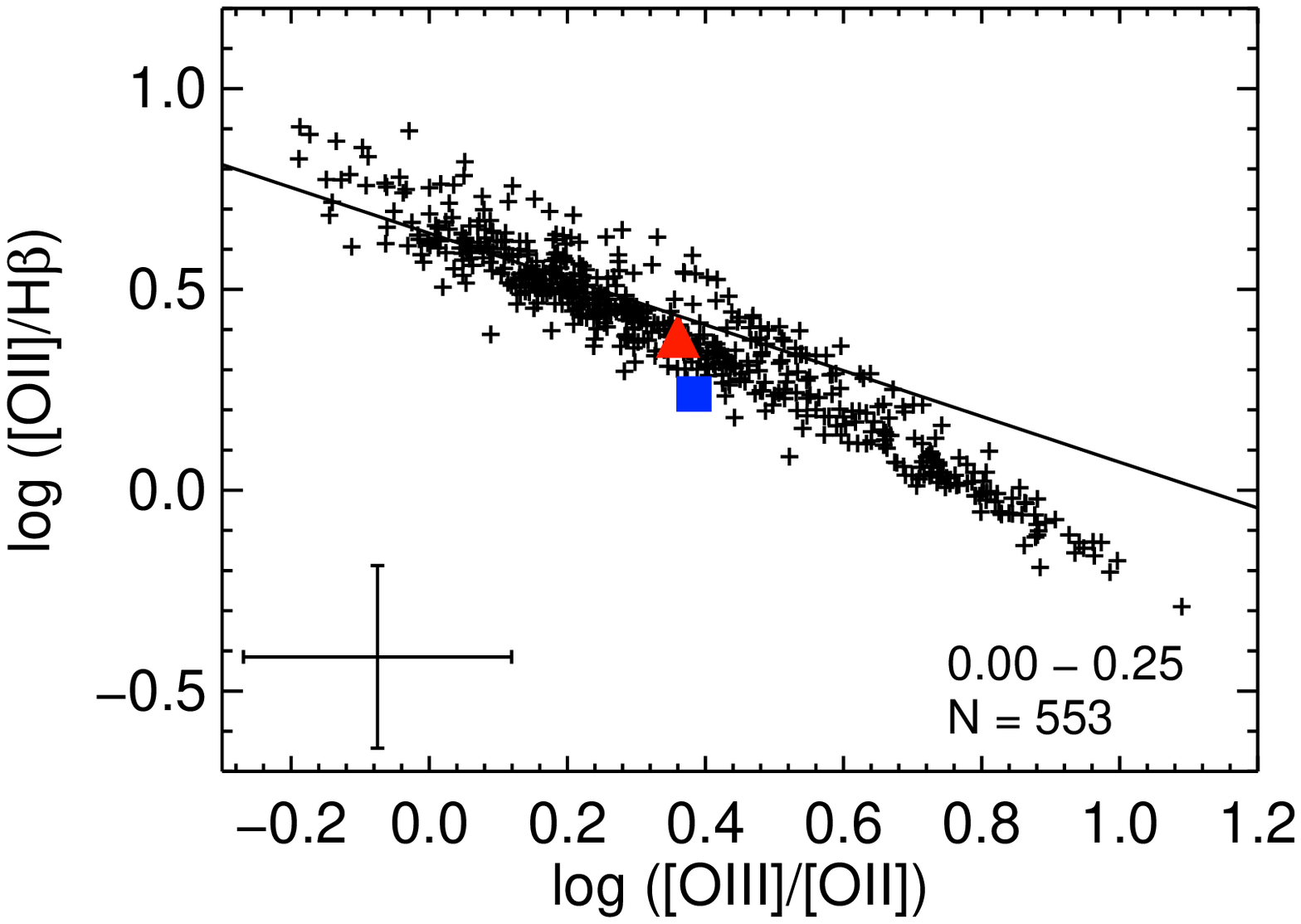}
\includegraphics[width=0.33\textwidth]{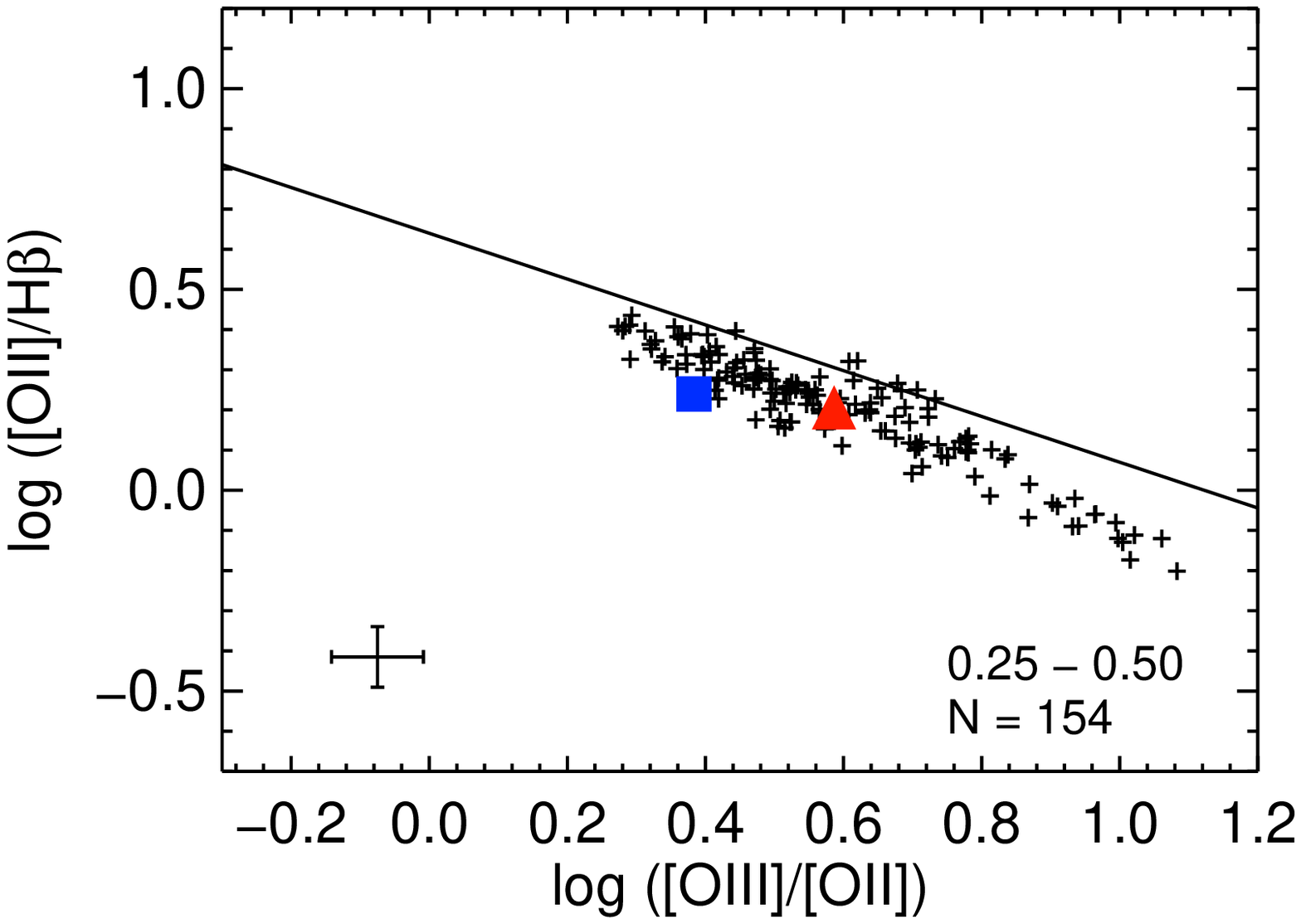}
\includegraphics[width=0.33\textwidth]{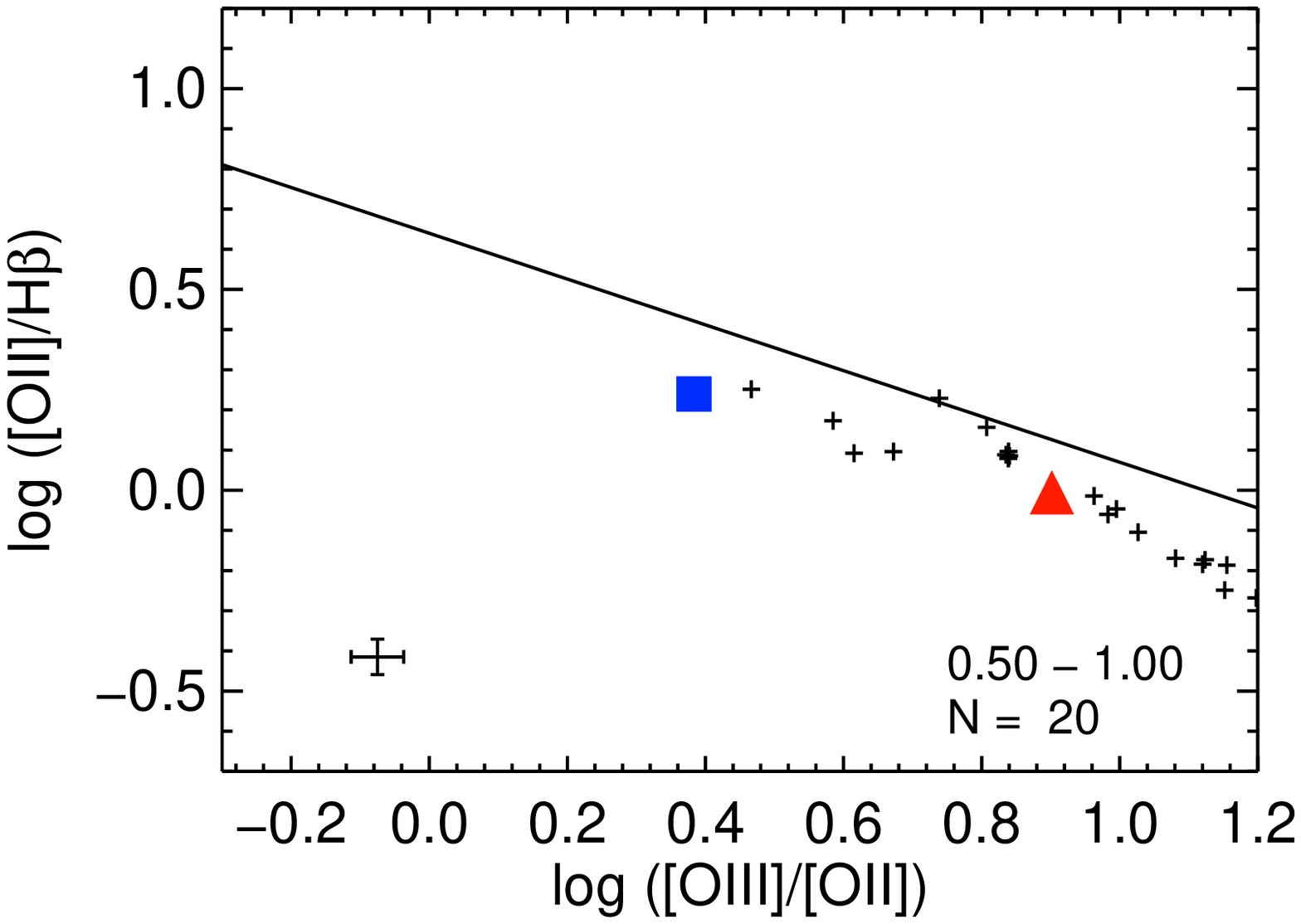}

\caption{Diagrams involving two $U$ sensitive line ratios. \emph{Upper row:} \oiiioii\ vs. \sha. \emph{Lower row:} \oiiioii\ vs. \oii/\hb.). The color/symbol code is as in Fig. \ref{diagdiag}. The number of considered data points as well as the  L(\hb)/L(\hb)$_{max}$ range are indicated in the right lower corner of the individual diagrams. The locus of equal estimated $U$ according to the relations proposed by \citet{dia00} is indicated with a black line.
\label{ucorrelas}}
\end{figure*}  

Diagnostic diagrams, where different areas of a given diagram are occupied by gas excited via different mechanisms, have been widely used to study the ionization conditions of the ISM. 
In the optical spectral range, the most popular are probably the BPT diagrams, first proposed by \citet{bal81} and later reviewed by \citet{vei87}.
%
Their wide use to study the ionization conditions in star-forming and starburst galaxies \citep[e.g.][]{alo10} is due to the fact that they involve emission lines that are relatively strong and line ratios that have (almost) no dependence on the extinction.

With the present 2D unbiased mapping, one can determine these line ratios  \emph{locally}.
In this way, it is possible to evaluate their dependence on the position within the GHIIR and relative surface brightness of the area under study. Moreover, one can make a comparison between integrated and local values.
%

We present the maps for the three available line ratios involved in the BPT diagrams  - namely \nha, \sha, and \ohb\ - in Fig. \ref{mapasionizacion}.
These maps show that the ionization structure in NGC~588 is complex.
The \nha\ and \sha\ maps present a rather similar structure. In both cases, the minimum is located neither at the peak of emission in \hb\ (i.e. ionized gas) nor at the one for the continuum (i.e. stars) but in the middle point between them. Then, line ratios increase outwards, following the ring structure of the region.

The \ohb\ map is roughly similar to the \nha\ and \sha\ maps  but its value varies in the opposite sense. Moreover, there are two main differences. The peak of \ohb\ (maximum for this line ratio) is broader than the peak for \nha\ and \sha\ (minimum for these ones). More relevant, there is an area of $\sim5^{\prime\prime}\times13^{\prime\prime}$ centred at $\sim[7\farcs0,-8\farcs0]$ of elevated \ohb\ values that do not show specially low \nha\ and \sha.

In order to better assess how the ionization conditions change in different parts of NGC~588, we divided our data in three luminosity bins, which sample the low, medium and high surface brightness areas of this GHIIR. In Fig. \ref{diagdiag}, we present the position of each individual spaxel in the BPT diagnostic diagrams together with the borders that separate \textsc{H\,ii} region-like ionization from ionization by other mechanisms according to several authors
\citep{vei87,kew01a,kau03,sta06}. As expected, all line ratios are
within the typical values expected for an \textsc{H\,ii} region-like ionization.

However, there are differences between the diagrams associated with the different luminosity bins.
%
Firstly, the range of observed values varies from $\sim$0.3 to $\sim$0.6~dex for \ohb, from $\sim$0.6 to $\sim$1.1~dex for \nha\ and from $\sim$0.5 to $\sim$1.1~dex for \sha, with larger ranges in those areas with lower surface brightness
Secondly, as indicated by the position of the red triangles, higher \nha\ and \sha\ ratios and lower \ohb\ ratios are detected in the areas of lower surface brightness.
%
This implies that the degree of ionization gets smaller with increasing distance from the ionizing source.

How do these results compare with our findings for NGC~595? In general, the tendencies in the differences between the integrated values and those for the individual spaxels in NGC~588 are similar to those found in NGC~595. However, the range of observed values in NGC~588 is smaller. This is seen in \sha\ and specially in \ohb, where the range for NGC~595 is twice as large and can be understood in terms of the different mapped area. While in NGC~588, we map just up to the border of the ring, in NGC~595, we were able to go into the very low surface brightness component, further away from the shell. 

An interesting result stands out in Fig. \ref{diagdiag} after comparing the mean line ratios (red triangles) for each bin with those measured for the integrated spectrum (blue squares): the case of NGC~588 shows that independently of the utilized line ratio, integrated values are more representative of the ionization conditions in the low surface brightness areas (L(\hb)$<25\%$L(\hb)$_{max}$),
which occupy $\sim$70\% of the region, than of those in the brightest parts. Moreover, \nha\ and \sha\ ratios derived for the integrated spectrum differ by $\sim$0.3~dex from those derived with long-slit (i.e. green diamonds and yellow inverted triangles).
Similar effects have been found in the few GHIIRs mapped up-to-date at such a level of detail \citep{pel10,rel10}.
However, when observing \textsc{H\,ii} regions with long-slits, these are usually the selected areas to be observed. Thus, this result should be taken into account when interpreting the ionization conditions in GHIIRs in distant star-forming galaxies, specially if calibrations derived from observations of local \textsc{H\,ii} regions are utilized.
As an example, at $\sim25-40$~Mpc, NGC~588 would occupy $\sim$1 arcsec$^2$ on sky, which is the typical size that nowadays IFS-based instruments can resolve under typical seeing conditions. At further distances, the situation would become even more uncertain since a typical spaxel would sample in addition some emission associated with the Diffuse Ionized Gas.

\subsubsection{Ionization parameter \label{sec_u}}

Line ratios presented in previous section are useful to study the ionization properties of a given region/galaxy. However, to have a more detailed view of the ionization structure, one should estimate how physical-chemical quantities like metallicity, relative abundances and ionization parameter vary within the region.
In this section, we will explore the behaviour of the different tracers of the ionization parameter ($U$).
%
This can be estimated from the ratio of lines of the same element that trace two different ionization states (e.g. \oiiioii). Assuming that the metallicity is known, one can also use the \ohb\ or the \sha\ line ratios. The  \sha\ is presented in Fig. \ref{mapasionizacion}, while the maps for \oiiioii\ and \oii/\hb\ appear in Fig. \ref{umap}. In all three cases the observed structure is the same: ratio values corresponding to high ionization parameters are found between the peak of emission in \hb\ and the main ionizing cluster, while 
ratio values typical of lower ionization parameters are found outwards, following the ring structure of the nebula.

This is better seen in Fig. \ref{ucorrelas} which shows a good correlation between the different tracers for the three flux bins under consideration. Also, as in the BPT diagrams, integrated line ratios are not dominated by the brightest zones of the gas but by the larger low surface brightness areas. In particular, differences in the \oiiioii\ ratio between the integrated values and those of the spaxels with high surface brightness can be of $\sim0.5$~dex on average and as high as $\sim0.8$~dex. 

Do these line ratios make consistent predictions of the ionization parameter? For the purpose of this discussion we will use the expressions provided by \citet{dia00} assuming a metallicity of 0.3~Z$_\odot$\footnote{We have employed 12 + $\log$(O/H)$_\odot$ = 8.66, from \citet{asp04} and the metallicity derived in Sec. \ref{sec_integrado}.}. Note that small variations of the metallicity (i.e. allowing for a range between 0.2 and 0.4~Z$_\odot$) would imply an offset in the estimated $\log U$ between -0.20 and 0.15~dex.
Also, in order to minimize the effect of the extinction, we utilized the \sha\ line ratio instead of the \textsc{[S\,ii]}$\lambda\lambda$6717,6731/\hb\ ratio and assumed \hb=\ha/2.86.
Fig. \ref{ucorrelas} also contains the locus of line ratios that trace the same ionization parameter and show that according to these relations, \oiiioii\ and \oii/\hb\ predict relatively consistent results while the \sha\ would correspond to smaller ionization parameters, even when considering the integrated spectrum. In any case, in the quantification of the ionization parameter we are not taken into account the 
fraction of ionizing photons leaking the \textsc{H\,ii} region which can be up to $\sim$50\% \citep{zur00,rel02}.

Both, the spatial variations of the different line ratios (see Fig. \ref{mapasionizacion}) and the  observed excess in the \sha\ ratio when compared with photoionization models (see Fig. \ref{ucorrelas}) are a direct consequence of the ionization structure of the GHIIR and constitute a nice observational counterpart to the 3D modelled structure of ionized regions.
These changes are obvious across the maps whereas the comparison of the different line ratios in Fig. \ref{ucorrelas} traces the ionization structure in the line of sight since areas of different degree of ionization are traced by different ions. Specifically, the expressions provided by \citet{dia00} were derived for the integrated spectra produced by one single ionizing star. However, the situation in a GHIIR like NGC~588, where the ionizing stars are distributed in 3D in an irregular manner, is much more complex.
In this scheme, the characteristic size of the different zones of the ionization structure will be determined by the architecture of the GHIIR (i.e. by the relative distribution of the ionizing sources). Thus the lower ionization species such as $S^+$, will delineate the more extended and common component while $O^{++}$ will be confined to different high ionization zones at the vicinity of the ionizing sources.
A model predicting the two-dimensional observable structure of the region designed to match our observations is in preparation (P\'erez-Montero et al. in prep.).


%

Using the \oiiioii\ as a baseline, we can compare our results with those for NGC~595 \citep{rel10}. NGC~588 presents higher values of \oiiioii, which would imply differences in $U$ ranging between $\sim$0.7 and $\sim$0.4~dex, being these differences larger when we are closer to the main ionizing cluster. 
At similar gas densities (as is the case for NGC~595 and NGC~588), the ionization parameter depends on the characteristics of the ionizing stars, the geometry of the region and the filling factor. Regarding the stars, the hotter these are, the higher number of ionizing photons they produce, and thus, a higher ionization parameter is locally expected. In general, the lower the metallicity of the region is and the younger the stellar population is, the larger number of hot stars is expected.
These tendencies can be seen by modelling of integrated spectra of \textsc{H\,ii} regions ionized by given stellar populations \citep[e.g.][]{lev10}.
Thus, the lower metallicity of NGC~588 and the youth of its stellar population with respect to NGC~595 can, at least partially, explain the difference between the observed \oiiioii\ line ratios. However, 2D detailed modelling also showed that the relative distribution of the ionizing sources is an important parameter \citep{erc07,jam08}: GHIIRs with more sparsely distributed ionizing sources have lower ionization parameter.
In that sense, a detailed modelling of the region will help to disentangle the relative role of geometry and filling factor (P\'erez-Montero et al. in prep.).

\begin{figure*}
\includegraphics[width=0.48\textwidth, clip=,bbllx=20, bblly=45,
  bburx=595, bbury=750]{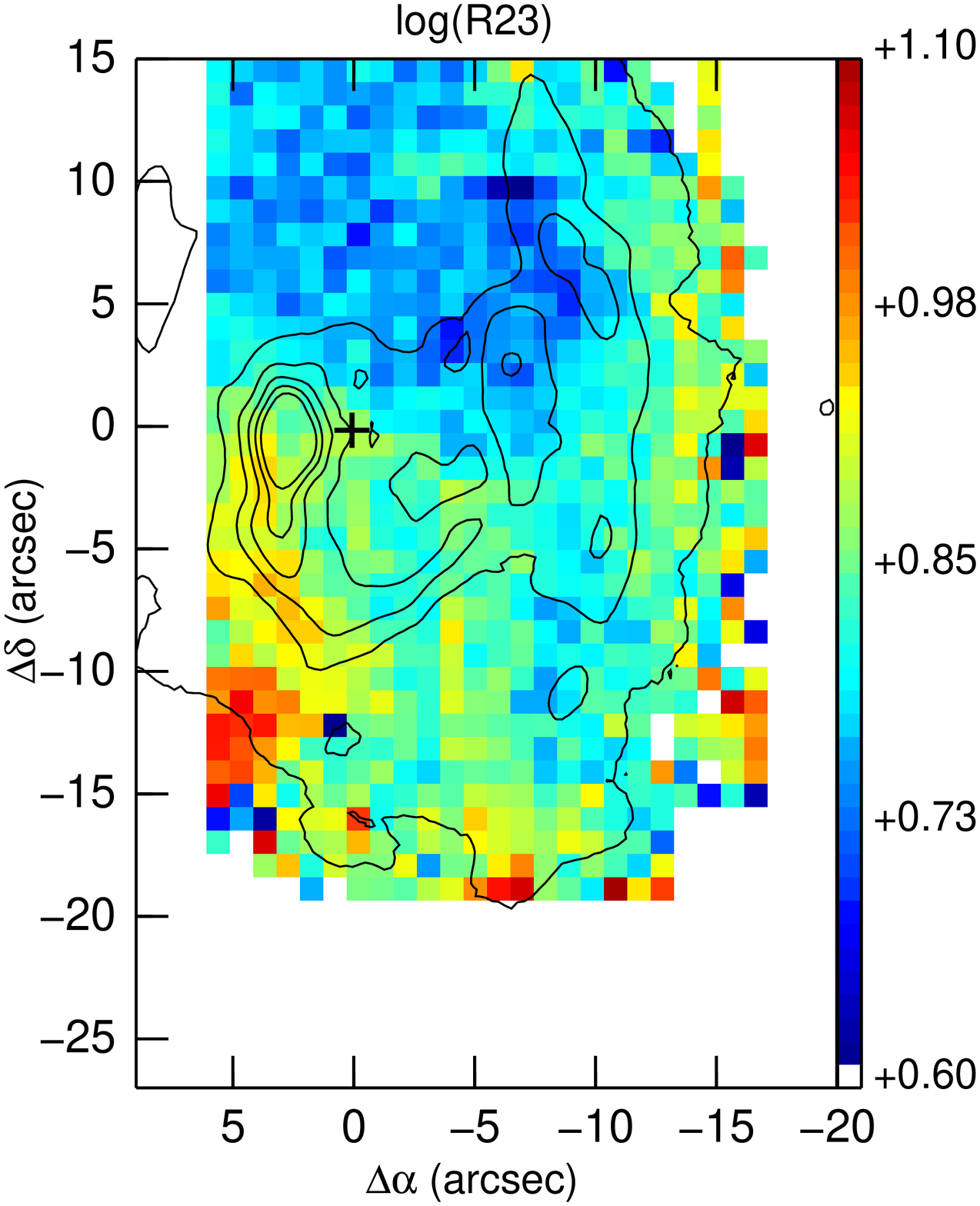}
\includegraphics[width=0.48\textwidth, clip=,bbllx=20, bblly=45,
  bburx=595, bbury=750]{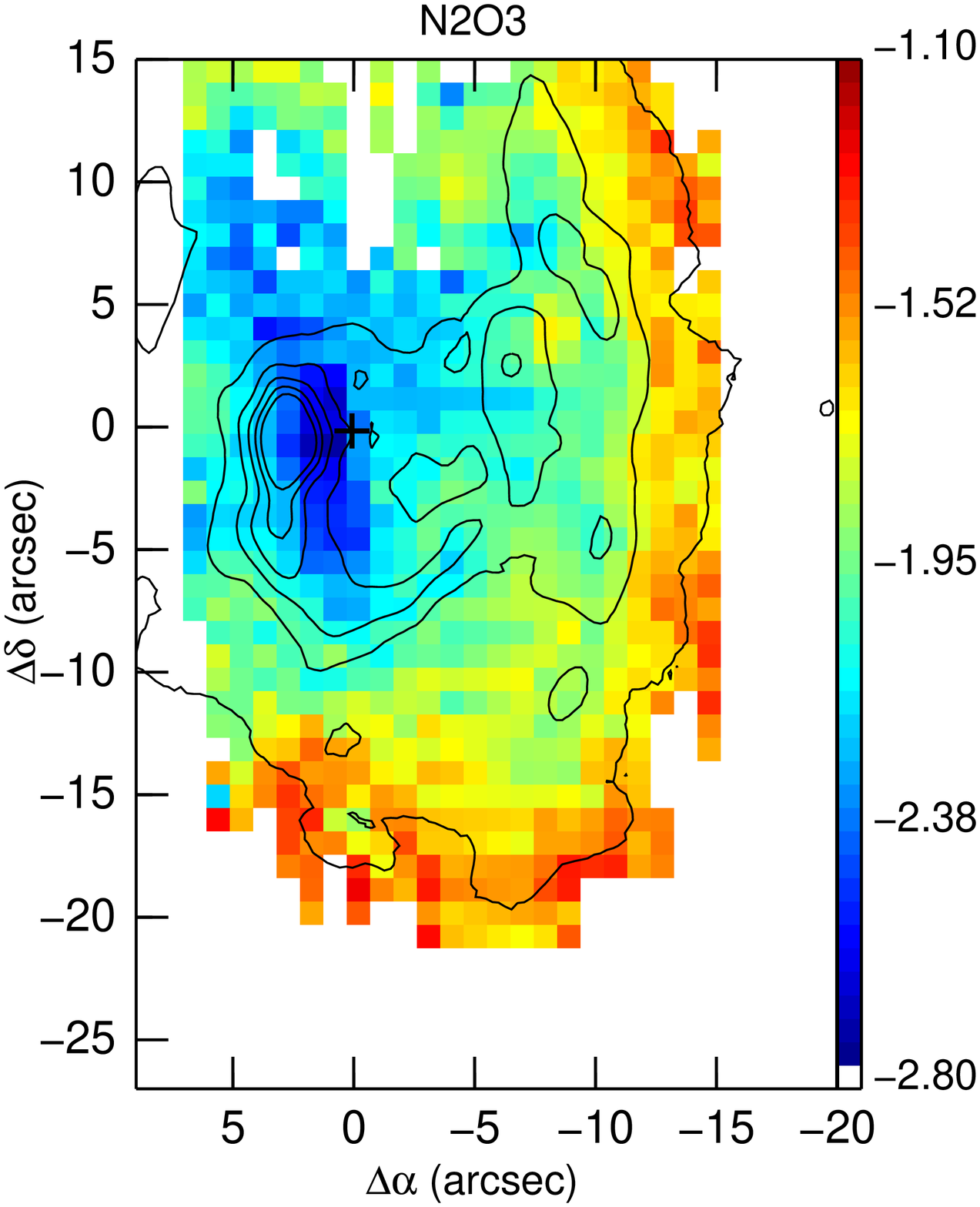}
\caption{Maps of the different observed line ratios that trace the metallicity: \emph{left:} the R23 parameter; \emph{right:} the N2O3 parameter.
 Contours correspond to the continuum subtracted
  \ha\ direct image from 
   NOAO Science Archive \citep{mas07}. The orientation is north up and
   east to the left. The main ionizing cluster, at coordinates
   RA(J2000): 1h32m45.7s, 
   Dec.(J2000): +30d38m55.1s, marks the origin of our coordinates
   system. \label{metalmap}}
\end{figure*}


\begin{figure*}
\includegraphics[width=0.33\textwidth]{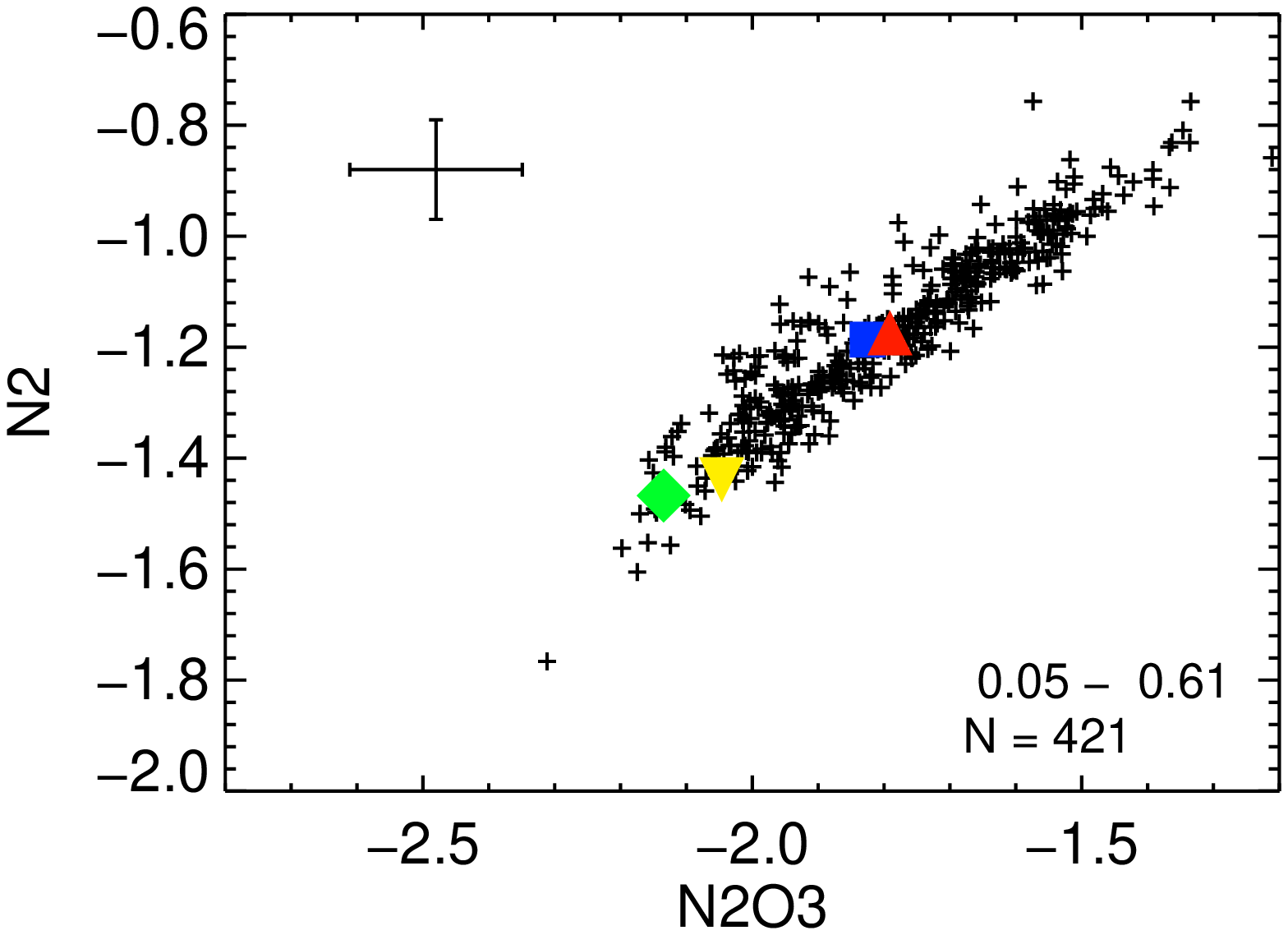}
\includegraphics[width=0.33\textwidth]{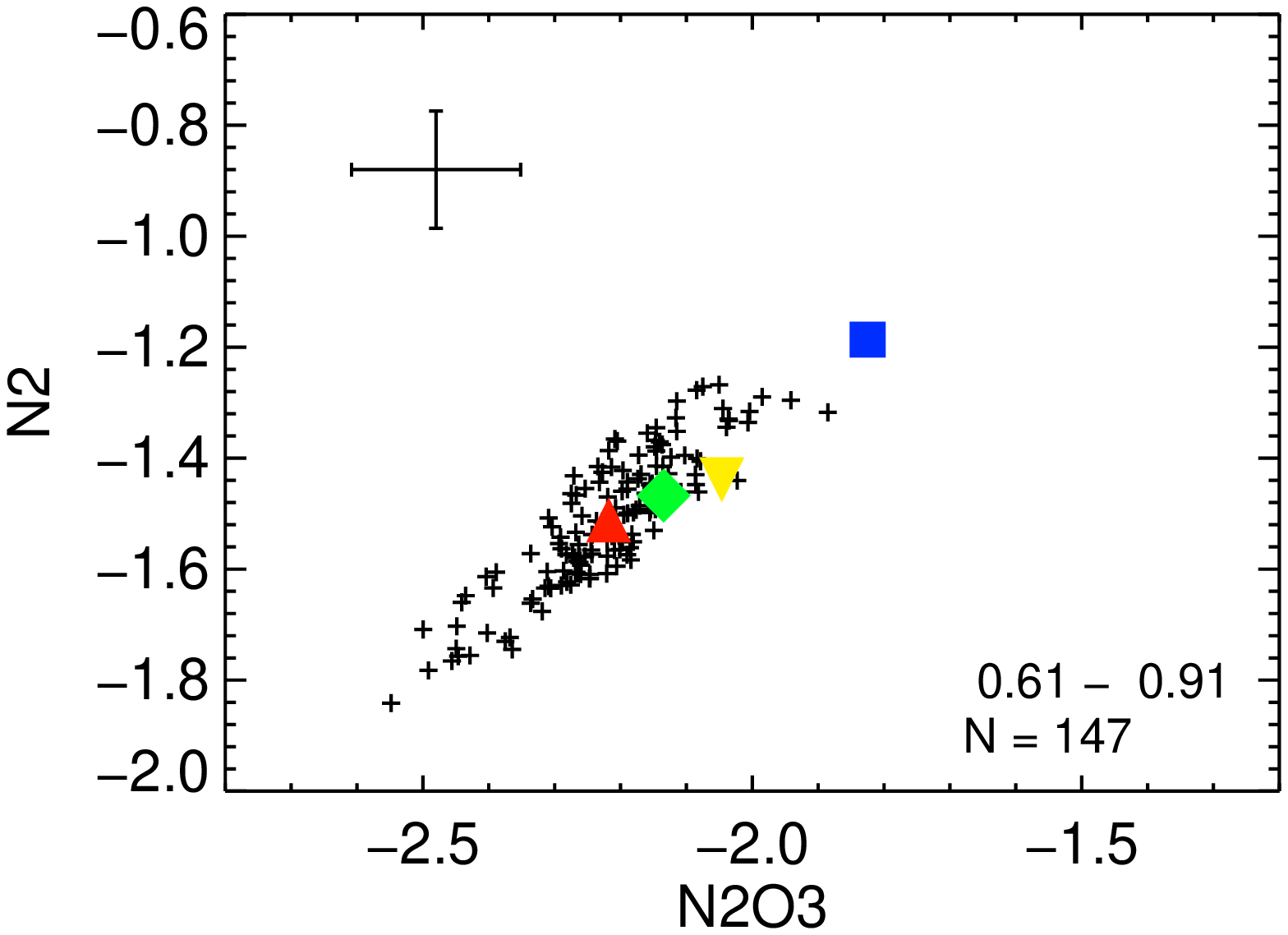}
\includegraphics[width=0.33\textwidth]{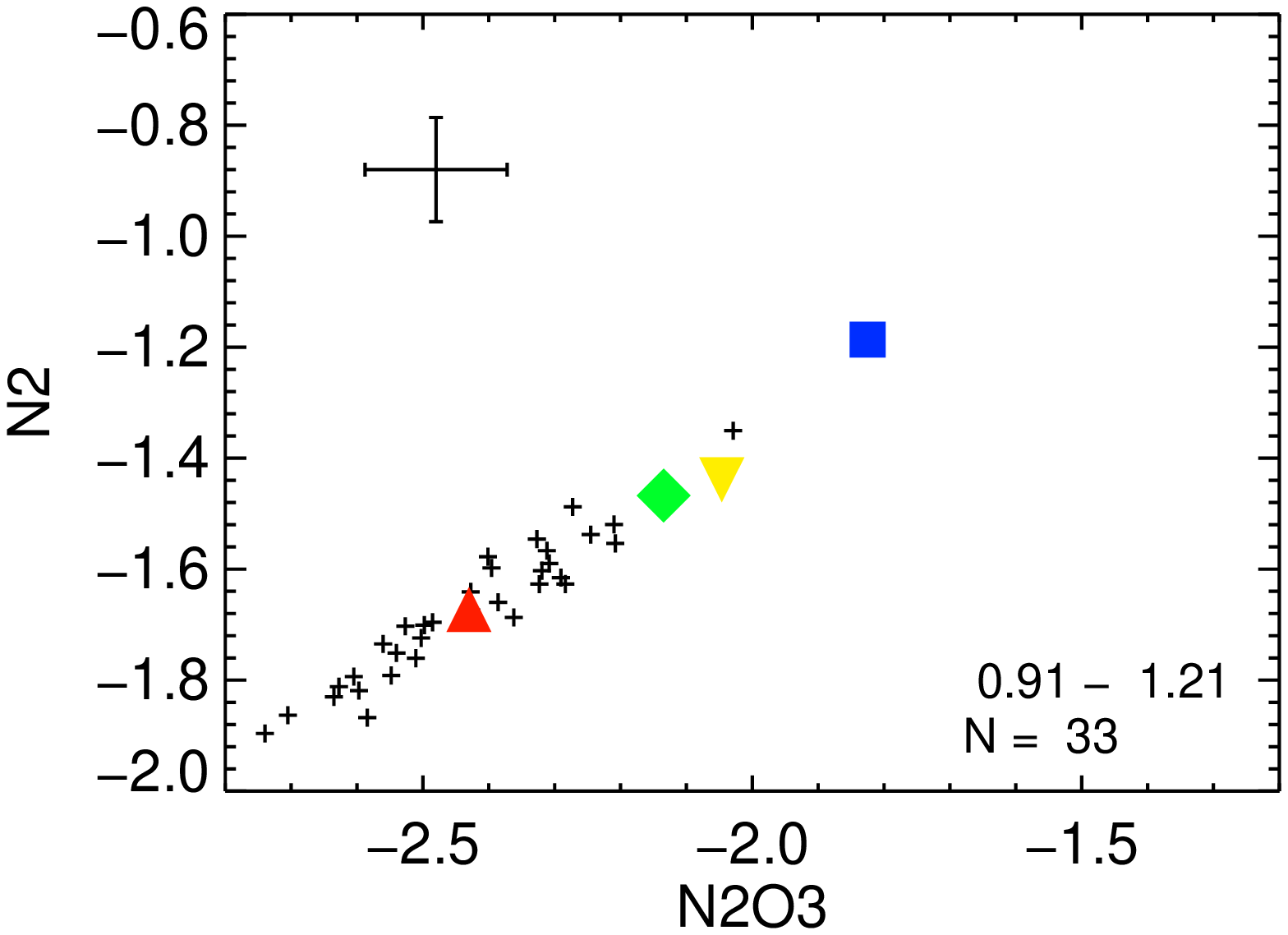}
\caption{\nha\ vs. $N2O3$.
    The color/symbol code is as in Fig. \ref{diagdiag}. The number of data points as well as the range of $\log$(\oiiioii) considered are indicated in the lower right corner of the individual diagrams. 
\label{n2havsr23}}
\end{figure*}

\begin{figure*}
\includegraphics[width=0.33\textwidth, clip=,bbllx=0, bblly=0,
  bburx=566, bbury=566]{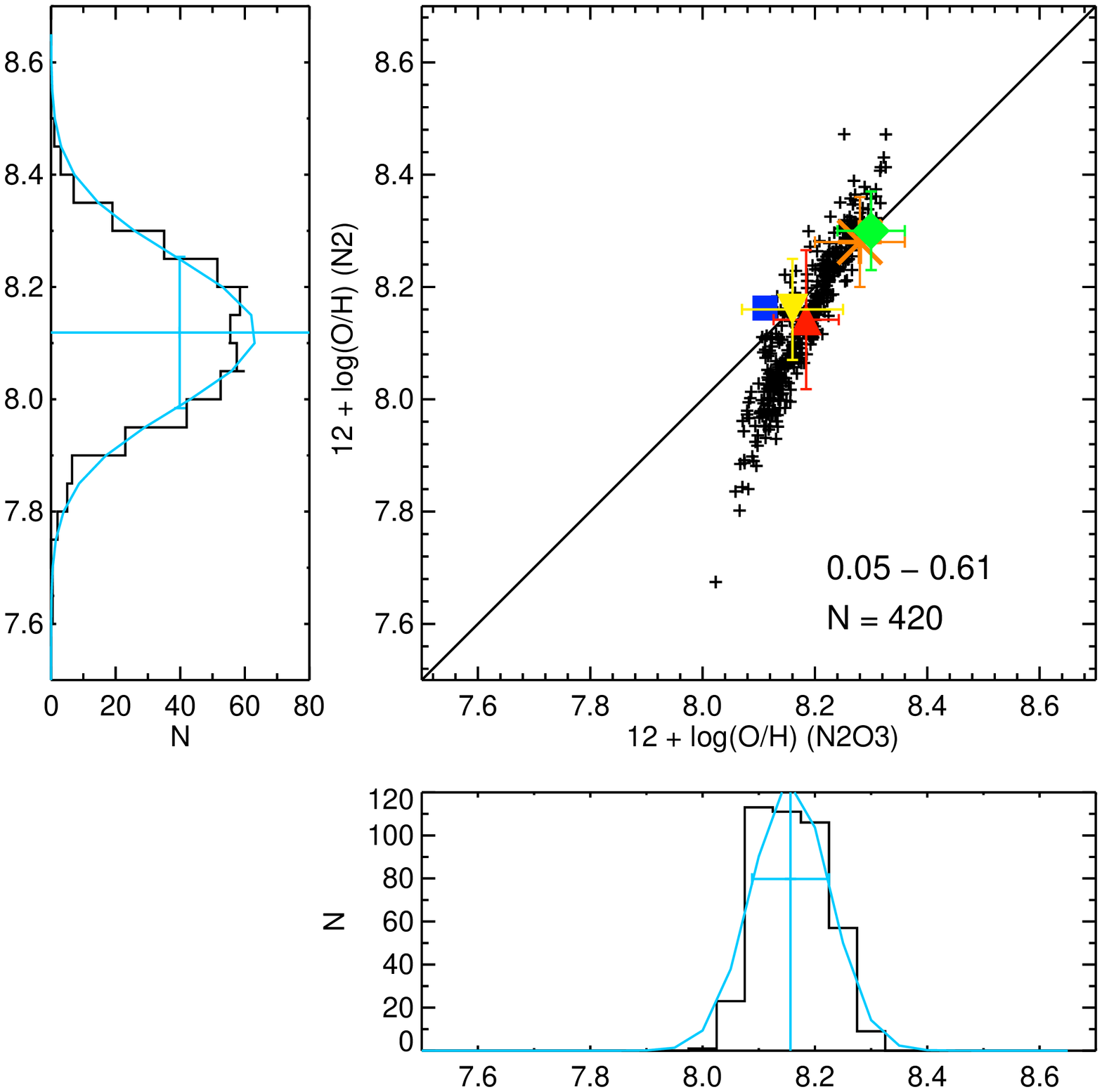}
\includegraphics[width=0.33\textwidth, clip=,bbllx=0, bblly=0,
  bburx=566, bbury=566]{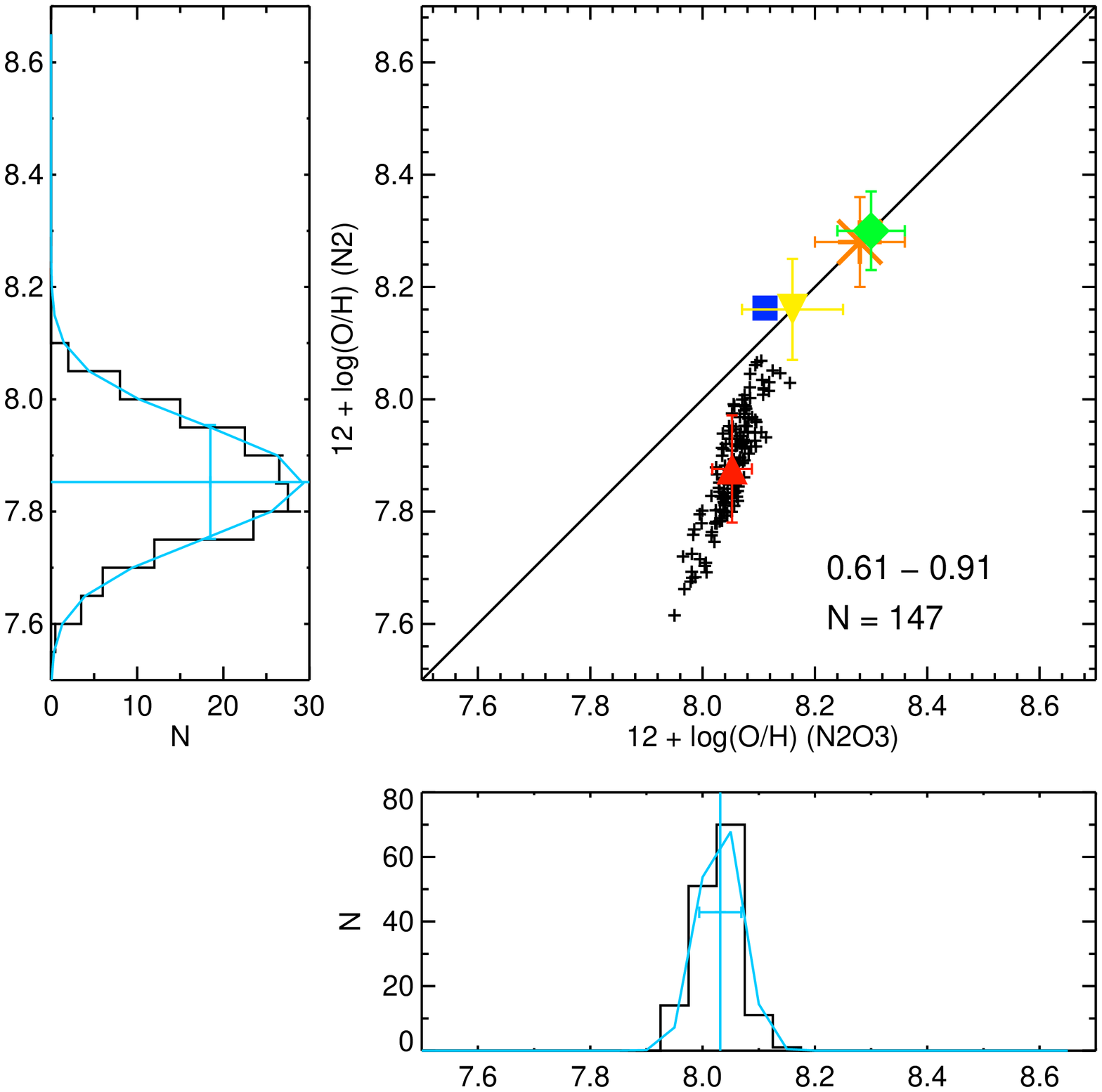}
\includegraphics[width=0.33\textwidth, clip=,bbllx=0, bblly=0,
  bburx=566, bbury=566]{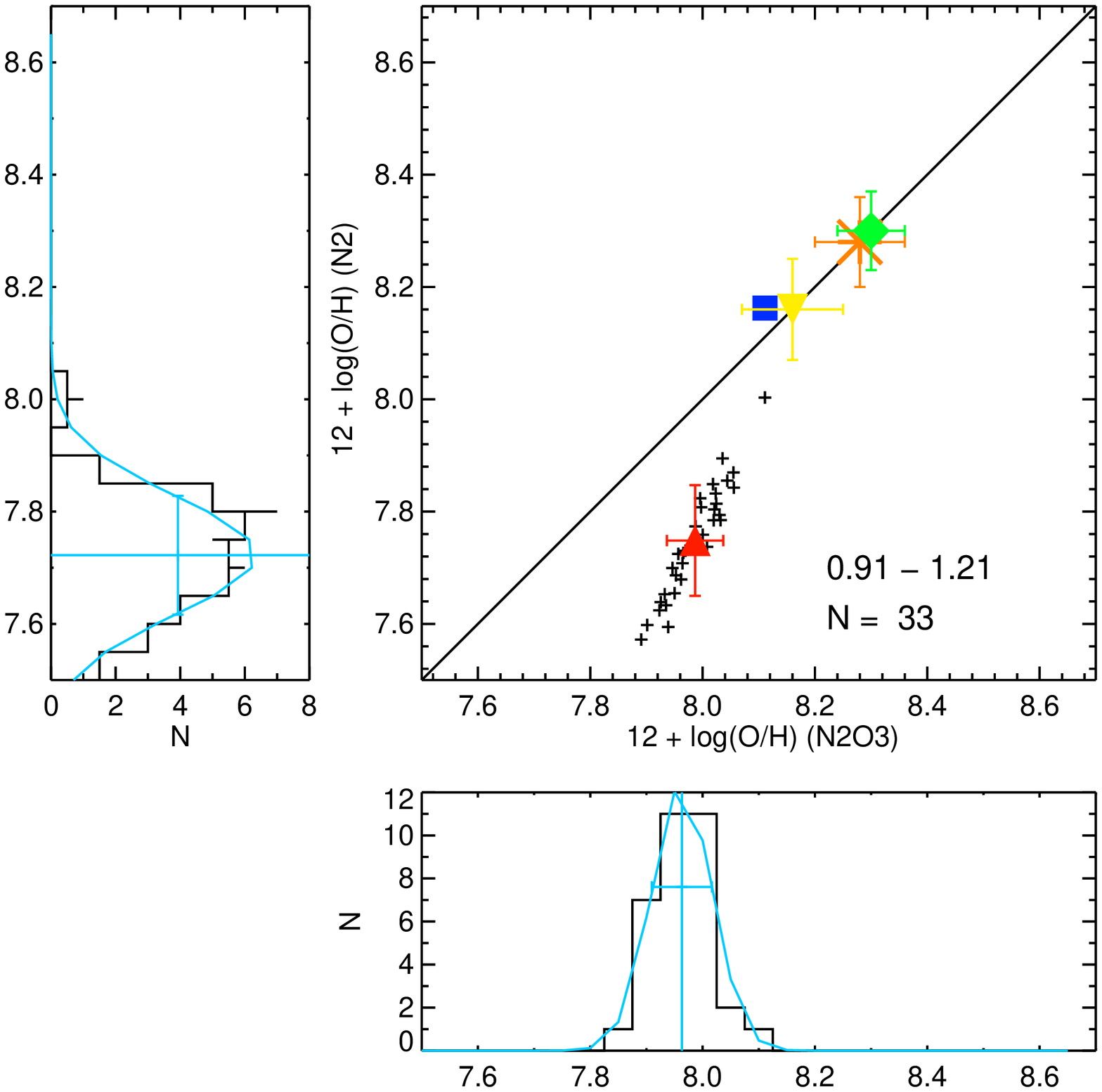}
\caption{Comparison of the metallicity estimates from N2O3 and N2.
    The color/symbol code is as in Fig. \ref{diagdiag}, however we include the metallicities from \citet{vil88} and \citet{jam05} derived using the direct method instead of those from the $N2$ and $N2O3$ line ratios. The red bars associated with the mean values indicate the standard deviation. In addition, the expected metallicity from the metallicity gradient for M~33 according to \citet{ros08} is plotted with an orange asterisk. The number of considered data points as well as the  $\log$(\oiiioii) range are indicated in the lower right corner of the individual diagrams. The locus of equal estimated metallicities is indicated with a black line. The data point distributions as well as their fit to a Gaussian are shown in the margins.
\label{zcompa}}
\end{figure*}  

\subsubsection{Metallicity tracers}

Ideally, metallicity is calculated in a direct manner. This requires the determination of the electron temperature via detection of the  - e.g. - faint \textsc{[O\,iii]}$\lambda$4363 line.
Another possibility is the use of certain combinations of strong emission lines for which empirical and/or theoretical calibrations have been established. Here, we will focus on those that can be evaluated using emission lines within our spectral range.  
In particular, we will see the spatial distribution of the metallicity tracers as well as explore their reliability as proxies of the metallicity.

The map for $N2=\log$(\nha) was presented in the upper left corner of Fig. \ref{mapasionizacion} while those for the $R23$ and the $N2O3$ parameters appear in Fig. \ref{metalmap}. None of them presents a uniform distribution. A comparison of the maps presented in Figs. \ref{metalmap} and \ref{umap} shows how the $N2O3$ and $N2$ parameters are modulated by $U$ (i.e. the higher values of $U$ correspond to the areas showing the lower $N203$ and $N2$ values).
%
The $R23$ parameter presents a different distribution due, in great measure, to the anomalous high \ohb\ values measured at the south-east of NGC~588 (see section \ref{sec_bpt}). This quadrant present typical values of $\log(R23)\gsim$0.9 very much at the turnover region of the $R23-Z$ relation while the other areas of NGC~588 have a relatively uniform distribution with typical values of $\log(R23)\sim$0.8-0.9.
As we stated in Sec. \ref{sec_integrado}, given the measured values and that the  $Z - R23$ relation is two-valued, $R23$ is not a reliable metallicity tracer for NGC~588 and will not be discussed further in this section.

The dependence of the remaining metallicity tracers on the ionization parameter is better seen in Fig. \ref{n2havsr23} where we grouped the data in three bins of degree of ionization using  \oiiioii\ tracer as baseline.
$N2O3$ varies from $-2.8$ to  $-1.3$~dex when going from high to low values of the ionization parameter while $N2$ varies from $-1.9$ to $-0.8$~dex.
Also, the comparison of the mean and integrated ratios shows how measurements of $N2O3$ and $N2$ are dominated by the areas with low $U$, which mostly correspond to zones of low surface brightness.

How these variations in the measured ratios translate into uncertainties in the metallicity determination?
To answer this question, we used the calibrations proposed by \citet{per09} for $N2$ and $N2O3$.
The relation between the predictions of the two parameters is shown in Fig. \ref{zcompa}.
The integrated spectrum gives consistent estimates of the metallicity for both tracers with a mean of $12+\log(O/H)=8.16$ (see Sec. \ref{sec_integrado}), and in agreement with those reported by \citet{jam05}. Since the utilized calibrations were derived empirically using data for integrated spectra, this is not particularly unexpected. Hereafter, this value will represent the reference metallicity for NGC~588.

The first result extracted from Fig. \ref{zcompa} is that independently of the utilized tracer, 
we derive different values for the metallicity which depend on the position in the \textsc{H\,ii} region at which the measurement is taken: the range of predicted metallicities covers $\sim$0.6~dex when using $N2O3$ and up to $\sim$1.0~dex for $N2$. Thus, $N2O3$ seems to be more reliable metallicity tracer than $N2$.

The second result is that the values of the metallicities predicted in the zones of
low excitation using the different indices show less variation between them than
those in the high excitation zones. Both indices predict rather low metallicities in the
latter, but the effect is bigger for $N2$, where the relative underestimate can reach
$\sim$0.5~dex. The ionization structure of the region gives a good explanation for this.
Where the \oiiioii\ ratios are low only gas with a relatively low degree of ionization emits along the line of sight, but where higher 
values of this ratio are found there is a significant contribution from zones with a
higher degree of ionization, closer to the ionizing stars. This leads to a deficiency in
the \textsc{[N\,ii]}  line intensity \citep[see e.g.][]{lev10} which leads to an
underestimate of the metallicity when $N2$ is used as the tracer.

\begin{figure*}
\includegraphics[width=0.48\textwidth, clip=,bbllx=20, bblly=45,
  bburx=595, bbury=750]{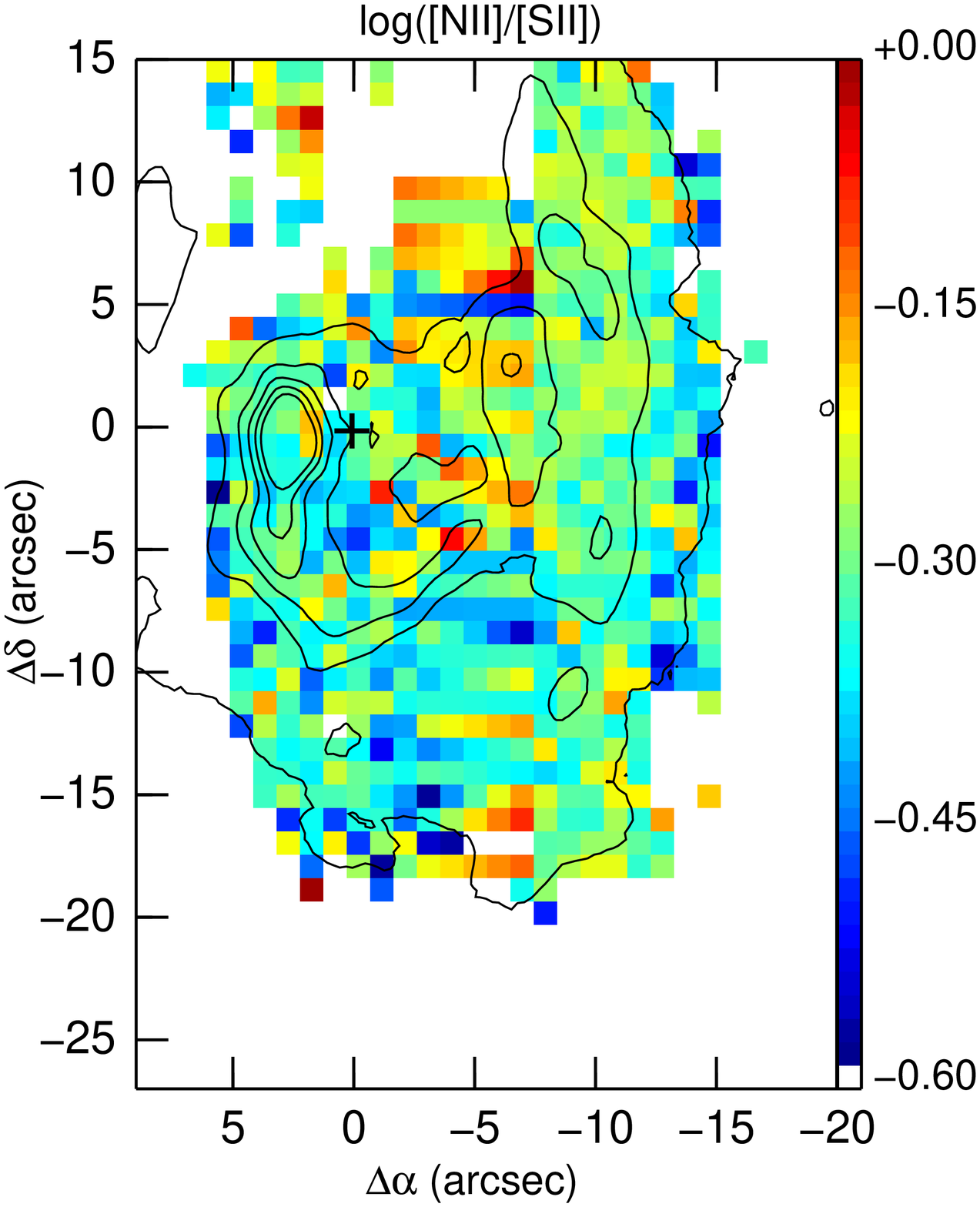}
\includegraphics[width=0.48\textwidth, clip=,bbllx=20, bblly=45,
  bburx=595, bbury=750]{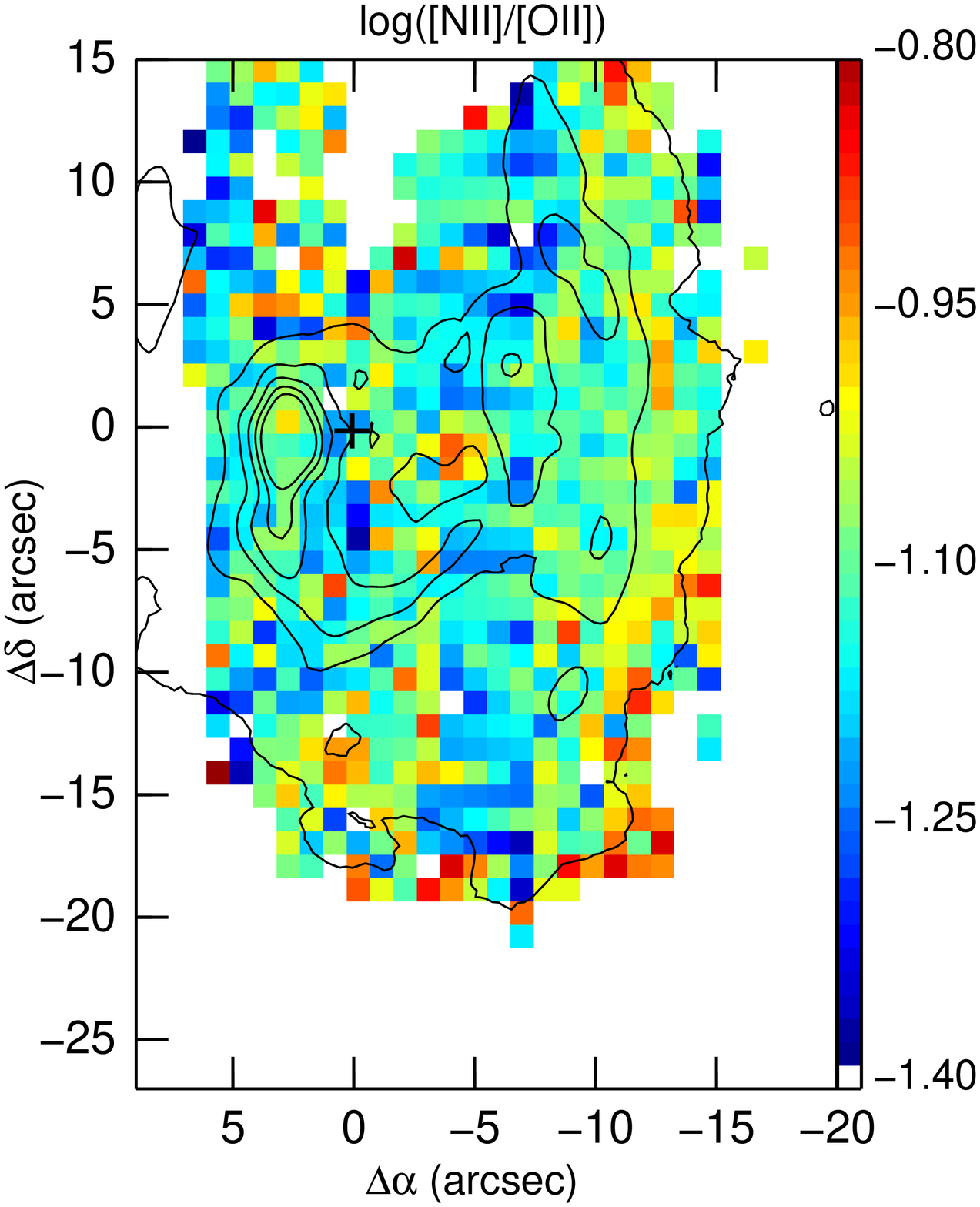}
\caption{Maps for line ratios tracing inhomogeneities in the abundances. \emph{Left:} \nii/\sii. \emph{Right:} \nii/\oii.
 Contours correspond to the continuum subtracted
  \ha\ direct image from 
   NOAO Science Archive \citep{mas07}. The orientation is north up and
   east to the left. The main ionizing cluster, at coordinates
   RA(J2000): 1h32m45.7s, 
   Dec.(J2000): +30d38m55.1s, marks the origin of our coordinates
   system. \label{abunmap}}
\end{figure*}

\subsubsection{Relative abundance tracers}

Massive stars, via their strong winds and supernovae explosions, eject all the heavy elements previously synthesized contributing to the metal-enrichment of the gas.
In particular, during their main sequence to earlier WR phases, 
material processed through the CNO cycle and enriched with nitrogen is poured
into the ISM. This contamination can be observationally spotted by the detection of areas in the ISM with higher relative abundance in nitrogen. The chances for these detections are a priori low, since the time scales involved in the dilution of this nitrogen are relatively quick \citep[e.g.][]{mon10}. However, this kind of data, that maps the ISM from the closest vicinity of the polluting stars to the more external parts of the GHIIR offer an invaluable opportunity to look for such inhomogeneities. Here we will look for areas of enhanced nitrogen abundance by means of the \nii/\sii\ and \nii/\oii\ ratios.

The maps for our two relative abundance tracers under consideration are displayed in Fig. \ref{abunmap}. No inhomogeneities are detected in any of them, not even in the vicinity of the two WR stars. The distributions for both line ratios (not shown) can be well reproduced by a single Gaussian function with mean($\pm$standard deviation) of $-0.34(\pm0.08)$ and $-1.27(\pm0.10)$ for the $\log$(\niisii) and $\log$(\niioii), respectively. Using the calibrations given by \citet{per09}, these values imply a $\log (N/O) = -1.29\pm0.10$. This is in good agreement, within the uncertainties with the nitrogen abundance reported by \citet{jam05} and higher by a factor of $\sim$1.7 to the one reported by \citet{vil88}.

\section{Summary and conclusions}
 \label{secondly}

This work presents a detailed analysis of NGC~588, a GHIIR in M33. The study is based on the joint analysis of more than 1\,000 optical spectra mapping in an un-biased manner an area of $\sim$120~pc$\times$180~pc, together with mid-IR images in the 8~$\mu$m and 24~$\mu$m \emph{Spitzer} bands. This allowed us to better understand the relationship between the different elements (i.e. stellar populations, ionized gas and dust) playing a role in the region. Our main results are:

1. Flux averaged properties of the region were derived by using the integrated spectrum. Differences in the relative line intensities from those previously reported (and usually sampling the brightest area of the region) range between the 8\% and the 50\%. The $n_e$ derived from the \textsc{[S\,ii]} line ratio is consistent with being within the low density limit, as reported in previous works. Our derived metallicity is in agreement with previous measurements as well as with the expected metallicity according the metallicity gradient for M~33.

2. This two-dimensional method reveals complex structure in the extinction distribution which had not been found with previous long-slit measurements. 
Our optical reddening map presents three maxima that correlate well with those for the dust emission at 24~$\mu$m and 8~$\mu$m. Moreover, the absorbed \ha\ luminosity map reproduces the structure observed in the 24~$\mu$m image from \emph{Spitzer} supporting this band as a good tracer of recent star formation.

3. Using the same methodology as in \citet{rel10}, we confirm the location of the two already detected WR stars in this region. This consolidates this technique as an efficient manner to find WRs. However, no new WR was detected. Since we mapped almost its whole surface, the census of the WR stellar content of NGC~588 can be considered complete.

4. We derived velocity maps from the strongest emission lines. No remarkable differences were found between them. We measured a velocity difference of $\sim50$~km~s$^{-1}$ between the areas of high and low surface brightness. 
This implies a smaller kinetic energy for the expanding shells than the one expected from the stellar population of NGC~588. The velocity field is consistent with NGC~588 being an evolved \textsc{H\,ii} region, in agreement with previous studies.

5. We measured the electron density over the whole face of the region using the \textsc{[S\,ii]}
emission line ratio and no significant variations for $n_e$ were found.
The estimated values for the \textsc{[S\,ii]} ratio are within the low 
density regime with a mean value  corresponding to $n_e$=250~cm$^{-3}$
for the assumed temperature.  This is in agreement within the uncertainties with the values
reported for specific areas. Using the EM of the region reported in the literature,
we estimate a $<n_e>$ of 5~cm$^{-3}$ for the whole GHIIR.

6. \nha\ and \sha\ present a radial structure with values increasing from the centre towards the outer parts of the region. In general, \ohb\ shows an inverted pattern. However, we detected an area towards the southeast of the region with high \ohb\ line ratios but not particularly low \nha\ and \sha\ ratios.
Line ratios involved in the BPT diagrams are typical of photoionization caused by massive stars. However, a larger range of observed line ratios is found at lower surface brightness, varying from  $\sim$0.5 to $\sim$1.2~dex for \nha, from $\sim$0.7 to $\sim$1.7~dex for \sha, and from $\sim$0.3 to $\sim$0.5~dex for \ohb. 

7. We studied the behaviour of three tracers of the ionization parameter. In all three cases the ratios corresponding to high ionization parameter are found between the peak of the emission in \hb\ and the main ionizing source decreasing when going outwards. Also, differences between the integrated and local values of the $U$ tracers can be as high as $\sim$0.8~dex, specially when using the \oiiioii\ ratio and in the high surface brightness spaxels. \ohb\ and \oiiioii\ predict similar local ionization parameters and consistent with those estimated for the integrated spectrum of an \textsc{H\,ii} region ionized by a single star. However, the \sha\ line ratio departs from these predictions reflecting the more complex ionization structure in GHIIRs.
The higher \oiiioii\ ratios in NGC~588 than those found in NGC~595, specially close to the main ionization source, can be partially explained by the lowest metallicity of NGC~588 and the youth of its stellar population. 

8. $R23$ presents a relatively uniform distribution, with variations within $\sim0.2$~dex. However, since it lies at the turnover of the $R23$-Z relation, $R23$ cannot be considered a good metallicity tracer for this particular region. Other metallicity tracers are modulated by the degree of ionization. The range of measured values for both the $N2O3$ and the $N2$ tracers is $\sim$1.5~dex. Both ratios vary with $U$ with variations of up to $\sim$0.5~dex on average when going from high to low ionization parameter. This implies an uncertainty in the determination of the metallicity associated with the position where the measurement is made of $\sim$0.6~dex and $\sim$1.0~dex for $N2O3$ and $N2$, respectively. Therefore, $N2O3$ seems to be a more accurate metallicity tracer than $N2$. 
 
9. The $N/O$ relative abundance is homogeneous over the whole face of the region according to the \niisii\ and \niioii\ line ratios with typical values of $\log (N/O) = -1.29\pm0.10$.

10. A common result emerges from our analysis of the line ratios involved in the BPT diagrams and those tracing metallicity and ionization parameter: the comparison between local and integrated values shows that the line ratios from GHIIR in galaxies at distances $\gsim$25~Mpc are likely to be dominated by the ionization conditions in their low surface brightness areas  (i.e. L(\hb)$<$25\%L(\hb)$_{max}$). This agrees with the results found by \citet{pel10} after analyzing 2D line ratio maps of 30~Doradus at the LMC.

\section*{Acknowledgments}

We would like to thank the referee, J. Beckman, for his careful reading and useful comments that have significantly improved the first submitted version of this paper. Also,
we thank C. Sandin for rapidly having modified p3d to work with
 the data of the new $4\mbox{k}\times4\mbox{k}$ CCD. 

AMI, EPM and JMV acknowledges partial funding through research 
projects AYA2007-67965-C03-03 from the Spanish PNAYA and
CSD2006-00070 1st Science with GTC of the MICINN.
MR is supported by a Marie Curie Intra European Fellowship within the 7th European Community Framework Programme.
CK, as a Humboldt Fellow, acknowledges support from the
Alexander von Humboldt Foundation, Germany.

This research draws upon data provided by The Resolved Stellar Content
of Local Group Galaxies Currently Forming Stars PI: Dr. Philip
Massey, as distributed by the NOAO Science Archive. NOAO is
operated by the Association of Universities for Research in Astronomy
(AURA), Inc., under a cooperative agreement with the
National Science Foundation.
%
This paper uses the plotting package JMAPLOT, developed by Jes\'us Ma\'{\i}z-Apell\'aniz
(available at http://dae45.iaa.csic.es:8080$\sim$jmaiz/software).



\bibliography{mybib_mnras}{}

\begin{thebibliography}{}

\bibitem[\protect\citeauthoryear{{Alonso-Herrero},
  {Garc{\'{\i}}a-Mar{\'{\i}}n}, {Monreal-Ibero}, {Colina}, {Arribas},
  {Alfonso-Garz{\'o}n} \& {Labiano}}{{Alonso-Herrero} et~al.}{2009}]{alo09}
{Alonso-Herrero} A.,  {Garc{\'{\i}}a-Mar{\'{\i}}n} M.,  {Monreal-Ibero} A.,
  {Colina} L.,  {Arribas} S.,  {Alfonso-Garz{\'o}n} J.,    {Labiano} A.,  2009,
  A\&A, 506, 1541

\bibitem[\protect\citeauthoryear{{Alonso-Herrero},
  {Garc{\'{\i}}a-Mar{\'{\i}}n}, {Rodr{\'{\i}}guez Zaur{\'{\i}}n},
  {Monreal-Ibero}, {Colina} \& {Arribas}}{{Alonso-Herrero}
  et~al.}{2010}]{alo10}
{Alonso-Herrero} A.,  {Garc{\'{\i}}a-Mar{\'{\i}}n} M.,  {Rodr{\'{\i}}guez
  Zaur{\'{\i}}n} J.,  {Monreal-Ibero} A.,  {Colina} L.,    {Arribas} S.,  2010,
  A\&A, 522, A7+

\bibitem[\protect\citeauthoryear{{Alonso-Herrero}, {Rieke}, {Rieke} \&
  {Scoville}}{{Alonso-Herrero} et~al.}{2002}]{alo02}
{Alonso-Herrero} A.,  {Rieke} G.~H.,  {Rieke} M.~J.,    {Scoville} N.~Z.,
  2002, AJ, 124, 166

\bibitem[\protect\citeauthoryear{{Asplund}, {Grevesse}, {Sauval}, {Allende
  Prieto} \& {Kiselman}}{{Asplund} et~al.}{2004}]{asp04}
{Asplund} M.,  {Grevesse} N.,  {Sauval} A.~J.,  {Allende Prieto} C.,
  {Kiselman} D.,  2004, A\&A, 417, 751

\bibitem[\protect\citeauthoryear{{Bacon}, {Accardo}, {Adjali} \& {et
  al.}}{{Bacon} et~al.}{2010}]{bac10}
{Bacon} R.,  {Accardo} M.,  {Adjali} L.,    {et al.} 2010, in SPIE Conference
  Series Vol.~7735, {The MUSE second-generation VLT instrument}

\bibitem[\protect\citeauthoryear{{Baldwin}, {Phillips}, {Terlevich} \&
  (BPT)}{{Baldwin} et~al.}{1981}]{bal81}
{Baldwin} J.~A.,  {Phillips} M.~M.,  {Terlevich} R.,    (BPT) 1981, PASP, 93, 5

\bibitem[\protect\citeauthoryear{{Bastian}, {Emsellem}, {Kissler-Patig} \&
  {Maraston}}{{Bastian} et~al.}{2006}]{bas06}
{Bastian} N.,  {Emsellem} E.,  {Kissler-Patig} M.,    {Maraston} C.,  2006,
  A\&A, 445, 471

\bibitem[\protect\citeauthoryear{{Calzetti}, {Wu}, {Hong} \& {et
  al.}}{{Calzetti} et~al.}{2010}]{cal10}
{Calzetti} D.,  {Wu} S.,  {Hong} S.,    {et al.} 2010, ApJ, 714, 1256

\bibitem[\protect\citeauthoryear{{Castellanos}, {D{\'{\i}}az} \&
  {Terlevich}}{{Castellanos} et~al.}{2002a}]{cas02b}
{Castellanos} M.,  {D{\'{\i}}az} A.~I.,    {Terlevich} E.,  2002a, MNRAS, 329,
  315

\bibitem[\protect\citeauthoryear{{Castellanos}, {D{\'{\i}}az} \&
  {Terlevich}}{{Castellanos} et~al.}{2002b}]{cas02}
{Castellanos} M.,  {D{\'{\i}}az} A.~I.,    {Terlevich} E.,  2002b, MNRAS, 337,
  540

\bibitem[\protect\citeauthoryear{{Conti} \& {Massey}}{{Conti} \&
  {Massey}}{1981}]{con81}
{Conti} P.~S.,  {Massey} P.,  1981, ApJ, 249, 471

\bibitem[\protect\citeauthoryear{{D{\'{\i}}az}, {Castellanos}, {Terlevich} \&
  {Luisa Garc{\'{\i}}a-Vargas}}{{D{\'{\i}}az} et~al.}{2000}]{dia00}
{D{\'{\i}}az} A.~I.,  {Castellanos} M.,  {Terlevich} E.,    {Luisa
  Garc{\'{\i}}a-Vargas} M.,  2000, MNRAS, 318, 462

\bibitem[\protect\citeauthoryear{{Dray}, {Dale}, {Beer}, {Napiwotzki} \&
  {King}}{{Dray} et~al.}{2005}]{dra05}
{Dray} L.~M.,  {Dale} J.~E.,  {Beer} M.~E.,  {Napiwotzki} R.,    {King} A.~R.,
  2005, MNRAS, 364, 59

\bibitem[\protect\citeauthoryear{{Drissen}, {Crowther}, {{\'U}beda} \&
  {Martin}}{{Drissen} et~al.}{2008}]{dri08}
{Drissen} L.,  {Crowther} P.~A.,  {{\'U}beda} L.,    {Martin} P.,  2008, MNRAS,
  389, 1033

\bibitem[\protect\citeauthoryear{{Ercolano}, {Bastian} \&
  {Stasi{\'n}ska}}{{Ercolano} et~al.}{2007}]{erc07}
{Ercolano} B.,  {Bastian} N.,    {Stasi{\'n}ska} G.,  2007, MNRAS, 379, 945

\bibitem[\protect\citeauthoryear{{Firpo}, {Bosch} \& {Morrell}}{{Firpo}
  et~al.}{2005}]{fir05}
{Firpo} V.,  {Bosch} G.,    {Morrell} N.,  2005, MNRAS, 356, 1357

\bibitem[\protect\citeauthoryear{{Fluks}, {Plez}, {The}, {de Winter},
  {Westerlund} \& {Steenman}}{{Fluks} et~al.}{1994}]{flu94}
{Fluks} M.~A.,  {Plez} B.,  {The} P.~S.,  {de Winter} D.,  {Westerlund} B.~E.,
    {Steenman} H.~C.,  1994, A\&As, 105, 311

\bibitem[\protect\citeauthoryear{{Freedman}, {Wilson} \& {Madore}}{{Freedman}
  et~al.}{1991}]{fre91}
{Freedman} W.~L.,  {Wilson} C.~D.,    {Madore} B.~F.,  1991, ApJ, 372, 455

\bibitem[\protect\citeauthoryear{{Garc{\'{\i}}a-Benito}, {D{\'{\i}}az},
  {H{\"a}gele}, {P{\'e}rez-Montero}, {L{\'o}pez}, {V{\'{\i}}lchez},
  {P{\'e}rez}, {Terlevich}, {Terlevich} \&
  {Rosa-Gonz{\'a}lez}}{{Garc{\'{\i}}a-Benito} et~al.}{2010}]{gar10}
{Garc{\'{\i}}a-Benito} R.,  {D{\'{\i}}az} A.,  {H{\"a}gele} G.~F.,
  {P{\'e}rez-Montero} E.,  {L{\'o}pez} J.,  {V{\'{\i}}lchez} J.~M.,
  {P{\'e}rez} E.,  {Terlevich} E.,  {Terlevich} R.,    {Rosa-Gonz{\'a}lez} D.,
  2010, MNRAS, pp 1242--+

\bibitem[\protect\citeauthoryear{{Garc{\'{\i}}a-Mar{\'{\i}}n}, {Colina},
  {Arribas} \& {Monreal-Ibero}}{{Garc{\'{\i}}a-Mar{\'{\i}}n}
  et~al.}{2009}]{gar09}
{Garc{\'{\i}}a-Mar{\'{\i}}n} M.,  {Colina} L.,  {Arribas} S.,
  {Monreal-Ibero} A.,  2009, A\&A, 505, 1319

\bibitem[\protect\citeauthoryear{{Giammanco}, {Beckman} \&
  {Cedr{\'e}s}}{{Giammanco} et~al.}{2005}]{gia05}
{Giammanco} C.,  {Beckman} J.~E.,    {Cedr{\'e}s} B.,  2005, A\&A, 438, 599

\bibitem[\protect\citeauthoryear{{Giammanco}, {Beckman}, {Zurita} \&
  {Rela{\~n}o}}{{Giammanco} et~al.}{2004}]{gia04}
{Giammanco} C.,  {Beckman} J.~E.,  {Zurita} A.,    {Rela{\~n}o} M.,  2004,
  A\&A, 424, 877

\bibitem[\protect\citeauthoryear{{Gratier}, {Braine}, {Rodriguez-Fernandez} \&
  {et al.}}{{Gratier} et~al.}{2010}]{gra10}
{Gratier} P.,  {Braine} J.,  {Rodriguez-Fernandez} N.~J.,    {et al.} 2010,
  A\&A, 522, A3+

\bibitem[\protect\citeauthoryear{{Israel}, {Hawarden}, {Geballe} \&
  {Wade}}{{Israel} et~al.}{1990}]{isr90}
{Israel} F.~P.,  {Hawarden} T.~G.,  {Geballe} T.~R.,    {Wade} R.,  1990,
  MNRAS, 242, 471

\bibitem[\protect\citeauthoryear{{James}, {Tsamis}, {Barlow}, {Westmoquette},
  {Walsh}, {Cuisinier} \& {Exter}}{{James} et~al.}{2009}]{jam09}
{James} B.~L.,  {Tsamis} Y.~G.,  {Barlow} M.~J.,  {Westmoquette} M.~S.,
  {Walsh} J.~R.,  {Cuisinier} F.,    {Exter} K.~M.,  2009, MNRAS, 398, 2

\bibitem[\protect\citeauthoryear{{Jamet} \& {Morisset}}{{Jamet} \&
  {Morisset}}{2008}]{jam08}
{Jamet} L.,  {Morisset} C.,  2008, A\&A, 482, 209

\bibitem[\protect\citeauthoryear{{Jamet}, {P{\'e}rez}, {Cervi{\~n}o},
  {Stasi{\'n}ska}, {Gonz{\'a}lez Delgado} \& {V{\'{\i}}lchez}}{{Jamet}
  et~al.}{2004}]{jam04}
{Jamet} L.,  {P{\'e}rez} E.,  {Cervi{\~n}o} M.,  {Stasi{\'n}ska} G.,
  {Gonz{\'a}lez Delgado} R.~M.,    {V{\'{\i}}lchez} J.~M.,  2004, A\&A, 426,
  399

\bibitem[\protect\citeauthoryear{{Jamet}, {Stasi{\'n}ska}, {P{\'e}rez},
  {Gonz{\'a}lez Delgado} \& {V{\'{\i}}lchez}}{{Jamet} et~al.}{2005}]{jam05}
{Jamet} L.,  {Stasi{\'n}ska} G.,  {P{\'e}rez} E.,  {Gonz{\'a}lez Delgado}
  R.~M.,    {V{\'{\i}}lchez} J.~M.,  2005, A\&A, 444, 723

\bibitem[\protect\citeauthoryear{{Kauffmann}, {Heckman}, {Tremonti},
  {Brinchmann}, {Charlot}, {White}, {Ridgway}, {Brinkmann}, {Fukugita}, {Hall},
  {Ivezi{\'c}}, {Richards} \& {Schneider}}{{Kauffmann} et~al.}{2003}]{kau03}
{Kauffmann} G.,  {Heckman} T.~M.,  {Tremonti} C.,  {Brinchmann} J.,  {Charlot}
  S.,  {White} S.~D.~M.,  {Ridgway} S.~E.,  {Brinkmann} J.,  {Fukugita} M.,
  {Hall} P.~B.,  {Ivezi{\'c}} {\v Z}.,  {Richards} G.~T.,    {Schneider} D.~P.,
   2003, MNRAS, 346, 1055

\bibitem[\protect\citeauthoryear{{Kehrig}, {V{\'{\i}}lchez}, {S{\'a}nchez},
  {Telles}, {P{\'e}rez-Montero} \& {Mart{\'{\i}}n-Gord{\'o}n}}{{Kehrig}
  et~al.}{2008}]{keh08}
{Kehrig} C.,  {V{\'{\i}}lchez} J.~M.,  {S{\'a}nchez} S.~F.,  {Telles} E.,
  {P{\'e}rez-Montero} E.,    {Mart{\'{\i}}n-Gord{\'o}n} D.,  2008, A\&A, 477,
  813

\bibitem[\protect\citeauthoryear{{Kelz}, {Verheijen}, {Roth}, {Bauer},
  {Becker}, {Paschke}, {Popow}, {S{\'a}nchez} \& {Laux}}{{Kelz}
  et~al.}{2006}]{kel06}
{Kelz} A.,  {Verheijen} M.~A.~W.,  {Roth} M.~M.,  {Bauer} S.~M.,  {Becker} T.,
  {Paschke} J.,  {Popow} E.,  {S{\'a}nchez} S.~F.,    {Laux} U.,  2006, PASP,
  118, 129

\bibitem[\protect\citeauthoryear{{Kennicutt} Jr.}{{Kennicutt}}{1984}]{ken84}
{Kennicutt} Jr. R.~C.,  1984, ApJ, 287, 116

\bibitem[\protect\citeauthoryear{{Kennicutt} Jr.}{{Kennicutt}}{1998}]{ken98}
{Kennicutt} Jr. R.~C.,  1998, ARA\&A, 36, 189

\bibitem[\protect\citeauthoryear{{Kewley}, {Dopita}, {Sutherland}, {Heisler} \&
  {Trevena}}{{Kewley} et~al.}{2001}]{kew01a}
{Kewley} L.~J.,  {Dopita} M.~A.,  {Sutherland} R.~S.,  {Heisler} C.~A.,
  {Trevena} J.,  2001, ApJ, 556, 121

\bibitem[\protect\citeauthoryear{{Kobulnicky}, {Kennicutt} Jr. \&
  {Pizagno}}{{Kobulnicky} et~al.}{1999}]{kob99}
{Kobulnicky} H.~A.,  {Kennicutt} Jr. R.~C.,    {Pizagno} J.~L.,  1999, ApJ,
  514, 544

\bibitem[\protect\citeauthoryear{{Leitherer}, {Schaerer}, {Goldader},
  {Delgado}, {Robert}, {Kune}, {de Mello}, {Devost} \& {Heckman}}{{Leitherer}
  et~al.}{1999}]{lei99}
{Leitherer} C.,  {Schaerer} D.,  {Goldader} J.~D.,  {Delgado} R.~M.~G.,
  {Robert} C.,  {Kune} D.~F.,  {de Mello} D.~F.,  {Devost} D.,    {Heckman}
  T.~M.,  1999, ApJS, 123, 3

\bibitem[\protect\citeauthoryear{{Levesque}, {Kewley} \& {Larson}}{{Levesque}
  et~al.}{2010}]{lev10}
{Levesque} E.~M.,  {Kewley} L.~J.,    {Larson} K.~L.,  2010, AJ, 139, 712

\bibitem[\protect\citeauthoryear{{L\'opez-S\'anchez}, {Mesa-Delgado},
  {L\'opez-Martin} \& {Esteban}}{{L\'opez-S\'anchez} et~al.}{2010}]{lop10}
{L\'opez-S\'anchez} A.~R.,  {Mesa-Delgado} A.,  {L\'opez-Martin} L.,
  {Esteban} C.,  2010, ArXiv e-prints/1010.1806

\bibitem[\protect\citeauthoryear{{Markwardt}}{{Markwardt}}{2009}]{mar09}
{Markwardt} C.~B.,  2009, in {D.~A.~Bohlender, D.~Durand, \& P.~Dowler} ed.,
  ASP Conference Series Vol.~411 of ASP Conference Series, {Non-linear
  Least-squares Fitting in IDL with MPFIT}.
pp 251--+

\bibitem[\protect\citeauthoryear{{Massey}, {Bianchi}, {Hutchings} \&
  {Stecher}}{{Massey} et~al.}{1996}]{mas96}
{Massey} P.,  {Bianchi} L.,  {Hutchings} J.~B.,    {Stecher} T.~P.,  1996, ApJ,
  469, 629

\bibitem[\protect\citeauthoryear{{Massey} \& {Conti}}{{Massey} \&
  {Conti}}{1983}]{mas83}
{Massey} P.,  {Conti} P.~S.,  1983, ApJ, 273, 576

\bibitem[\protect\citeauthoryear{{Massey}, {McNeill}, {Olsen}, {Hodge},
  {Blaha}, {Jacoby}, {Smith} \& {Strong}}{{Massey} et~al.}{2007}]{mas07}
{Massey} P.,  {McNeill} R.~T.,  {Olsen} K.~A.~G.,  {Hodge} P.~W.,  {Blaha} C.,
  {Jacoby} G.~H.,  {Smith} R.~C.,    {Strong} S.~B.,  2007, AJ, 134, 2474

\bibitem[\protect\citeauthoryear{{Melnick}}{{Melnick}}{1979}]{mel79}
{Melnick} J.,  1979, ApJ, 228, 112

\bibitem[\protect\citeauthoryear{{Melnick}, {Moles}, {Terlevich} \&
  {Garc\'{\i}a-Pelayo}}{{Melnick} et~al.}{1987}]{mel87}
{Melnick} J.,  {Moles} M.,  {Terlevich} R.,    {Garc\'{\i}a-Pelayo} J.,  1987,
  MNRAS, 226, 849

\bibitem[\protect\citeauthoryear{{Monreal-Ibero}, {Colina}, {Arribas} \&
  {Garc{\'{\i}}a-Mar{\'{\i}}n}}{{Monreal-Ibero} et~al.}{2007}]{mon07}
{Monreal-Ibero} A.,  {Colina} L.,  {Arribas} S.,
  {Garc{\'{\i}}a-Mar{\'{\i}}n} M.,  2007, A\&A, 472, 421

\bibitem[\protect\citeauthoryear{{Monreal-Ibero}, {V{\'{\i}}lchez}, {Walsh} \&
  {Mu{\~n}oz-Tu{\~n}{\'o}n}}{{Monreal-Ibero} et~al.}{2010}]{mon10}
{Monreal-Ibero} A.,  {V{\'{\i}}lchez} J.~M.,  {Walsh} J.~R.,
  {Mu{\~n}oz-Tu{\~n}{\'o}n} C.,  2010, A\&A, 517, A27+

\bibitem[\protect\citeauthoryear{{Mu\~noz-Tu\~non}, {Tenorio-Tagle},
  {Casta\~neda} \& {Terlevich}}{{Mu\~noz-Tu\~non} et~al.}{1996}]{mun96}
{Mu\~noz-Tu\~non} C.,  {Tenorio-Tagle} G.,  {Casta\~neda} H.~O.,    {Terlevich}
  R.,  1996, AJ, 112, 1636

\bibitem[\protect\citeauthoryear{{Osterbrock} \& {Flather}}{{Osterbrock} \&
  {Flather}}{1959}]{ost59}
{Osterbrock} D.,  {Flather} E.,  1959, ApJ, 129, 26

\bibitem[\protect\citeauthoryear{{Osterbrock} \& {Ferland}}{{Osterbrock} \&
  {Ferland}}{2006}]{ost06}
{Osterbrock} D.~E.,  {Ferland} G.~J.,  2006, {Astrophysics of gaseous nebulae
  and active galactic nuclei}

\bibitem[\protect\citeauthoryear{{Pagel}, {Edmunds}, {Blackwell}, {Chun} \&
  {Smith}}{{Pagel} et~al.}{1979}]{pag79}
{Pagel} B.~E.~J.,  {Edmunds} M.~G.,  {Blackwell} D.~E.,  {Chun} M.~S.,
  {Smith} G.,  1979, MNRAS, 189, 95

\bibitem[\protect\citeauthoryear{{Pellegrini}, {Baldwin} \&
  {Ferland}}{{Pellegrini} et~al.}{2010}]{pel10}
{Pellegrini} E.~W.,  {Baldwin} J.~A.,    {Ferland} G.~J.,  2010, ApJS, 191, 160

\bibitem[\protect\citeauthoryear{{Pellerin}}{{Pellerin}}{2006}]{pel06}
{Pellerin} A.,  2006, AJ, 131, 849

\bibitem[\protect\citeauthoryear{{P{\'e}rez-Montero} \&
  {Contini}}{{P{\'e}rez-Montero} \& {Contini}}{2009}]{per09}
{P{\'e}rez-Montero} E.,  {Contini} T.,  2009, MNRAS, 398, 949

\bibitem[\protect\citeauthoryear{{P\'erez-Montero}, {Rela\~no}, {V\'{\i}lchez}
  \& {Monreal-Ibero}}{{P\'erez-Montero} et~al.}{2010}]{per10}
{P\'erez-Montero} E.,  {Rela\~no} M.,  {V\'{\i}lchez} J.~M.,    {Monreal-Ibero}
  A.,  2010, ArXiv e-prints/1010.6102

\bibitem[\protect\citeauthoryear{{Pettini} \& {Pagel}}{{Pettini} \&
  {Pagel}}{2004}]{pet04}
{Pettini} M.,  {Pagel} B.~E.~J.,  2004, MNRAS, 348, L59

\bibitem[\protect\citeauthoryear{{Plucinsky}, {Williams}, {Long},  \& {et
  al.}}{{Plucinsky} et~al.}{2008}]{plu08}
{Plucinsky} P.~P.,  {Williams} B.,  {Long} K.~S.,     {et al.} 2008, ApJS, 174,
  366

\bibitem[\protect\citeauthoryear{{Rela{\~n}o}, {Beckman}, {Zurita}, {Rozas} \&
  {Giammanco}}{{Rela{\~n}o} et~al.}{2005}]{rel05}
{Rela{\~n}o} M.,  {Beckman} J.~E.,  {Zurita} A.,  {Rozas} M.,    {Giammanco}
  C.,  2005, A\&A, 431, 235

\bibitem[\protect\citeauthoryear{{Rela{\~n}o} \& {Kennicutt}}{{Rela{\~n}o} \&
  {Kennicutt}}{2009}]{rel09}
{Rela{\~n}o} M.,  {Kennicutt} R.~C.,  2009, ApJ, 699, 1125

\bibitem[\protect\citeauthoryear{{Rela{\~n}o}, {Monreal-Ibero},
  {V{\'{\i}}lchez} \& {Kennicutt}}{{Rela{\~n}o} et~al.}{2010}]{rel10}
{Rela{\~n}o} M.,  {Monreal-Ibero} A.,  {V{\'{\i}}lchez} J.~M.,    {Kennicutt}
  R.~C.,  2010, MNRAS, 402, 1635

\bibitem[\protect\citeauthoryear{{Rela{\~n}o}, {Peimbert} \&
  {Beckman}}{{Rela{\~n}o} et~al.}{2002}]{rel02}
{Rela{\~n}o} M.,  {Peimbert} M.,    {Beckman} J.,  2002, ApJ, 564, 704

\bibitem[\protect\citeauthoryear{{Rieke}, {Alonso-Herrero}, {Weiner},
  {P{\'e}rez-Gonz{\'a}lez}, {Blaylock}, {Donley} \& {Marcillac}}{{Rieke}
  et~al.}{2009}]{rie09}
{Rieke} G.~H.,  {Alonso-Herrero} A.,  {Weiner} B.~J.,  {P{\'e}rez-Gonz{\'a}lez}
  P.~G.,  {Blaylock} M.,  {Donley} J.~L.,    {Marcillac} D.,  2009, ApJ, 692,
  556

\bibitem[\protect\citeauthoryear{{Rieke} \& {Lebofsky}}{{Rieke} \&
  {Lebofsky}}{1985}]{rie85}
{Rieke} G.~H.,  {Lebofsky} M.~J.,  1985, ApJ, 288, 618

\bibitem[\protect\citeauthoryear{{Rosolowsky} \& {Simon}}{{Rosolowsky} \&
  {Simon}}{2008}]{ros08}
{Rosolowsky} E.,  {Simon} J.~D.,  2008, ApJ, 675, 1213

\bibitem[\protect\citeauthoryear{{Roth}, {Fechner}, {Wolter}, {Sandin}, {Kelz},
  {Bauer}, {Popow}, {Monreal-Ibero}, {Kehrig} \& {Streicher}}{{Roth}
  et~al.}{2010}]{rot10}
{Roth} M.~M.,  {Fechner} T.,  {Wolter} D.,  {Sandin} C.,  {Kelz} A.,  {Bauer}
  S.~M.,  {Popow} E.,  {Monreal-Ibero} A.,  {Kehrig} C.,    {Streicher} O.,
  2010, in Society of Photo-Optical Instrumentation Engineers (SPIE) Conference
  Series Vol.~7742 of Society of Photo-Optical Instrumentation Engineers (SPIE)
  Conference Series, {Commissioning of the CCD231 4K$\times$4K detector for
  PMAS}

\bibitem[\protect\citeauthoryear{{Roth}, {Kelz}, {Fechner}, {Hahn}, {Bauer},
  {Becker}, {B{\"o}hm}, {Christensen}, {Dionies}, {Paschke}, {Popow}, {Wolter},
  {Schmoll}, {Laux} \& {Altmann}}{{Roth} et~al.}{2005}]{rot05}
{Roth} M.~M.,  {Kelz} A.,  {Fechner} T.,  {Hahn} T.,  {Bauer} S.-M.,  {Becker}
  T.,  {B{\"o}hm} P.,  {Christensen} L.,  {Dionies} F.,  {Paschke} J.,  {Popow}
  E.,  {Wolter} D.,  {Schmoll} J.,  {Laux} U.,    {Altmann} W.,  2005, PASP,
  117, 620

\bibitem[\protect\citeauthoryear{{Rozas}, {Beckman} \& {Knapen}}{{Rozas}
  et~al.}{1996}]{roz96}
{Rozas} M.,  {Beckman} J.~E.,    {Knapen} J.~H.,  1996, A\&A, 307, 735

\bibitem[\protect\citeauthoryear{{Sabbadin}, {Rafanelli} \&
  {Bianchini}}{{Sabbadin} et~al.}{1980}]{sab80}
{Sabbadin} F.,  {Rafanelli} P.,    {Bianchini} A.,  1980, A\&AS, 39, 97

\bibitem[\protect\citeauthoryear{{S\'anchez}}{{S\'anchez}}{2006}]{san06}
{S\'anchez} S.~F.,  2006, Astronomische Nachrichten, 327, 850

\bibitem[\protect\citeauthoryear{{Sandin}, {Becker}, {Roth}, {Gerssen},
  {Monreal-Ibero}, {B{\"o}hm} \& {Weilbacher}}{{Sandin} et~al.}{2010}]{san10}
{Sandin} C.,  {Becker} T.,  {Roth} M.~M.,  {Gerssen} J.,  {Monreal-Ibero} A.,
  {B{\"o}hm} P.,    {Weilbacher} P.,  2010, A\&A, 515, A35+

\bibitem[\protect\citeauthoryear{{Schlegel}, {Finkbeiner} \&
  {Davis}}{{Schlegel} et~al.}{1998}]{sch98}
{Schlegel} D.~J.,  {Finkbeiner} D.~P.,    {Davis} M.,  1998, ApJ, 500, 525

\bibitem[\protect\citeauthoryear{{Shaw} \& {Dufour}}{{Shaw} \&
  {Dufour}}{1995}]{sha95}
{Shaw} R.~A.,  {Dufour} R.~J.,  1995, PASP, 107, 896

\bibitem[\protect\citeauthoryear{{Shields}}{{Shields}}{1990}]{shi90}
{Shields} G.~A.,  1990, ARA\&A, 28, 525

\bibitem[\protect\citeauthoryear{{Stasi{\'n}ska}, {Cid Fernandes}, {Mateus},
  {Sodr{\'e}} \& {Asari}}{{Stasi{\'n}ska} et~al.}{2006}]{sta06}
{Stasi{\'n}ska} G.,  {Cid Fernandes} R.,  {Mateus} A.,  {Sodr{\'e}} L.,
  {Asari} N.~V.,  2006, MNRAS, 371, 972

\bibitem[\protect\citeauthoryear{{Storey} \& {Hummer}}{{Storey} \&
  {Hummer}}{1995}]{sto95}
{Storey} P.~J.,  {Hummer} D.~G.,  1995, MNRAS, 272, 41

\bibitem[\protect\citeauthoryear{{{\'U}beda} \& {Drissen}}{{{\'U}beda} \&
  {Drissen}}{2009}]{ube09}
{{\'U}beda} L.,  {Drissen} L.,  2009, MNRAS, 394, 1847

\bibitem[\protect\citeauthoryear{{van den Bergh}}{{van den
  Bergh}}{2000}]{van00}
{van den Bergh} S.,  2000, {The Galaxies of the Local Group}.
Cambridge

\bibitem[\protect\citeauthoryear{{Veilleux} \& {Osterbrock}}{{Veilleux} \&
  {Osterbrock}}{1987}]{vei87}
{Veilleux} S.,  {Osterbrock} D.~E.,  1987, ApJS, 63, 295

\bibitem[\protect\citeauthoryear{{Verley}, {Corbelli}, {Giovanardi} \&
  {Hunt}}{{Verley} et~al.}{2010}]{ver10}
{Verley} S.,  {Corbelli} E.,  {Giovanardi} C.,    {Hunt} L.~K.,  2010, A\&A,
  510, A64+

\bibitem[\protect\citeauthoryear{{Viallefond} \& {Goss}}{{Viallefond} \&
  {Goss}}{1986}]{via86}
{Viallefond} F.,  {Goss} W.~M.,  1986, A\&A, 154, 357

\bibitem[\protect\citeauthoryear{{V\'{\i}lchez}, {Pagel}, {D\'{\i}az},
  {Terlevich} \& {Edmunds}}{{V\'{\i}lchez} et~al.}{1988}]{vil88}
{V\'{\i}lchez} J.~M.,  {Pagel} B.~E.~J.,  {D\'{\i}az} A.~I.,  {Terlevich} E.,
   {Edmunds} M.~G.,  1988, MNRAS, 235, 633

\bibitem[\protect\citeauthoryear{{Yang}, {Chu}, {Skillman} \&
  {Terlevich}}{{Yang} et~al.}{1996}]{yan96}
{Yang} H.,  {Chu} Y.,  {Skillman} E.~D.,    {Terlevich} R.,  1996, AJ, 112, 146

\bibitem[\protect\citeauthoryear{{Zurita}, {Rozas} \& {Beckman}}{{Zurita}
  et~al.}{2000}]{zur00}
{Zurita} A.,  {Rozas} M.,    {Beckman} J.~E.,  2000, A\&A, 363, 9

\end{thebibliography}
\bibliographystyle{./mn2e}




\end{document}